\documentclass[11pt]{article}
\newtheorem{Def}{Def.}[section]
\newtheorem{Thm}[Def]{Theorem}
\newtheorem{Prp}[Def]{Proposition}
\newtheorem{Lemma}[Def]{Lemma}
\newcommand{\Proof}{{\em{Proof: }}}
\newcommand{\QED}{\ \hfill $\FBox$ \\[1em]}

\newcommand{\spc}{\;\;\;\;\;\;\;\;\;\;}
\newcommand{\bra}{\mbox{$< \!\!$ \nolinebreak}}
\newcommand{\ket}{\mbox{\nolinebreak $>$}}
\newcommand{\C}{\:\mbox{\rm I \hspace{-1.25 em} {\bf C}}}
\newcommand{\R}{\mbox{\rm I \hspace{-.8 em} R}}
\newcommand{\1}{\mbox{\rm 1 \hspace{-1.05 em} 1}}
\newcommand{\Z}{\mbox{\rm \bf Z}}

\newcommand{\sR}{\mbox{\rm \scriptsize I \hspace{-.8 em} R}}
\newcommand{\N}{\mbox{\rm I \hspace{-.8 em} N}}

\newcommand{\Pdd}{\mbox{$\partial$ \hspace{-1.2 em} $/$}}
\newcommand{\Aslsh}{\mbox{ $\!\!A$ \hspace{-1.2 em} $/$}}
\newcommand{\slsh}{\mbox{ \hspace{-1.1 em} $/$}}

\newcommand{\FBox}{\rule{2mm}{2.25mm}}
\newcommand{\OBox}{\raisebox{.6ex}{\fbox{}}\,}
\newcommand{\Pexp}{\mbox{\rm{Pexp}}}
\newcommand{\Pe}{\mbox{\rm{Pe}}}

\begin{document}

\title{Light-Cone Expansion of the Dirac Sea in the Presence of Chiral 
and Scalar Potentials}
\author{Felix Finster\thanks{Mailing address after October 1998: 
Max-Planck-Institut f\"ur Mathematik in den Naturwissenschaften,
Inselstr.\ 22-26, 04103 Leipzig, Germany.}\\
Department of Mathematics, Harvard University}
\date{September/November 1998}
\maketitle

\begin{abstract}
We study the Dirac sea in the presence of external chiral and 
scalar/pseudoscalar potentials. In preparation, a method is developed 
for calculating the advanced and retarded Green's functions in an 
expansion around the light cone. For this, we first expand all 
Feynman diagrams and then explicitly sum up the perturbation 
series. The light-cone expansion expresses the Green's functions as 
an infinite sum of line integrals over the external potential and its 
partial derivatives.

The Dirac sea is decomposed into a causal and a non-causal contribution. 
The causal contribution has a light-cone expansion which is closely 
related to the light-cone expansion of the Green's functions; it 
describes the singular behavior of the Dirac sea in terms of 
nested line integrals along the light cone.
The non-causal contribution, on the other hand, is, to every order in
perturbation theory, a smooth function in position space.
\end{abstract}

\section{Introduction}
\setcounter{equation}{0}
In relativistic quantum mechanics with interaction, the fermionic wave 
functions $\Psi$ are solutions of a Dirac equation of the form
\begin{equation}
        (i \Pdd + {\cal{B}} - m) \:\Psi \;=\; 0 \;\;\; ,
        \label{l:01}
\end{equation}
where ${\cal{B}}$ is composed of the classical bosonic potentials.
According to the common conception, the Dirac sea of the system is
built up of all the negative-energy solutions of the Dirac
equation. We can describe it with the so-called {\em{fermionic projector}}
$\tilde{P}$ \cite{F1}. On the non-rigorous level of this introduction, 
the fermionic projector is given by the formal 
sum of the projectors on all these solutions; i.e.
\begin{equation}
        \tilde{P}(x,y) \;\stackrel{\mbox{\scriptsize{formally}}}{=}\;
        \sum \!\!\!\!\!\!\!\int_a \Psi_a(x) \: 
        \overline{\Psi_a(y)} \;\;\; ,
        \label{l:02}
\end{equation}
where the index `$a$' runs over all the quantum numbers of the 
negative-energy states. We want to analyze how the fermionic 
projector depends on the bosonic potentials in (\ref{l:01}). According 
to the decomposition (\ref{l:02}) into the individual states, this 
dependence can be regarded as a collective effect of all the fermions of
the Dirac sea moving in the external potential ${\cal{B}}$. Following Dirac's
original concept that the completely filled Dirac sea describes the vacuum, we 
can also say that we are interested in how the fermionic 
vacuum is influenced by the bosonic fields. Our aim is to describe this 
physical effect in a detailed and explicit way.

It turns out that the dependence of the fermionic projector on the 
external potential has a complicated non-local structure. In order to 
simplify the problem, we shall study $\tilde{P}(x,y)$ in an 
expansion about the light cone, which is called {\em{light-cone expansion}}.
The light cone around a space-time point $x$ consists of all points 
which can be reached from $x$ with a light ray. In flat Minkowski 
space, which we will consider here, the light cone is given by all pairs of
points $(x,y)$ of Minkowski space whose Lorentzian distance
$(y-x)^2 \equiv (y-x)_j \:(y-x)^j$ is zero.
In the simplest case of a smooth function $A(x,y)$, the 
light-cone expansion is just an expansion in powers of $(y-x)^2$, i.e.
a representation of the form
\begin{equation}
        A(x,y) \;=\; \sum_{j=0}^\infty (y-x)^{2j} \:A_j(x,y)
        \label{l:03}
\end{equation}
with smooth functions $A_j(x,y)$. Since the expansion parameter
$(y-x)^2$ vanishes on the light cone, the coefficients $A_j(x,y)$ give
approximations of $A(x,y)$ in a neighborhood of the light cone
(i.e., $A_0(x,y)$ coincides with $A(x,y)$ on the light cone,
$A_1(x,y)$ gives the first order behavior of $A(x,y)$ 
for pairs $(x,y)$ which are close to the light cone, etc.).
The important point is that the $A_j(x,y)$ are approximations of
$A(x,y)$ even for points $x, y$ which are far apart. We only need that 
the pair $(x,y)$ is close to the light cone, which is an unbounded 
hypersurface in $\R^4 \times \R^4$. In this sense, the light-cone 
expansion is a {\em{non-local}} expansion.
The major advantage over local approximation techniques
(like e.g.\ Taylor expansions in the space-time coordinates) is that
the light-cone expansion gives a much more detailed description
of the fermionic projector in position space.
Furthermore, since the light cone is the boundary of the domain of 
causal dependence, all the effects related to the causality of the 
Dirac equation occur near the light cone. Thus the light-cone expansion describes
the fermionic projector precisely in the region which is most interesting
physically. In this paper, we will develop an efficient method for performing
the light-cone expansion of the fermionic projector.

After this simplified and very qualitative introduction, we briefly 
discuss the difficulties and methods of the more detailed study. First 
of all, it is not obvious how to characterize the ``negative-energy 
solutions'' of the Dirac equation in the case with general interaction.
In other words, one problem is to find the right quantum numbers for the
$a$-summation in (\ref{l:02}).
As explained in \cite{F1}, this problem can only be solved if the notion of 
``negative-energy states'' is given up and replaced by a causality 
principle for the Dirac sea; this gives a unique definition of
$\tilde{P}$ in terms of a power series in the external potential.
Our task is to convert this formal definition into 
explicit formulas for the fermionic projector in position space.
The basic technique is to construct solutions of the inhomogeneous 
Klein-Gordon and Dirac equations and to show that these solutions 
coincide with the contributions to the perturbation expansion of $\tilde{P}$.
For the contribution to first order in the external potential, a 
similar technique was already used in \cite{F2}, which also contains a 
general discussion of the method.
In the following, we will first generalize this technique to higher order
perturbation theory. Then we will explicitly sum up the
light-cone expansions of all Feynman diagrams, which will finally yield exact
formulas for the light-cone expansion of the fermionic projector without 
the restriction for the external potential to be (in any sense) 
``small.'' We shall use the notation, the definitions, and the results of 
\cite{F1} throughout. Since the method of \cite{F2} had to be refined 
considerably for the analysis of higher order Feynman diagrams,
we will develop the light-cone expansion from the very beginning.
Thus this paper can be considered as being independent of \cite{F2}.
Nevertheless, the more elementary approach in \cite{F2} is a preparation 
and might be helpful for the understanding.
We will use the so-called {\em{residual argument}} to deduce the 
light-cone expansion of the Dirac sea from that of the advanced and 
retarded Green's functions. This allows us to bypass the 
explicit Fourier transformations in \cite{F2}. However, the residual 
argument has its limitations; making it mathematically precise leads 
to the decomposition of the Dirac sea into a causal and a non-causal 
contribution.

In the remainder of this section, we specify our problem in mathematical
terms. Since a realistic physical system consists of several types of fermions,
we describe empty space by the {\em{fermionic projector of the vacuum}},
which was introduced in \cite{F1} as the direct sum of $f\geq1$ Dirac seas.
The {\em{chiral asymmetry matrix}} $X$ and the {\em{mass 
matrix}} $Y$ are considered as a-priori given. The reader who is only
interested in the light-cone expansion of a single Dirac sea may specialize
to $f=1$ and $X=\1=Y$. On the wave functions, we consider the indefinite
scalar product
\begin{equation}
\bra \Psi \:|\: \Phi \ket \;=\; \int_{\sR^4} \sum_{l=1}^f 
\overline{\Psi_l(x)} \:\Phi_l(x) \;d^4x
        \label{l:3s}
\end{equation}
with the adjoint spinor $\overline{\Psi}=\Psi^* \gamma^0$.
Similar to (\ref{l:01}), the interaction is described by a 
perturbation ${\cal{B}}$ of the Dirac operator. We allow ${\cal{B}}$ to be
composed of chiral and scalar/pseudoscalar potentials,
\begin{equation}
        {\cal{B}}(x) \;=\; \chi_L \:\Aslsh_R(x) \:+\: \chi_R \:\Aslsh_L(x) \:+\: 
        \Phi(x) \:+\: i \gamma^5 \:\Xi(x) \;\;\; ,
        \label{l:1}
\end{equation}
where $\chi_{L\!/\!R}=\frac{1}{2}(\1 \mp \gamma^5)$ are the chiral 
projectors and where we use a matrix notation in the Dirac sea 
index,
\begin{equation}
        A_{L\!/\!R} \;=\; (A^l_{L\!/\!R \;m})_{l,m=1,\ldots,f} 
        \;\;\;,\;\;\;\;\; \Phi \;=\; (\Phi^l_m)_{l,m=1,\ldots,f} \;\;\;,\;\;\;\;\;
        \Xi \;=\; (\Xi^l_m)_{l,m=1,\ldots,f}
        \label{l:4c}
\end{equation}
(so the potentials may be non-diagonal on the Dirac seas and thus 
describe a general interaction of all the fermions). Furthermore, the 
perturbation ${\cal{B}}$ shall be Hermitian with respect to the 
scalar product (\ref{l:3s}).
We assume the Dirac operator to be {\em{causality compatible}} with 
$X$, i.e.
\[ X^* \:(i \Pdd + {\cal{B}} - m Y) \;=\;
(i \Pdd + {\cal{B}} - m Y) \:X \;\;\; . \]
This assumption is crucial; if it was violated, unbounded line integrals
would occur in the light-cone expansion, making it impossible to carry out
the sum over all Feynman diagrams (see the calculations in \cite{F0} for more 
details).
The form of the chiral decomposition $\chi_L \Aslsh_R + \chi_R 
\Aslsh_L$ is useful because, as we will see later, the left and right 
handed components of the fermions couple to $A_L$ and $A_R$, 
respectively.
An interesting feature of our system is that, as a consequence of the 
non-diagonal form (\ref{l:4c}) of the potentials on the Dirac seas, the
potentials and the mass matrix do in general not commute with each other,
\begin{equation}
        [A_L(x), A_{L\!/\!R}(y)] \neq 0 \;\;\;,\;\;\;\;\; [A_L(x), \Phi(y)] \neq 0
        \;\;\;,\;\;\;\;\; [A_L(x), Y] \neq 0 \;\;,\ldots \;\;\;.
        \label{l:4b}
\end{equation}
Compared to the situation in \cite{F2}, the bilinear potential 
$H_{jk} \sigma^{jk}$ is missing in (\ref{l:1}). It leads to 
complications when the sum over all Feynman diagrams is carried out.
These complications are not serious, but in order to keep the expansion
reasonably simple, the bilinear potential was left out.
Furthermore, we do not consider the gravitational field. The reason is 
that the higher order contributions in the gravitational potential 
become more and more singular on the light cone. This leads to 
technical problems which we will not deal here.
Despite these simplifications, the considered ansatz for ${\cal{B}}$ includes 
arbitrary left and right handed Yang-Mills potentials and is general 
enough for a description of e.g.\ the interactions of the standard model.
The {\em{fermionic projector in the presence of external fields}},
$\tilde{P}(x,y)$, is defined via the perturbation series in 
\cite{F1}, which is a formal sum of operator products of the form
\[ \tilde{P} \;=\; \sum_{n=0}^\infty 
\sum_{\ldots}
{\mbox{const}}(n,\ldots) \;C_{n,\ldots} \:{\cal{B}}\: C_{n-1,\ldots}
\:{\cal{B}}\:\cdots\:{\cal{B}}\:C_{0,\ldots} \;\;\; , \]
where the factors $C_{j,\ldots}$ coincide either with the spectral projectors
$k$, $p$, or with the Green's function $s$ (the index `\ldots' is a 
short notation for the different configurations of these factors).

In the language of Feynman diagrams,
the perturbation series for $\tilde{P}$ only consists of tree diagrams.
These tree diagrams are all finite; this is not difficult to prove if we 
assume suitable regularity of the potential:
\begin{Lemma}
\label{l:lemma0}
Let $(C_j)$, $0 \leq j \leq n$, be a choice of operators $C_j=k, \;p$, 
or $s$. If the external potential (\ref{l:1}) is smooth and decays so fast at
infinity that the functions ${\cal{B}}(x)$, $x^i {\cal{B}}(x)$, and $x^i x^j 
{\cal{B}}(x)$ are integrable, then the operator product
\begin{equation}
        (C_n \:{\cal{B}} \:C_{n-1}\: {\cal{B}} \cdots {\cal{B}} \:C_0)(x,y)
        \label{l:2}
\end{equation}
is a well-defined tempered distribution on $\R^4 \times \R^4$.
\end{Lemma}
{\Proof}
Calculating the Fourier transform of (\ref{l:2}) gives the formal 
expression
\begin{eqnarray}
\lefteqn{M(q_2,q_1) \;:=\; \int \frac{d^4 p_1}{(2 \pi)^4}
\cdots \int \frac{d^4 p_{n-1}}{(2 \pi)^4}} \nonumber \\
&\times&\!\!\!\!\!
        C_n(q_2) \:\tilde{\cal{B}}(q_1-p_{n-1}) \:C_{n-1}(p_{n-1})\:
        \tilde{\cal{B}}(p_{n-1}-p_{n-2})
        \cdots C_1(p_1) \:\tilde{\cal{B}}(p_1-q_1) \:C_0(q_1) \; , \;\;\;\;\;\;\;\;
        \label{l:3}
\end{eqnarray}
where we consider the $C_j$ as multiplication operators in momentum 
space and where $\tilde{\cal{B}}$ denotes the Fourier transform of 
${\cal{B}}$. It is more convenient to work in momentum space because 
the operators $C_j$ then have a simpler form. We will show that
$M(q_2,q_1)$ is a well-defined tempered distribution; our Lemma then
immediately follows by transforming back to position space.

The assumptions on ${\cal{B}}$ yield 
that $\tilde{\cal{B}}$ is $C^2$ and has rapid decay at infinity, i.e.
\[ \sup_{q \in \sR^4, \;|\kappa| \leq 2} |q^{i_1} \cdots q^{i_n} \:
\partial_\kappa {\tilde{\cal{B}}}(q)| \;<\; \infty \]
for all $n$, tensor indices $i_1,\ldots, i_n$, and multi-indices 
$\kappa$ (so $\kappa=(\kappa^1,\ldots,\kappa^p)$, $|\kappa|:=p$).
As is verified explicitly in momentum space, the 
distributions $k$, $p$, and $s$ are bounded in the Schwartz norms of the test
functions involving derivatives of only first order, more precisely
\[ |C(f)| \;\leq\; {\mbox{const}}\: \|f\|_{4,1} \;\;\;\;\;
{\mbox{with $C=k,p,$ or $s$ and $f \in {\cal{S}}$,}} \]
where the Schwartz norms are as usual defined by
\[ \|f\|_{p,q} \;=\; \max_{|I| \leq p,\; |J| \leq q} \;\;\sup_{x \in \sR^4}
|x^I \:\partial_J f(x)| \;\;\; . \]
As a consequence, we can apply $k$, $p$, and $s$ even to functions with rapid
decay which are only $C^1$. Furthermore, we can form the convolution of such 
functions with $k$, $p$, or $s$; this gives continuous functions (which will no
longer have rapid decay, however). A convolution decreases the order of
differentiability of the functions by one.

We consider the combination of multiplication and convolution
\begin{equation}
        F(p_2) \;:=\; \int \frac{d^4p_1}{(2 \pi)^4} \;
        f(p_2-p_1) \:C(p_1) \:g(p_1) \;\;\; ,
        \label{l:3a}
\end{equation}
where we assume that $f \in C^2$ has rapid decay and $g \in C^1$ is 
bounded together with its first derivatives, $\|g\|_{0,1}<\infty$.
For any fixed $p_2$,
the integral in (\ref{l:3a}) is well-defined and finite because $f(p_2-.) \:g(.)$
is $C^1$ and has rapid decay. The resulting function $F$ is $C^1$ and 
bounded together with its first derivatives, more precisely
\begin{eqnarray}
\|F\|_{0,1} &\leq& {\mbox{const}} \;\|f\|_{4,2} \:\|g\|_{0,1} \;\;\; .
        \label{l:3c}
\end{eqnarray}

After these preparations, we can estimate the integrals in (\ref{l:3}) from 
the right to the left: We choose two test functions $f, g \in 
{\cal{S}}(\R^4, \C^{4f})$ and introduce the functions
\begin{eqnarray}
        F_1(p_1) &=& \int \frac{d^4q_2}{(2 \pi)^4} \; \tilde{\cal{B}}(p_1-q_1)
        \:C_0(q_1)\:g(q_1) \label{l:4z} \\
        F_j(p_j) &=& \int \frac{d^4p_{j-1}}{(2 \pi)^4} 
        \:\tilde{\cal{B}}(p_j-p_{j-1})
        \:C_{j-1}(p_{j-1}) \:F_{j-1}(p_{j-1}) \;\;, \;\;\; 1 < j \leq n
        \;. \label{l:4a}
\end{eqnarray}
The integral (\ref{l:4z}) is of the form (\ref{l:3a}) and satisfies the 
above considered assumptions on the integrand. Using the bound 
(\ref{l:3c}), we can proceed inductively in (\ref{l:4a}).
Finally, we perform the $q_2$-integration,
\[ M(f,g) \;=\; \int \frac{d^4q_2}{(2 \pi)^4} \:f(q_2) \:C_n(q_2) 
\:F_n(q_2) \;\;\; . \]
We conclude that $M$ is a linear functional on
${\cal{S}}(\R^4,\C^{4f}) \times {\cal{S}}(\R^4, \C^{4f})$, which is 
bounded in the Schwartz norm $\|.\|_{4,1}$ of the test functions.
\QED
Clearly, the existence of the perturbation expansion to every order 
does not imply the convergence of the perturbation series, and we will 
not study this problem here. Our method is to first perform the 
light-cone expansion of the individual Feynman diagrams. For the resulting
formulas, it will then be possible to sum up the perturbation series.
Since the Feynman diagrams are only defined as distributions, we must 
generalize (\ref{l:03}) in a way which allows for the possibility that
$A(x,y)$ is singular on the light cone.
\begin{Def}
\label{l:def1}
A tempered distribution $A(x,y)$ is of the order 
${\cal{O}}((y-x)^{2p})$, $p \in \Z$, if the product
\[ (y-x)^{-2p} \: A(x,y) \]
is a regular distribution (i.e. a locally integrable function).
It has the {\bf{light-cone expansion}}
\begin{equation}
        A(x,y) \;=\; \sum_{j=g}^{\infty} A^{[j]}(x,y)
        \label{l:6a}
\end{equation}
with $g \in \Z$ if the distributions $A^{[j]}(x,y)$ are of the order
${\cal{O}}((y-x)^{2j})$ and if $A$ is approximated by the partial sums
in the way that
\begin{equation}
        A(x,y) \;-\; \sum_{j=g}^p A^{[j]}(x,y) \spc
        {\mbox{is of the order}} \spc {\cal{O}}((y-x)^{2p+2})
        \label{l:6b}
\end{equation}
for all $p \geq g$.
\end{Def}
The lowest summand $A^{[g]}(x,y)$ gives the leading order of $A(x,y)$ 
on the light cone. If $A$ is singular on the light cone, $g$ will be 
negative. The light-cone expansion (\ref{l:03}) of a smooth function
is recovered as a special case by 
setting $g=0$ and $A^{[j]}(x,y) = (y-x)^{2j} \:A_j(x,y)$.
Notice that the definition of the light-cone expansion does not 
include the convergence of the infinite sum in (\ref{l:6a}), which is 
only a convenient notation for the approximation of $A(x,y)$ by the 
partial sums (\ref{l:6b}).

\section{The Light-Cone Expansion of the Green's Functions}
\setcounter{equation}{0}
\label{l:sec_2}
In this section, we shall perform the light-cone expansion for the 
advanced and retarded Green's functions $\tilde{s}^\vee$, 
$\tilde{s}^\wedge$, which are defined in \cite{F1} by the 
perturbation series
\begin{equation}
        \tilde{s}^\vee \;=\; \sum_{k=0}^\infty (-s^\vee \:{\cal{B}})^k 
        s^\vee \;\;\;,\spc \tilde{s}^\wedge \;=\; \sum_{k=0}^\infty
        (-s^\wedge \:{\cal{B}})^k s^\wedge \;\;\; .
        \label{l:9}
\end{equation}
These perturbation expansions are causal in the sense that 
$\tilde{s}^\vee(x,y)$ and $\tilde{s}^\wedge(x,y)$ only depend on the 
potential ${\cal{B}}$ in the ``diamond'' $(L^\vee_x \cap L^\wedge_y) 
\cup (L^\wedge_x \cap L^\vee_y)$, where
\begin{equation}
L^{\vee}_{x} \;=\; \left\{ y \:|\: (y-x)^{2} \geq 0 ,\;
     y^{0}-x^{0} \geq 0 \right\} \;\;,\;\;\;
   L^{\wedge}_{x} \;=\; \left\{ y \:|\: (y-x)^{2} \geq 0 ,\;
     y^{0}-x^{0} \leq 0 \right\} \label{l:n18}
\end{equation}
denote the future and past light cones around $x$, respectively.
Since this ``diamond'' is (for fixed $x$ and $y$) a bounded region of 
space-time, the assumptions of Lemma \ref{l:lemma0} on the decay of the external 
potential at infinity are not necessary in this section; it suffices 
to assume that ${\cal{B}}$ is smooth. Furthermore, the chiral 
asymmetry matrix $X$ and the causality compatibility condition will 
not be needed in this section.

In order to get a first idea of how the light-cone expansion can be 
carried out, we look at the free advanced Green's function $s^\vee_m$ of a 
single Dirac sea in position space: It is convenient to pull the Dirac
matrices out of $s^\vee_m$ by setting
\begin{equation}
        s^\vee_m(x,y) \;=\; (i \Pdd_x + m) \: S^\vee_{m^2}(x,y) \;\;\; ,
        \label{l:10}
\end{equation}
where $S^\vee_{m^2}$ is the advanced Green's function of the 
Klein-Gordon operator,
\begin{equation}
        S^\vee_{m^2}(x,y) \;=\; \lim_{0<\varepsilon \rightarrow 0} \int 
        \frac{d^4p}{(2 \pi)^4} \:\frac{1}{p^2-m^2-i \varepsilon p^0} \:
        e^{-ip(x-y)} \;\;\; .
        \label{l:11}
\end{equation}
This Fourier integral can be calculated explicitly; we expand the 
resulting Bessel function $J_1$ in a power series,
\begin{eqnarray}
S^\vee_{m^2}(x,y) &=& -\frac{1}{2 \pi} \:\delta((y-x)^2) \:
\Theta(y^0 - x^0) \nonumber \\
&&+\:\frac{m^2}{4 \pi} \:\frac{J_1(\sqrt{m^2 
\:(y-x)^2}}{\sqrt{m^2 \:(y-x)^2}} \:\Theta((y-x)^2) \:\Theta(y^0 - 
x^0) \nonumber \\
&=& -\frac{1}{2 \pi} \:\delta((y-x)^2) \:
\Theta(y^0 - x^0) \nonumber \\
&&+\:\frac{m^2}{8 \pi}
\sum_{j=0}^\infty \frac{(-1)^j}{j! \:(j+1)!} \: \frac{(m^2 (y-x)^2)^j}{4^j} \:
\Theta((y-x)^2) \:\Theta(y^0 - x^0)
\label{l:12}
\end{eqnarray}
($\Theta$ is the Heaviside function $\Theta(t)=1$ for $t \geq 0$ 
and $\Theta(t)=0$ otherwise).
This computation shows that $S^\vee_{m^2}(x,y)$ has a $\delta((y-x)^2)$-like
singularity on the light cone. Furthermore, one sees that $S^\vee_{m^2}$
is a power series in $m^2$. The important point for the following is 
that the higher order contributions in $m^2$ contain more 
factors $(y-x)^2$ and are thus of higher order on the light cone. 
More precisely,
\begin{equation}
\left( \frac{d}{dm^2} \right)^n S^\vee_{m^2 \:|m^2=0}(x,y) \spc 
{\mbox{is of the order $\;\;\;{\cal{O}}((y-x)^{2n-2})$.}}
        \label{l:24b}
\end{equation}
According to (\ref{l:10}), the Dirac Green's function is obtained by 
computing the first partial derivatives of (\ref{l:12}). Therefore 
$s^\vee_m(x,y)$ has a singularity on the light cone which is even
$\sim \delta^\prime((y-x)^2)$.
The higher order contributions in $m$ are again of increasing order on 
the light cone. This means that we can view the Taylor expansion of 
(\ref{l:10}) in $m$,
\[ s^\vee_m(x,y) \;=\; \sum_{n=0}^\infty (i \Pdd + m) \;\frac{1}{n!}
\left( \frac{d}{dm^2} \right)^n S^\vee_{m^2 \:|\: m^2=0}(x,y) \;\;\; , \]
as a light-cone expansion of the free Green's function. Our idea is to
generalize this formula to the case with interaction. More precisely, we want
to express the perturbed Green's function in the form
\begin{equation}
        \tilde{s}^\vee(x,y) \;=\; \sum_{n=0}^\infty F_n(x,y) \: \left( 
        \frac{d}{dm^2} \right)^n S^\vee_{m^2 \:|\: m^2=0}(x,y)
        \label{l:14a}
\end{equation}
with factors $F_n$ which depend on the external potential.
We will see that this method is very convenient; especially, we can in 
this way avoid working with the rather complicated explicit formula (\ref{l:12}).
Apart from giving a motivation for the desired form (\ref{l:14a}) of the
formulas of the light-cone expansion, the mass expansion (\ref{l:12}) leads
to the conjecture
that even the higher order contributions in the mass to the {\em{perturbed}}
Green's functions might be of higher order on the light cone.
If this conjecture was true, it would be a good idea to expand the
perturbation expansion for $\tilde{s}$ with respect to the parameter $m$.
Therefore our basic strategy is to first expand (\ref{l:9}) with respect to
the mass and to try to express the contributions to the resulting expansion
in a form similar to (\ref{l:14a}).

The expansion of (\ref{l:9}) with respect to $m$ gives a double 
sum over the orders in the mass parameter and in the external 
potential. It is convenient to combine these two expansions in a single 
perturbation series. For this, we first write the mass matrix and the 
scalar/pseudoscalar potentials together by setting
\begin{equation}
        Y_L(x) \;=\; Y \:-\: \frac{1}{m} \:(\Phi(x) + i \Xi(x)) \;\;\;,\spc
        Y_R(x) \;=\; Y \:-\: \frac{1}{m} \:(\Phi(x) - i \Xi(x)) \;\;\; .
        \label{l:n20}
\end{equation}
The matrices $Y_{L\!/\!R}(x)$ are called {\em{dynamic mass matrices}};
notice that $Y_L^*=Y_R$. With this notation, we can rewrite the Dirac 
operator in the form
\begin{eqnarray}
i \Pdd + {\cal{B}} - m Y  &=& i \Pdd + B \spc {\mbox{with}} \\
B &=& \chi_L \:(\Aslsh_R -m\:Y_R) \:+\: \chi_R \:(\Aslsh_L - m\:Y_L) \;\;\; .
\label{l:18a}
\end{eqnarray}
For the light-cone expansion of the Green's functions, we will always
view $B$ as the perturbation of 
the Dirac operator. This has the advantage that the free theory consists only
of zero-mass fermions; the Green's functions of the free Dirac operator
have the simple form
\begin{equation}
        s^\vee(x,y) \;=\; i \Pdd_x \:S^\vee_{m^2=0}(x,y) \;\;\;,\spc
        s^\wedge(x,y) \;=\; i \Pdd_x \:S^\wedge_{m^2=0}(x,y)
        \label{l:11a}
\end{equation}
(to be very precise, we should write $s^\vee=i \Pdd S^\vee_0 \otimes 
\1$, where $\1=(\1^l_m)_{l,m=1,\ldots,f}$ is the identity on the 
Dirac seas). The Green's functions with interaction are given by the 
perturbation series
\begin{equation}
        \tilde{s}^\vee \;=\; \sum_{k=0}^\infty (-s^\vee \:B)^k 
        s^\vee \;\;\;,\spc \tilde{s}^\wedge \;=\; \sum_{k=0}^\infty
        (-s^\wedge \:B)^k s^\wedge \;\;\; .
        \label{l:11b}
\end{equation}

The constructions of the following subsections are exactly the same for
the advanced and retarded Green's functions. In order to treat both 
cases at once, we will in the remainder of this section omit all 
superscripts `$^\vee$', `$^\wedge$'. The formulas for the advanced and 
retarded Green's functions are obtained by either adding `$^\vee$' or
`$^\wedge$' to all factors $s$, $S$.

\subsection{Inductive Light-Cone Expansion of All Feynman Diagrams}
\label{l:sec_21}
In this subsection, we explain how the individual contributions 
to the perturbation expansion (\ref{l:11b}) can be written similar to 
the right side of (\ref{l:14a}) as a sum of terms of increasing order 
on the light cone. For the mass expansion of $S_{m^2}$, we set $a=m^2$ and
use the notation
\begin{equation}
        S^{(l)} \;=\; \left( \frac{d}{da} \right)^l S_{a | a=0} \;\;\; .
        \label{l:23b}
\end{equation}
In preparation, we derive some computation rules for the $S^{(l)}$:
$S_a$ satisfies the defining equation of a Klein-Gordon Green's function
\[ (-\OBox_x - a) \:S_a(x,y) \;=\; \delta^4(x-y) \;\;\; . \]
Differentiating with respect to $a$ yields
\begin{equation}
        -\OBox_x S^{(l)}(x,y) \;=\; \delta_{l,0} \:\delta^4(x-y)
        \:+\: l \:S^{(l-1)}(x,y) \;\;\;,\spc l \geq 0 . \label{l:5}
\end{equation}
(For $l=0$, this formula does not seem to make sense because $S^{(-1)}$ 
is undefined. The expression is meaningful, however, if one keeps 
in mind that in this case the factor $l$ is zero, and thus the whole 
second summand vanishes. We will also use this convention in the following 
calculations.) Next, we differentiate the formulas for $S_a$ in momentum space,
\begin{equation}
        S_a^\vee(p)\;=\; \frac{1}{p^2-a-i \varepsilon p^0} \;\;\;,\spc
        S_a^\wedge(p)\;=\; \frac{1}{p^2-a+i \varepsilon p^0} \;\;\; ,
        \label{l:21x}
\end{equation}
with respect to both $p$ and $a$. Comparing the results gives the 
relation
\[ \frac{\partial}{\partial p^k} S_{a}(p) \;=\; -2p_k \:\frac{d}{da} 
S_a(p) \;\;\; , \]
or, after expanding in the parameter $a$,
\begin{equation}
        \frac{\partial}{\partial p^k} S^{(l)}(p) \;=\; -2 p_k
        \:S^{(l+1)}(p) \;\;\;,\spc l \geq 0 .
        \label{l:21a}
\end{equation}
This formula also determines the derivatives of $S^{(l)}$ in position space;
namely
\begin{eqnarray}
\lefteqn{ \frac{\partial}{\partial x^k} S^{(l)}(x,y) \;=\;
\int \frac{d^4p}{(2 \pi)^4} \:S^{(l)}(p) \:(-i p_k) \:e^{-ip(x-y)} }
\nonumber \\
         & \stackrel{(\ref{l:21a})}{=} & \frac{i}{2} \int \frac{d^4p}{(2 \pi)^4} \: 
         \frac{\partial}{\partial p^k}
         S^{(l-1)}(p) \; e^{-ip(x-y)}
         \;=\; -\frac{i}{2} \int \frac{d^4p}{(2 \pi)^4} \: 
         S^{(l-1)}(p) \;\frac{\partial}{\partial p^k} e^{-ip(x-y)} \nonumber \\
         &=&\frac{1}{2} \: (y-x)_k \: S^{(l-1)}(x,y) \;\;\;,\spc l \geq 1 .
        \label{l:7}
\end{eqnarray}
We iterate this relation to calculate the Laplacian,
\begin{eqnarray*}
        -\OBox_x S^{(l)}(x,y) & = & -\frac{1}{2} \:\frac{\partial}{\partial 
        x^k} \left( (y-x)^k \:S^{(l-1)}(x,y) \right) \\
        &=& 2 \:S^{(l-1)}(x,y) \:-\: \frac{1}{4} \:(y-x)^2 \:S^{(l-2)}(x,y)
        \;\;\;,\spc l \geq 2 .
\end{eqnarray*}
After comparing with (\ref{l:5}), we conclude that
\begin{equation}
        (y-x)^2 \:S^{(l)}(x,y) \;=\; -4l\: S^{(l+1)}(x,y)\;\;\;,\spc l \geq 0 .
        \label{l:22a}
\end{equation}
Finally, $S^{(l)}(x,y)$ is only a function of $(y-x)$, which 
implies that
\begin{equation}
        \frac{\partial}{\partial x^k} S^{(l)}(x,y) \;=\;
        -\frac{\partial}{\partial y^k} S^{(l)}(x,y) \;\;\;,\spc l \geq 0 .
        \label{l:20a}
\end{equation}

The following lemma gives the light-cone expansion of an operator product which
is linear in the external potential. We will later use it for the 
iterative light-cone expansion of more complicated operator products;
in this case, the potential will be a composite 
expression in $B$ and its partial derivatives. In order to 
avoid confusion then, we denote the external potential by $V$.
\begin{Lemma}{\bf{(light-cone expansion to first order)}}
\label{l:lemma1}
The operator product $S^{(l)} \:V\: S^{(r)}$ with $l, r \geq 
0$ has the light-cone expansion
\begin{eqnarray}
\lefteqn{ (S^{(l)} \:V\: S^{(r)})(x,y) } \nonumber \\
&=& \sum_{n=0}^\infty 
        \frac{1}{n!} \int_0^1 \alpha^{l} \:(1-\alpha)^{r} \:
        (\alpha - \alpha^2)^n \: (\OBox^n V)_{|\alpha y + (1-\alpha) x} \:d\alpha \;
        S^{(n+l+r+1)}(x,y) \;\;\;\;\;. \spc
        \label{l:4}
\end{eqnarray}
\end{Lemma}
{\Proof}
The method of the proof is to first compute the Laplacian of both 
sides of (\ref{l:4}). The resulting formulas will be
of similar structure, which allows us to proceed inductively.

On the left side of (\ref{l:4}), we calculate the Laplacian with the 
help of (\ref{l:5}) to
\begin{equation}
        -\OBox_x (S^{(l)} \:V\: S^{(r)})(x,y) \;=\; \delta_{l,0} 
        \:V(x) \: S^{(r)}(x,y) \:+\:
        l \: (S^{(l-1)} \:V\: S^{(r)})(x,y) \;\;\; .
        \label{l:6}
\end{equation}

The Laplacian of the integral on the right side of (\ref{l:4}) can be 
computed with (\ref{l:7}) and (\ref{l:5}),
\begin{eqnarray}
\lefteqn{-\OBox_x \int_0^1 \alpha^{l} \:(1-\alpha)^{r} \:(\alpha-\alpha^2)^n
\: (\OBox^n V)_{|\alpha y + (1-\alpha) x} \:d\alpha \;
S^{(n+l+r+1)}(x,y) } \label{l:8} \\
         & = & -\int_0^1 \alpha^{l} \:(1-\alpha)^{r+2} \:(\alpha-\alpha^2)^n \:
         (\OBox^{n+1} V)_{|\alpha y + (1-\alpha) x} \:d\alpha \;
S^{(n+l+r+1)}(x,y) \nonumber \\
&&-\int_0^1 \alpha^{l} \:(1-\alpha)^{r+1} \:(\alpha-\alpha^2)^n \:
(\partial_k \OBox^{n} V)_{|\alpha y + (1-\alpha) x}
\:d\alpha \; (y-x)^k \:
S^{(n+l+r)}(x,y) \nonumber \\
&&+(n+l+r+1) \int_0^1 \alpha^{l} \:(1-\alpha)^{r} \:(\alpha-\alpha^2)^n \:
(\OBox^{n} V)_{|\alpha y + (1-\alpha) x}
\:d\alpha \;S^{(n+l+r)}(x,y) \;\;\; . \nonumber
\end{eqnarray}
In the second summand, we rewrite the partial derivative as a
derivative with respect to $\alpha$ and integrate by parts,
\begin{eqnarray*}
\lefteqn{\int_0^1 \alpha^{l} \:(1-\alpha)^{r+1} \:(\alpha-\alpha^2)^n \:
(\partial_k \OBox^{n} V)_{|\alpha y + (1-\alpha) x}
\:d\alpha \; (y-x)^k } \\
&=& \int_0^1 \alpha^{l} \:(1-\alpha)^{r+1} \:(\alpha-\alpha^2)^n \:
\frac{d}{d\alpha}(\OBox^{n} V)_{|\alpha y + (1-\alpha) x}
\:d\alpha \\
&=& -\delta_{n,0} \:\delta_{l,0}\:V(x) \:-\: (n+l)
         \int_0^1 \alpha^{l} \:(1-\alpha)^{r+2} \:(\alpha-\alpha^2)^{n-1} \:
(\OBox^{n} V)_{|\alpha y + (1-\alpha) x} \:d\alpha \\
&&+(n+r+1) \int_0^1 \alpha^{l} \:(1-\alpha)^{r} \:(\alpha-\alpha^2)^n \:
(\OBox^{n} V)_{|\alpha y + (1-\alpha) x} \:d\alpha \\
&=& -\delta_{n,0} \:\delta_{l,0}\:V(x) \\
&&-n \int_0^1 \alpha^{l} \:(1-\alpha)^{r+2} \:(\alpha-\alpha^2)^{n-1} \:
(\OBox^{n} V)_{|\alpha y + (1-\alpha) x} \:d\alpha \\
&&+(n+l+r+1) \int_0^1 \alpha^{l} \:(1-\alpha)^{r} \:(\alpha-\alpha^2)^n \:
(\OBox^{n} V)_{|\alpha y + (1-\alpha) x} \:d\alpha \\
&&-l \int_0^1 \alpha^{l-1} \:(1-\alpha)^{r} \:(\alpha-\alpha^2)^{n} \:
(\OBox^{n} V)_{|\alpha y + (1-\alpha) x} \:d\alpha \;\;\; .
\end{eqnarray*}
We substitute back into the original equation and obtain
\begin{eqnarray*}
        (\ref{l:8}) & = & \delta_{n,0} \:\delta_{l,0} \: V(x) \: 
        S^{(r)}(x,y) \\
&&+l \int_0^1 \alpha^{l-1} \:(1-\alpha)^{r} \:(\alpha-\alpha^2)^{n} \:
(\OBox^{n} V)_{|\alpha y + (1-\alpha) x} 
\:d\alpha \; S^{(n+l+r)}(x,y) \\
&&-\int_0^1 \alpha^{l} \:(1-\alpha)^{r+2} \:(\alpha-\alpha^2)^{n} \:
(\OBox^{n+1} V)_{|\alpha y + (1-\alpha) x} 
\:d\alpha \; S^{(n+l+r+1)}(x,y) \\
&&+n\int_0^1 \alpha^{l} \:(1-\alpha)^{r+2} \:(\alpha-\alpha^2)^{n-1} \:
(\OBox^{n} V)_{|\alpha y + (1-\alpha) x} 
\:d\alpha \; S^{(n+l+r)}(x,y) \;\;\; .
\end{eqnarray*}
After dividing by $n!$ and summation over $n$, the last two summands 
are telescopic and cancel each other. Thus one gets
\begin{eqnarray}
\lefteqn{-\OBox \sum_{n=0}^\infty \frac{1}{n!} \int_0^1 
\alpha^{l} \:(1-\alpha)^{r} \:(\alpha-\alpha^2)^n \: (\OBox^n 
V)_{|\alpha y + (1-\alpha) x} \:d\alpha \; S^{(n+l+r+1)}(x,y) } 
\nonumber \\
         & = & \delta_{l, 0} \:V(x)\:S^{(r)}(x,y) \nonumber \\
         && +l \sum_{n=0}^\infty \frac{1}{n!} \int_0^1 
\alpha^{l-1} \:(1-\alpha)^{r} \:(\alpha-\alpha^2)^n \: (\OBox^n 
V)_{|\alpha y + (1-\alpha) x} \:d\alpha \; S^{(n+l+r)}(x,y) \;\;\;.
\spc \label{l:9a}
\end{eqnarray}

We now compare the formulas (\ref{l:6}) and (\ref{l:9a}) for the Laplacian of 
both sides of (\ref{l:4}). In the special case $l=0$, these formulas
coincide, and we can use a uniqueness argument for the solutions of the 
wave equation to prove (\ref{l:4}): We assume that
we consider the advanced Green's function (for the 
retarded Green's function, the argument is analogous). For given $y$,
we denote the difference of both sides of (\ref{l:4}) by $F(x)$.
Since the support of $F(x)$ is in the past light cone $x \in
L^\wedge_y$, $F$ vanishes in a neighborhood of the 
hypersurface ${\cal{H}}=\{z \in \R^4 \:|\: z^0 = y^0 + 1\}$. 
Moreover, the Laplacian of $F$ is identically zero according to 
(\ref{l:6}) and (\ref{l:9a}). We conclude that
\[      \OBox F \;=\; 0 \spc {\mbox{and}} \spc F_{|{\cal{H}}}\;=\; \partial_k 
        F_{|{\cal{H}}} \;=\; 0 \;\;\; . \]
Since the wave equation has a unique solution for given initial data 
on the Cauchy surface ${\cal{H}}$, $F$ vanishes identically.

The general case follows by induction in $l$: Suppose that 
(\ref{l:4}) holds for given $\hat{l}$ (and arbitrary $r$).
Then, according to (\ref{l:6}), (\ref{l:9a}), and the induction hypothesis,
the Laplacian of both sides of (\ref{l:4}) coincides for $l = \hat{l} + 1$.
The above uniqueness argument for the solutions of the wave equation again 
gives (\ref{l:4}).
\QED
We recall for clarity that, according to (\ref{l:24b}), the higher 
$a$-derivatives of $S_a(x,y)$ are of higher order on the light cone. 
Thus the summands in (\ref{l:4}) are of increasing order on the light 
cone, and the infinite sum makes mathematical sense in terms of Def. 
\ref{l:def1} via the approximation by the partial sums (\ref{l:6b}).

The representation (\ref{l:4}) of an operator product as an infinite 
series of line integrals has some similarity with the formal 
light-cone expansion \cite[Theorem 3.3]{F2}. In 
order to make it easier for the reader to see the connection between 
these two expansions, we briefly discuss the analogy and the 
differences: First of all, we are here considering the 
Green's functions instead of the negative-energy solutions of the 
Klein-Gordon equation. The causality of the Green's functions 
(i.e.\ ${\mbox{supp }} S(x,.) \subset L_x$) simplifies the construction 
considerably. We could use it to avoid the explicit Fourier transformations
of the proofs in \cite{F2}; furthermore, it makes the resummation of 
non-local contributions unnecessary. In this paper, the complications 
related to the non-causality of the negative-energy states will reappear 
in the light-cone expansion of the Dirac sea in Section \ref{l:sec_dirac}.
We point out that the light-cone expansion in 
\cite{F2} is more general in the way that the mass parameter $a=m^2$ need not 
be zero. This is sometimes more convenient, 
because then the mass is just a parameter of the Green's functions 
instead of occurring in the line integrals of the light-cone expansion.
With our concept of dynamic mass matrices, however, an expansion 
around $a=0$ is more appropriate.
The important generalization in (\ref{l:4}) is that the two factors $S$ on 
the left side may be derivatives with respect to $a$. On the right 
side of equation (\ref{l:4}), this is taken into account by the additional
factors $\alpha^{l} (1-\alpha)^{r}$ in the integrand and by the higher
order $(n+l+r+1)$ of the $a$-derivative of $S$.
This generalization is the basis for the following iterations.

Lemma \ref{l:lemma1} can be used for the light-cone expansion of 
more complicated operator products. To explain the method, we look at 
the simplest example of three factors $S^{(0)}$ and two 
potentials $V$, $W$,
\begin{equation}
(S^{(0)} \:V\: S^{(0)} \:W\: S^{(0)})(x,y) \;=\; \int d^4z \: 
S^{(0)}(x,z) \: V(z)\; (S^{(0)}\:W\:S^{(0)})(z,y) \;\;\; .
\label{l:28a}
\end{equation}
Having split up the operator product in this form, we can apply 
Lemma \ref{l:lemma1} to the factor $S^{(0)} W S^{(0)}$,
\[ \;=\; \sum_{n=0}^\infty \frac{1}{n!} \int d^4z \;
S^{(0)}(x,z) \left\{ V(z) \int_0^1 (\alpha-\alpha^2)^n \:
(\OBox^n W)_{|\alpha y + (1-\alpha) z} \:d\alpha \right\} S^{(n+1)}(z,y) 
\;\;\; . \]
Now we rewrite the $z$-integral as the operator product
$(S^{(0)} g_y S^{(0)})(x,y)$, where $g_y(z)$ is the function in the 
curly brackets. The $y$-dependence of $g_y$ causes no problems because we can 
view $y$ as a fixed parameter throughout the expansion. Thus we can simply
apply Lemma \ref{l:lemma1} once again and obtain
\begin{eqnarray*}
&=& \sum_{m,n=0}^\infty \frac{1}{m!\:n!} \int_0^1 d\beta 
\;(1-\beta)^{n+1} \:(\beta-\beta^2)^m \:\int_0^1 d\alpha 
\;(\alpha-\alpha^2)^n \\
&&\hspace*{1.5cm}
\times\; \OBox^m_z \left( V(z) \:(\OBox^n W)_{|\alpha y + (1-\alpha) 
z} \right)_{|z=\beta y + (1-\beta) x} \; S^{(m+n+2)}(x,y) \;\;\; .
\end{eqnarray*}
The Laplacian $\OBox^m_z$ could be further computed with the Leibniz
rule. Notice that the manipulations of the infinite sums are not 
problematic because the number of terms to every order on the light cone is
actually finite (the situation would be more difficult if we studied the 
convergence of the sum (\ref{l:6a}), but, as pointed out earlier, the 
light-cone expansion is defined merely via the partial sums).

We want to iteratively perform the light-cone expansion 
of the operator products in (\ref{l:11b}). This is not directly possible 
with the method just described, because (\ref{l:11b}) contains the Dirac 
Green's function $s$ (instead of $S$). We must think about how to deal 
with this problem. Relation (\ref{l:11a}) allows us to replace the
factors $s$ by $S$, but this gives additional partial derivatives in 
the operator product. These derivatives can be
carried out after each iteration step using the Leibniz rule and 
the differentiation rule (\ref{l:7}). In the simplest example, we have
\begin{eqnarray*}
\lefteqn{ (s^{(0)} \:V\: S^{(0)})(x,y) \;=\; (i \Pdd_x) 
(S^{(0)}\:V\:S^{(0)})(x,y) } \\
&=& i \Pdd_x \sum_{n=0}^\infty \frac{1}{n!} \int_0^1 
(\alpha-\alpha^2)^n \:(\OBox^nV)_{|\alpha y + (1-\alpha) x} \; 
S^{(n+1)}(x,y) \\
&=& i \sum_{n=0}^\infty \frac{1}{n!} \int_0^1 
(1-\alpha)\:(\alpha-\alpha^2)^n \:(\Pdd\:\OBox^nV)_{|\alpha y + (1-\alpha) x}
\; S^{(n+1)}(x,y) \\
&&+\frac{i}{2}  \:\sum_{n=0}^\infty \frac{1}{n!} \int_0^1 
(\alpha-\alpha^2)^n \:(\OBox^nV)_{|\alpha y + (1-\alpha) x} \;
(y-x)_j \gamma^j \; S^{(n)}(x,y) \;\;\; .
\end{eqnarray*}
The only problem with this method is that the partial derivatives
might hit a factor $S^{(0)}$, in which case rule (\ref{l:7}) cannot be 
applied. In order to resolve this problem, we extend our 
constructions in a way which allows us to use all previous formulas 
also in this special case. For this, we take (\ref{l:7}) as the defining
equation for $(y-x)_k \:S^{(-1)}(x,y)$,
\begin{equation}
(y-x)_k \:S^{(-1)}(x,y) \;:=\; 2 \:\frac{\partial}{\partial x^k}
        S^{(0)}(x,y) \label{l:29z}
\end{equation}
(notice that $S^{(-1)}$ itself remains undefined, only the combination
$(y-x)_k \:S^{(-1)}(x,y)$ makes mathematical sense as the partial 
derivative of the distribution $S^{(0)}$).
With this definition, the computation rule (\ref{l:22a}) becomes also valid 
for $l=-1$: According to (\ref{l:22a}), the distribution $(y-x)^2 
\:S^{(0)}(x,y)$ vanishes identically, and thus
\begin{eqnarray*}
0 &=& \OBox_x \left( (y-x)^2 \:S^{(0)}(x,y) \right) \\
&=& \left( \OBox_x (y-x)^2 \right) S^{(0)}(x,y) \:-\: 4 \:(y-x)^j 
\:\frac{\partial}{\partial x^j} S^{(0)}(x,y) \:+\: (y-x)^2 \;\OBox_x 
S^{(0)}(x,y) \;\;\; .
\end{eqnarray*}
In the last summand, we substitute (\ref{l:5}), which gives $-(y-x)^2 
\:\delta^4(x-y) = 0$. Using the identity $\OBox_x (y-x)^2 = 8$ and the
definition (\ref{l:29z}), we conclude that
\begin{equation}
        (y-x)^2 \:S^{(-1)}(x,y) \;=\; 4 \:S^{(0)}(x,y) \;\;\; .
        \label{l:25a}
\end{equation}

The following lemma extends the result of Lemma \ref{l:lemma1} to the case
when the right factor of the operator product is $S^{(-1)}$.
\begin{Lemma} {\bf{(light-cone expansion to first order for $r=-1$)}}
\label{l:lemma2}
The operator product $(S^{(l)} \:.\: S^{(-1)})$, $l \geq 0$, has the 
light-cone expansion
\begin{eqnarray}
\lefteqn{ \int d^4z \; S^{(l)}(x,z) \:V(z) \:(y-z)_k 
\:S^{(-1)}(z,y) \;=\; \sum_{n=0}^\infty \frac{1}{n!} \int_0^1 \alpha^{l}
\:(1-\alpha)^{-1} \:(\alpha-\alpha^2)^n } \nonumber \\
&& \spc\times \: \OBox^n_z \left( V(z) \: (y-z)_k 
\right)_{|z=\alpha y + (1-\alpha) x} \: d\alpha \; S^{(n+l)}(x,y) 
\;\;\;. \spc\spc \label{l:21}
\end{eqnarray}
\end{Lemma}
{\Proof}
We deduce the light-cone expansion for the left side of 
(\ref{l:21}) from (\ref{l:4}) by pulling one $y$-derivative outside,
\begin{eqnarray}
\lefteqn{ \int d^4z \:S^{(l)}(x,z) \:V(z) \:(y-z)_k\:S^{(-1)}(z,y)
\;\stackrel{(\ref{l:29z}),(\ref{l:20a})}{=}\; -2\:\frac{\partial}{\partial y^k} (S^{(l)} \:V\: 
S^{(0)})(x,y) } \nonumber \\
&=& -2\:\frac{\partial}{\partial y^k} \sum_{n=0}^\infty \frac{1}{n!}
\int_0^1 \alpha^{l}\: (\alpha-\alpha^2)^n \:(\OBox^n V)_{|\alpha 
y + (1-\alpha) x} \:d\alpha \; S^{(n+l+1)}(x,y) \nonumber \\
&\stackrel{(\ref{l:20a}),(\ref{l:7})}{=}&\sum_{n=0}^\infty \frac{1}{n!} \int_0^1
\alpha^{l}\: (\alpha-\alpha^2)^n \: (\OBox^n V)_{|\alpha y + (1-\alpha) x}
\:d\alpha \:(y-x)_k \; S^{(n+l)}(x,y) \label{l:30a} \\
&&-2\sum_{n=0}^\infty \frac{1}{n!}
\int_0^1 \alpha^{l+1}\: (\alpha-\alpha^2)^n \:(\partial_k \OBox^n V)_{|\alpha 
y + (1-\alpha) x} \:d\alpha \; S^{(n+l+1)}(x,y) \;\;\; . \spc \label{l:22}
\end{eqnarray}
In (\ref{l:22}), we substitute the identity
\begin{equation}
        \partial_k \OBox^n V(z) \;=\; -\frac{1}{2 (n+1)} \:\OBox^{n+1}_z
        \left(V(z) \:(y-z)_k \right) \:+\: \frac{1}{2 (n+1)} \:\left(\OBox^{n+1}_z
        V(z) \right) \:(y-z)_k \;\;\; .
        \label{l:33}
\end{equation}
The contribution of the second summand in (\ref{l:33}) cancels all the
summands $n=1,2,\ldots$ of (\ref{l:30a}), and we get
\begin{eqnarray*}
\lefteqn{ \int d^4z \:S^{(l)}(x,z) \:V(z) \:(y-z)_k\:S^{(-1)}(z,y)
\;=\; \int_0^1 \alpha^{l}\: V_{|\alpha y + (1-\alpha) x} \:d\alpha
\:(y-x)_k \;S^{(l)}(x,y) } \\
&&\hspace*{-.5cm} +\sum_{n=0}^\infty \frac{1}{(n+1)!} \int_0^1
\alpha^{l+1}\:(\alpha-\alpha^2)^n \:\OBox^{n+1}_z\left(V(z) \:(y-z)_k
\right)_{|z=\alpha y + (1-\alpha) x} \:d\alpha \;S^{(n+1+l)}(x,y) \;\;\; .
\end{eqnarray*}
After shifting the summation index, this coincides with (\ref{l:21}).
\QED
Notice that the pole of the factor $(1-\alpha)^{-1}$ for $\alpha=1$ 
does not lead to problems in (\ref{l:21}): In the case $n=0$, it disappears
since $(1-\alpha)^{-1} (y-z)=y-x$; for $n>0$, it is compensated by the 
zero of the factor $(\alpha-\alpha^2)^n$.

After these preparations, we come to the light-cone expansion of 
general Feynman diagrams.
For the line integrals of Lemma \ref{l:lemma1} and Lemma 
\ref{l:lemma2}, we introduce the short notation
\begin{equation}
        \int_x^y [l,r\:|\: n] \:dz \; f(z) \;:=\; \int_0^1 d\alpha \;
        \alpha^{l}\:(1-\alpha)^{r} \:(\alpha-\alpha^2)^n \; f(\alpha 
        y + (1-\alpha) x) \;\;\; .
        \label{l:29x}
\end{equation}
If $l$, $r$, and $n$ all vanish, we sometimes omit the bracket, i.e.\
\[ \int_x^y dz\; f(z) \;:=\; \int_0^1 d\alpha \; f(\alpha y + 
(1-\alpha)x) \;\;\; . \]
Furthermore, we abbreviate the following products with multi-indices
\[ \partial_z^J \;:=\; \frac{\partial}{\partial z^{j_1}} \cdots
\frac{\partial}{\partial z^{j_l}} \;\;\;,\;\;\;\;
(y-x)^J \;:=\; (y-x)^{j_1} \cdots (y-x)^{j_l} \;\;\;,\;\;\;\;
\gamma^J \;:=\; \gamma^{j_1} \cdots \gamma^{j_l} \;\;\; , \]
where $J=(j_1, \ldots, j_l)$.
\begin{Thm} {\bf{(inductive light-cone expansion)}}
\label{l:thm1}
The light-cone expansion of the $k^{\mbox{\scriptsize{th}}}$ order 
contribution $((-s \:B)^k \:s)(x,y)$ to the perturbation series 
(\ref{l:11b}) can be written as an infinite sum of expressions, each of which 
has the form
\begin{eqnarray}
&&\chi_c \:C \:(y-x)^I \int_x^y [l_1, r_1 \:|\: n_1] \:dz_1 \; 
\partial_{z_1}^{I_1}\: \OBox_{z_1}^{p_1} \:V^{(1)}_{J_1, c_1}(z_1)
\int_{z_1}^y [l_2, r_2 \:|\: n_2] \:dz_2 \; \partial_{z_2}^{I_2}\:
\OBox_{z_2}^{p_2}\: V^{(2)}_{J_2, c_2}(z_2) \nonumber \\
&& \hspace*{1cm} \cdots \int_{z_{k-1}}^y [l_k, r_k \:|\: n_k] \:dz_k \; 
\partial_{z_k}^{I_k}\: \OBox_{z_k}^{p_k}\: V^{(k)}_{J_k, c_k}(z_k) \;
\gamma^J \;S^{(h)}(x,y) \;\;\; . \label{l:l1}
\end{eqnarray}
In this formula, $C$ is a complex number and the parameters
$l_a$, $r_a$, $n_a$, and $p_a$ are non-negative integers; the indices
$c$ and $c_a$ can take the two values $L$ or $R$.
The functions $V^{(a)}_{J_a, c_a}$ 
coincide with any of the individual potentials in (\ref{l:18a}) with 
chirality $c_a$, i.e.
\begin{eqnarray}
V^{(a)}_{c_a} &=& A_{c_a} \;\;\;\: \;\;\:\;\;\;\;\;
{\mbox{(in which case $|J_a|=1$)}} \spc {\mbox{or}} \nonumber \\
V^{(a)}_{c_a} &=& m Y_{c_a} \;\;\;\;\;\;\;\;\; {\mbox{(in which case 
$|J_a|=0$)}} \;\;\; .
\label{l:42e}
\end{eqnarray}
The chirality $c_a$ of the potentials is determined by the following rules:
\begin{description}
\item[\rm{\em{(i)}}] The chirality $c_1$ of the first potential 
coincides with the chirality of the projector $\chi_c$.
\item[\rm{\em{(ii)}}] The chirality of the potentials is reversed at 
every mass matrix, i.e.
\[ {\mbox{$c_a$ and $c_{a+1}$}}\;\;\;\left\{ \begin{array}{cl} 
\mbox{coincide} & \mbox{if $V^{(a)}_{c_a}=A_{c_a}$} \\
\mbox{are opposite} & \mbox{if $V^{(a)}_{c_a}=m Y_{c_a} \;\;\; .$}
\end{array} \right. \]
\end{description}
The tensor indices of the multi-indices in (\ref{l:l1}) are all
contracted with each other, according to the following rules:
\begin{description}
\item[\rm{\em{(a)}}] No two tensor indices of the same multi-index are
contracted with each other.
\item[\rm{\em{(b)}}] The tensor indices of the factor $\gamma^{J}$ are 
all contracted with different multi-indices, in the order of their appearance in
the product (\ref{l:l1}) (i.e., for $J=(j_1,\ldots,j_l)$ and $1 \leq a < b 
\leq l$, the multi-index with which $j_a$ is contracted must stand to the 
left of the multi-index corresponding to $j_b$).
\end{description}
The parameter $h$ is given by
\begin{equation}
        2h \;=\; k - 1 - |I| + \sum_{a=1}^k (|I_a| + 2 p_a) \;\;\; .
        \label{l:l3}
\end{equation}
The number of factors $(y-x)$ is bounded by
\begin{equation}
        |I| \;\leq\; k+1-\sum_{a=1}^k |I_a| \;\;\; .
        \label{l:l3a}
\end{equation}
\end{Thm}
Basically, this theorem states that the light-cone expansion of the
$k^{\mbox{\scriptsize{th}}}$ order Feynman diagrams can be written with $k$ 
nested line integrals. Notice that, according to (\ref{l:4b}), the potentials
$V^{(a)}(z_a)$ do in general not commute with each other, so that the order
of multiplication is important in (\ref{l:l1}).
In order to avoid misunderstandings, we point out that the derivatives
$\partial^{I_a}_{z_a}$ and $\OBox_{z_a}^{p_a}$ do not only act on $V^{(a)}(z_a)$, but also on 
all the following factors $V^{(a+1)}(z_{a+1})$, $V^{(a+2)}(z_{a+2})$,\ldots\
(note that the variables $z_{a+1}$, $z_{a+2}$,\ldots\ implicitly depend 
on $z_a$ via the inductive definition of the line integrals).
Clearly, these derivatives could be carried out further with the
Leibniz rule, but it is easier not to do this at the moment.
The restrictions {\em{(a)}} and {\em{(b)}} on the possible contractions of
the tensor indices were imposed in order to avoid an abuse of our
multi-index notation. More precisely, {\em{(a)}} prevents factors
$(y-x)^2$ in $(y-x)^I$, an unnecessary
large number of $\gamma$-matrices in $\gamma^J$, and ``hidden'' Laplacians in
the partial derivatives $\partial_{z_a}^{I_a}$.
The rule {\em{(b)}}, on the other hand, prevents factors $(y-x)^2$ and
hidden Laplacians in combinations of the form
$(y-x)_i \:(y-x)_j \: \gamma^i \:\gamma^j$ and
$\partial_{ij} V^{(a)}_{J_a} \: \gamma^i \:\gamma^j$, respectively.
Our ordering condition for the $\gamma$-matrices is just a matter of
convenience. Relation (\ref{l:l3}) is very useful because it
immediately tells for any configuration of the line integrals and 
potentials in (\ref{l:l1}) what the corresponding order on the light cone is.
Notice that (\ref{l:l3}) and (\ref{l:l3a}) imply the inequality
\begin{equation}
         h \;\geq\; -1 + \sum_{a=1}^k |I_a| + p_a \;\;\; .
        \label{l:45o}
\end{equation}
Especially, one sees that $h \geq -1$. In the case $h=-1$, (\ref{l:l3}) yields
that $|I|>0$, so that (\ref{l:l1}) must contain at least one factor $(y-x)$.
Therefore the factor $S^{(h)}$ in (\ref{l:l1}) is always well-defined
by either (\ref{l:23b}) or (\ref{l:29z}).

We point out that, although the total number of summands (\ref{l:l1}) 
is infinite, the number of summands for any given value of the 
parameter $h$ is finite. This is clear because, for fixed $h$, the 
relations (\ref{l:l3}) and (\ref{l:l3a}) only allow a finite number of 
possibilities to choose the parameters $|I|$, $|I_a|$, and $p_a$; 
thus one only gets a finite combinatorics for all expressions of the 
form (\ref{l:l1}). Since, according to (\ref{l:24b}), the contributions
for higher values of $h$ are of higher order on the light cone, we 
conclude that the number of summands (\ref{l:l1}) is finite to every 
order on the light cone. Therefore the light-cone expansion of 
Theorem \ref{l:thm1} makes mathematical sense in terms of Def.\ \ref{l:def1}.
\\[.5em]
{\em{Proof of Theorem \ref{l:thm1}: }}
We proceed inductively in $k$. For $k=0$, the assumption is true 
because we can write the free Dirac Green's function in the 
required form (\ref{l:l1}),
\begin{equation}
        s(x,y) \;\stackrel{(\ref{l:11a}),(\ref{l:29z})}{=}\; (\chi_L + \chi_R) 
        \:\frac{i}{2} \:(y-x)^j \gamma_j \:S^{(-1)}(x,y) \;\;\; .
        \label{l:233a}
\end{equation}
The conditions {\em{(i)}}, {\em{(ii)}}, {\em{(a)}}, {\em{(b)}}, and the
relations (\ref{l:l3}), (\ref{l:l3a}) are clearly satisfied.

Assume that the theorem holds for a given $k$. With the formula
\begin{equation}
        ((-s \:B)^{k+1}\:s)(x,y) \;\stackrel{(\ref{l:11a})}{=}\; -i 
        \Pdd_x \int d^4 z \; S^{(0)}(x,z) \:B(z) \:
        ((-s \:B)^k\:s)(z,y) \;\;\; ,
        \label{l:i4}
\end{equation}
we can express the $(k+1)^{\mbox{\scriptsize{st}}}$ order 
contribution to the perturbation series
(\ref{l:11b}) in terms of the $k^{\mbox{\scriptsize{th}}}$ order
contribution. We must show that (\ref{l:i4}) can again 
be written as a sum of expressions of the form (\ref{l:l1}) (with $k$ 
replaced by $k+1$), and that {\em{(i)}}, {\em{(ii)}}, {\em{(a)}}, {\em{(b)}}, 
and (\ref{l:l3}), (\ref{l:l3a}) are satisfied.
This is done in several construction steps:
\begin{description}
\item[\rm{\em{1) Chiral decomposition:}}]
We substitute the induction hypothesis (\ref{l:l1}) into (\ref{l:i4}).
This gives a sum of expressions of the form
\begin{eqnarray}
&& C \:i \Pdd_x \int d^4z \; S^{(0)}(x,z) \left\{ (y-z)^I \:
B(z) \:\chi_c \int_z^y [l_1, r_1 \:|\: n_1] \:dz_1 \;
\partial_{z_1}^{I_1} \:\OBox^{p_1} \:V^{(1)}_{J_1, c_1}(z_1) \right.
\nonumber \\
&& \hspace*{1cm}
\cdots \left. \int_{z_{k-1}}^y [l_k, r_k \:|\: n_k] \:dz_k \;
\partial_{z_k}^{I_k}\: \OBox^{p_k}\: V^{(k)}_{J_k, c_k}(z_k)
\;\gamma^J \right\} S^{(h)}(z,y) \;\;\; .
\label{l:45x}
\end{eqnarray}
We insert the special form of the potential $B$, (\ref{l:18a}), and expand.
Using the commutation rule $\gamma^i \:\chi_{L\!/\!R} =
\chi_{R\!/\!L} \:\gamma^i$, we bring all chiral projectors to the 
very left, where they can be combined with the formula $\chi_c 
\chi_d=\delta_{cd}\:\chi_c$ to a single chiral projector.
Next, we bring the $\gamma$-matrices of $B$ to the right and write 
them together with the factor $\gamma^J$ in
(\ref{l:45x}) (notice that the Dirac matrices commute with the potentials
$V^{(a)}_{c_a}$, which are only non-diagonal in the Dirac sea index).
Denoting the individual potentials of the factor $B$ in 
(\ref{l:45x}) by $V^{(0)}_{J_0, c_0}$, we thus get for (\ref{l:45x})
a sum of expressions of the form
\begin{eqnarray}
&& \chi_c\:C \:i \Pdd_x \int d^4z \; S^{(0)}(x,z) \left\{ (y-z)^I
\: V^{(0)}_{J_0, c_0}(z) \int_z^y [l_1, r_1 \:|\: n_1] \:dz_1 \;
\partial_{z_1}^{I_1} \:\OBox^{p_1} \:V^{(1)}_{J_1, c_1}(z_1) \right.
\nonumber \\
&& \hspace*{1cm}
\cdots \left. \int_{z_{k-1}}^y [l_k, r_k \:|\: n_k] \:dz_k \;
\partial_{z_k}^{I_k}\: \OBox^{p_k}\: V^{(k)}_{J_k, c_k}(z_k)
\;\gamma^J \right\} S^{(h)}(z,y) \;\;\; .
\label{l:35u}
\end{eqnarray}
The chiral decomposition (\ref{l:18a}) and the induction 
hypothesis {\em{(i)}} yield that the chirality $c_0$ coincides with 
$c$, whereas $c_1$ coincides with $c_0$ if and only if 
$V^{(0)}_{c_0}=A_{c_0}$. Since the chiralities $c_2$, $c_3$,\ldots
satisfy the induction hypothesis {\em{(ii)}}, we 
conclude that the rules {\em{(i)}} and {\em{(ii)}} are also satisfied 
in (\ref{l:35u}) (after relabeling the indices in an obvious way).
The chirality of the potentials will not be 
affected in all the following construction steps; to simplify the
notation, we will omit the indices $c_a$ from now on.
\item[\rm{\em{2) Light-cone expansion with Lemma \ref{l:lemma1},
Lemma \ref{l:lemma2}:}}]
Since $y$ can be considered as a fixed parameter, we can in (\ref{l:35u})
apply either Lemma \ref{l:lemma1} or Lemma \ref{l:lemma2} with $V$ 
given by the expression in the curly brackets,
\begin{eqnarray}
\lefteqn{ (\ref{l:35u}) \;=\; \chi_c\:C \:i \Pdd_x \sum_{n=0}^\infty
\frac{1}{n!} \int_x^y [0, h \:|\: n] \:dz } \nonumber \\
&&\times\; \OBox^n_z \left( (y-z)^I \:
V^{(0)}_{J_0}(z) \int_z^y [l_1, r_1 \:|\: n_1] \:dz_1 \;
\partial_{z_1}^{I_1} \:\OBox^{p_1}\: V^{(1)}_{J_1}(z_1)
\right. \nonumber \\
&& \hspace*{.7cm} \left. \cdots
\int_{z_{k-1}}^y [l_k, r_k \:|\: n_k] \:dz_k \;
\partial_{z_k}^{I_k} \:\OBox^{p_k} \:V^{(k)}_{J_k}(z_k) \right)
\:\gamma^J \; S^{(n+h+1)}(x,y) \;\;\;\;\; . \spc \label{l:33n}
\end{eqnarray}
\item[\rm{\em{3) Computation of the Laplacian $\OBox^n_z$:}}]
We carry out the $z$-derivatives in (\ref{l:33n}) inductively with the
Leibniz rule. Each derivative can act either on the factors $(y-z)^I$
or on the functions $V^{(a)}$. In the first case, one of the 
factors $(y-z)$ disappears. Thus we get a sum of expressions of the form
\begin{eqnarray}
\lefteqn{ \chi_c\:C \:i \Pdd_x \int_x^y [0, h \:|\: n] \:dz \; (y-z)^{\hat{I}}
\:\partial_{z}^{I_0} \:\OBox^{p_0}_z \:V^{(0)}_{J_0}(z)
\int_{z}^y [l_1, r_1 \:|\: n_1] \:dz_1 \;
\partial_{z_1}^{I_1} \:\OBox^{p_1} \:V^{(1)}_{J_1}(z_1) } \nonumber \\
&& \hspace*{1cm} \cdots
\int_{z_{k-1}}^y [l_k, r_k \:|\: n_k] \:dz_k \;
\partial_{z_k}^{I_k} \:\OBox^{p_k} \:V^{(k)}_{J_k}(z_k)
\:\gamma^J \; S^{(n+h+1)}(x,y) \spc \spc \label{l:34j}
\end{eqnarray}
with $|\hat{I}| \leq |I|$ and
\begin{equation}
        2n=|I|-|\hat{I}|+|I_0|+2p_0 \;\;\; .
        \label{l:47x}
\end{equation}
We can assume that no tensor indices of $\partial_z^{I_0}$ are contracted 
with each other (otherwise we rewrite the corresponding partial 
derivatives as additional Laplacians).
Then all the partial derivatives $\partial_z$ in (\ref{l:34j}) were 
generated in the case when one derivative of a Laplacian
$\OBox_z$ in (\ref{l:33n}) hit a factor $(y-z)$ whereas the other derivative
acted on the $V^{(a)}$.
Thus the number of factors $(y-z)$ which disappeared by carrying out 
the Laplacians in (\ref{l:33n}) is larger or equal than the number of partial 
derivatives $\partial_z$,
\begin{equation}
        |I| - |\hat{I}| \;\geq\; |I_0| \;\;\; .
        \label{l:45a}
\end{equation}
\item[\rm{\em{4) Extraction of the factors $(y-x)$:}}]
In (\ref{l:34j}), we iteratively apply the identity
\begin{equation}
        \int_x^y [0, r \:|\: n] \:dz \; (y-z) \cdots \;=\;
        (y-x) \int_x^y [0, r+1 \:|\: n] \:dz \; \cdots \;\;\; .
        \label{l:49x}
\end{equation}
This gives $(k+1)$ nested line integrals of the form
\begin{eqnarray}
(\ref{l:34j}) &=& \chi_c\:C \:i \Pdd_x (y-x)^{\hat{I}} \; S^{(\hat{h})}(x,y)
\int_x^y [l_0, r_0 \:|\: n_0] \:dz_0 \;
\partial_{z_0}^{I_0} \:\OBox^{p_0} \:V^{(0)}_{J_0}(z_0) \nonumber \\
&& \hspace*{1cm} \cdots
\int_{z_{k-1}}^y [l_k, r_k \:|\: n_k] \:dz_k \;
\partial_{z_k}^{I_k} \:\OBox^{p_k} \:V^{(k)}_{J_k}(z_a)
\:\gamma^J \label{l:34g}
\end{eqnarray}
with
\begin{eqnarray}
l_0 &=& 0 \;\;\;,\spc r_0 = h+|\hat{I}| \;\;\;,\spc n_0=n \label{l:42f} \\
0 &\leq& 2 \hat{h} \;=\; 2(n + h + 1) \;\stackrel{(\ref{l:47x})}{=}\;
2 h + 2 + |I| - |\hat{I}| + |I_0| + 2 p_0 \;\;\; . \spc \label{l:46a}
\end{eqnarray}
We can arrange that the parameters $l_0$, $r_0$, and $n_0$ are all 
positive: The only parameter which might be negative is $r_0$; in this 
case, $h=-1$, $|\hat{I}|=0$, and thus $r_0=-1$. The induction 
hypothesis (\ref{l:l3}) yields that $|I|>0$. Thus $|I|>|\hat{I}|$, and 
relation (\ref{l:47x}) gives that $(n_0=)n>0$. Therefore we can 
apply the identity
\[ [l_0, r_0 \:|\: n_0] \;=\; [l_0+1, r_0+1 \:|\: n_0-1] \]
to make all the parameters in this bracket positive.
\item[\rm{\em{5) Computation of the partial derivative $\Pdd_x$:}}]
The $x$-derivative in (\ref{l:34g}) can act on the factors
$S^{(\hat{h})}$, $(y-x)^{\hat{I}}$, or $V^{(a)}(z_a)$. The first case can be
computed with the rules (\ref{l:7}) or (\ref{l:29z}); it decreases $\hat{h}$ 
by one and gives one additional factor $(y-x)$. In the second case, 
one factor $(y-x)$ disappears, and thus $|\hat{I}|$ is decremented.
The last case can be handled with the  rule
\begin{equation}
        \frac{\partial}{\partial x^k} \int_x^y [l,r \:|\: n] \:dz\; f(z,y) 
        \;=\; \int_x^y [l,r+1 \:|\: n] \:\frac{\partial}{\partial z^k} 
        f(z,y) \;\;\; ,
        \label{l:34h}
\end{equation}
which increases $|I_0|$ by one. As is immediately verified in each of these
cases, equation (\ref{l:46a}) transforms into
\begin{equation}
        2 \hat{h} \;=\; 2h + 1 + |I| - |\hat{I}| + |I_0| + 2 p_0 \;\;\; , \label{l:47a}
\end{equation}
whereas inequality (\ref{l:45a}) must be weakened to
\begin{equation}
        |\hat{I}| \;\leq\; 1 + |I| - |I_0| \;\;\; . \label{l:47b}
\end{equation}
Finally, we combine the $\gamma$-matrix of the factor $\Pdd_x$ with $\gamma^J$.
\end{description}
After these transformations, the $(k+1)^{\mbox{\scriptsize{st}}}$ order 
Feynman diagram consists of a sum of terms of the form
\begin{eqnarray}
\lefteqn{ \chi_c\:C \:(y-x)^{\hat{I}} \int_x^y [l_0, r_0 \:|\: n_0]
\:dz_0 \; \partial_{z_0}^{I_0}\: \OBox_{z_0}^{p_0} \:V^{(0)}_{J_0}(z_0) }
\nonumber \\
&& \hspace*{1cm} \cdots \int_{z_{k-1}}^y \!\!\!\![l_k, r_k \:|\: n_k] \:dz_k
\; \partial_{z_k}^{I_k}\: \OBox_{z_k}^{p_k}\: V^{(k)}_{J_k}(z_k) \;
\gamma^J \;S^{(\hat{h})}(x,y) \;\;\; .
\label{l:54a}
\end{eqnarray}
Notice that the parameters $I_a, p_a$, $a=1,\ldots, k$, were not 
changed by the above construction steps; they are still the same as 
in the induction hypothesis (\ref{l:l1}).
After renaming the indices and the integration variables, (\ref{l:54a}) 
is of the required form (\ref{l:l1}). The conditions {\em{(a)}} and {\em{(b)}}
for the contractions of the tensor indices, however, will in general be
violated. Therefore we need two further computation steps:
\begin{description}
\item[\rm{\em{6) Simplification of the $\gamma$-matrices:}}]
If any two of the tensor indices of the factor $\gamma^J$ 
are contracted with each other, we reorder the $\gamma$-matrices with 
the anti-commutation relations
\begin{equation}
        \{ \gamma^i,\:\gamma^j \} \;=\; 2 \:g^{ij}\:\1
        \label{l:ar1}
\end{equation}
until the corresponding matrices are next to each other. Applying the
identity $\gamma^i \gamma_i = 4\:\1$, both Dirac matrices disappear.
We iterate this procedure until no tensor indices of $\gamma^J$ are 
contracted with each other (notice that the iteration comes to an end
because the number of $\gamma$-factors is decreased by two in each 
step). Again using the anti-commutation rule (\ref{l:ar1}), we reorder the
Dirac matrices until they are in the same order in which the factors 
to which their tensor indices are contracted appear in the product
(\ref{l:54a}). If any two of the $\gamma$-matrices are 
contracted with the same multi-index, these $\gamma$-matrices are
next to each other, and we can use the symmetry in the tensor
indices to eliminate them both, more precisely
\begin{eqnarray}
(y-x)_i \:(y-x)_j \cdots \gamma^i \gamma^j &=& (y-x)^2 \cdots \1 \\
\partial_{ij} V^{(a)} \cdots \gamma^i \gamma^j &=& \OBox V^{(a)} 
        \cdots \1 \;\;\; .
        \label{l:58e}
\end{eqnarray}
After all these transformations, condition {\em{(b)}} is satisfied.

Notice that the parameters $|I_a|$ and $p_a$ are in general changed in 
this construction step. More precisely, each transformation 
(\ref{l:58e}) modifies the parameters according to
\begin{equation}
|I_a| \:\rightarrow\: |I_a| - 2 \;\;\;,\spc
p_a \:\rightarrow\: p_a + 1 \;\;\; .
        \label{l:25t}
\end{equation}
\item[\rm{\em{7) Handling of the new contractions:}}]
If any two tensor indices of a factor $\partial_{z_a}^{I_a}$ are contracted
with each other, we rewrite the corresponding partial derivatives as a 
Laplacian; this changes the parameters $|I_a|$ and $p_a$ according to
(\ref{l:25t}).
If two tensor indices of the factor $(y-x)^{\hat{I}}$ are contracted with
each other, this gives a factor $(y-x)^2$. Using the identity
(\ref{l:22a}) or (\ref{l:25a}), we inductively absorb the factors $(y-x)^2$
into $S^{(\hat{h})}(x,y)$, which transforms $\hat{h}$ and $|\hat{I}|$ as
\begin{equation}
\hat{h} \:\rightarrow\: \hat{h}+1 \;\;\;,\spc
|\hat{I}| \:\rightarrow\: |\hat{I}| - 2 \;\;\; .
        \label{l:25u}
\end{equation}
After these transformations, condition {\em{(a)}} is also satisfied.
\end{description}
After all these construction steps, the 
$(k+1)^{\mbox{\scriptsize{th}}}$ order Feynman diagram is a sum of 
terms of the form (\ref{l:54a}) satisfying the conditions {\em{(a)}} and 
{\em{(b)}}. It remains to show that the relations (\ref{l:l3}) and (\ref{l:l3a})
remain valid in our inductive construction: As mentioned earlier, the 
parameters $I_a$, $p_a$, $a=1,\ldots,k$ are not changed in the 
construction steps {\em{1)}} to {\em{5)}}. In the steps {\em{6)}} and 
{\em{7)}}, the transformations (\ref{l:25t}) and (\ref{l:25u}) preserve 
both the induction hypothesis (\ref{l:l3}),(\ref{l:l3a}) and the relations 
(\ref{l:47a}),(\ref{l:47b}), as is immediately verified. By substituting (\ref{l:47a})
and (\ref{l:47b}) into (\ref{l:25t}),(\ref{l:25u}), we conclude that
\[ \spc 2 \hat{h} \;=\; (k+1) - 1 - |\hat{I}| + \sum_{a=0}^k |I_a| + 2 p_a 
\;\;\;,\spc |\hat{I}| \;\leq\; (k+1) + 1 - \sum_{a=0}^k |I_a| 
\;\;\; . \hspace*{0.7cm} \FBox \]
Note that this proof is constructive in the sense
that it gives a procedure with which the light-cone expansion of every Feynman 
diagram can be carried out explicitly. The disadvantage of this
procedure, however, is that the resulting formulas become more and more 
complicated to higher order perturbation theory. Therefore it is
essential to rearrange and collect the contributions of all 
Feynman diagrams in a way which makes it clear how $\tilde{s}(x,y)$ looks like
to every order on the light cone. In preparation for this analysis, 
which will be the task of the following Subsection \ref{l:sec_22}, we 
shall now simplify the light-cone expansion of Theorem \ref{l:thm1} a 
little bit. More precisely, we want to eliminate from (\ref{l:l1}) the 
partial derivatives in direction of the line integrals, i.e.\ those 
derivatives $\partial_{z_l}$ which are contracted with a factor 
$(y-x)$. We call these derivatives {\em{tangential}}. The following 
combinatorial lemma controls the number of tangential derivatives.
\begin{Lemma}
\label{l:lemma3}
The contributions (\ref{l:l1}) to the light-cone expansion of 
Theorem \ref{l:thm1} satisfy for $a=1,\ldots,k$ the inequalities
\begin{equation}
        l_a + n_a \;\geq\; t_a-1 \spc {\mbox{and}} \spc r_a + n_a \;\geq\; 
        \sum_{b=a}^k t_b \;\;\; ,
        \label{l:73f}
\end{equation}
where $t_a$ denotes the number of tensor indices of the multi-index 
$I_a$ which are contracted with the factor $(y-x)^I$.
\end{Lemma}
{\Proof}
As in the proof of Theorem \ref{l:thm1}, we proceed inductively in the
order $k$ of the perturbation theory.
For $k=0$, the inequalities (\ref{l:73f}) are trivially satisfied according
to (\ref{l:233a}).
Assume that (\ref{l:73f}) is true for a given $k$. We go through the construction 
steps {\em{1)}} to {\em{7)}} of Theorem \ref{l:thm1} and check that the
inequalities (\ref{l:73f}) then also hold in (\ref{l:54a}) for $a=0,\ldots, k$.

We first consider the case $a>0$. The parameters $l_a$, $r_a$, and $n_a$ remain
unchanged in all the construction steps of Theorem \ref{l:thm1}.
Furthermore, it is obvious that the 
parameters $t_a$ are not affected in the steps  {\em{1)}}, {\em{2)}}, 
{\em{4)}}, and {\em{7)}}. In the steps {\em{3)}} and {\em{5)}}, the computation 
of the derivatives $\OBox^n_z$ and $\Pdd_x$ might annihilate some of 
the factors $(y-x)$ which were contracted with the factors 
$\partial_{z_a}^{I_a}$; this may decrease the parameters $t_a$. For 
the analysis of step {\em{6)}}, note that all $\gamma$-matrices which
are contracted with factors $(y-x)$ stand to the left of those 
$\gamma$-matrices which are contracted with the 
$\partial_{z_a}^{I_a}$, $a=1,\ldots, k$ (this follows from the ordering 
condition {\em{(b)}} in the induction hypothesis and the fact that
additional factors $(y-x)^j \cdots\gamma_j$ are only generated during the 
construction if the partial derivative $\Pdd_x$ 
hits $S^{(\hat{h})}$ in step {\em{5)}}; in this case, the corresponding 
$\gamma$-matrix stands at the very left in $\gamma^J$). Therefore the 
commutations of the Dirac matrices do not lead to additional contractions
between factors $(y-x)$ and $\partial_{z_a}^{I_a}$, which implies that
the parameters $t_a$ remain unchanged in step {\em{6)}}.
We conclude that the $l_a$, $r_a$, and $n_a$ remain unchanged whereas the $t_a$
may only decrease, and thus (\ref{l:73f}) holds for $a=1,\ldots, k$
throughout all the construction steps.

It remains to show that the inequalities (\ref{l:73f}) hold in (\ref{l:54a}) for
$a=0$. We first look at the situation after step {\em{4)}}
in (\ref{l:34g}): The values (\ref{l:42f}) for $l_0$, $r_0$, and $n_0$ give in 
combination with (\ref{l:47x}) the equations
\begin{eqnarray}
l_0 + n_0 &=& \frac{1}{2} \left( |I| \:-\: |\hat{I}| \:+\: |I_0| \:+\:
2 p_0 \right) \label{l:ln1} \\
r_0 + n_0 &=& h \:+\: \frac{1}{2} \left( |I| \:+\: |\hat{I}| \:+\: 
|I_0| \:+\: 2 p_0 \right) \;\;\; . \label{l:rn1}
\end{eqnarray}
Moreover, the number of tangential derivatives $t_0$ at the first 
potential is clearly bounded by the total number of derivatives there,
\begin{equation}
        |I_0| \;\geq\; t_0 \;\;\; .
        \label{l:bn1}
\end{equation}
Furthermore, the total number of tangential derivatives is smaller 
than the number of factors $(y-x)$,
\begin{equation}
|\hat{I}| \;\geq\; \sum_{a=0}^k t_a \;\;\; .
        \label{l:bn2}
\end{equation}
Substituting (\ref{l:45a}) and (\ref{l:bn1}) into (\ref{l:ln1}) yields the 
inequalities
\begin{equation}
        l_0 + n_0 \;\geq\; |I_0| + p_0 \;\geq\; t_0 \;\;\; .
        \label{l:5t}
\end{equation}
In order to get a bound for $r_0+n_0$, we must distinguish two cases. 
If $h \geq 0$, we substitute (\ref{l:45a}) into (\ref{l:rn1}) and get 
with (\ref{l:bn2}) the inequality
\begin{equation}
        r_0 + n_0 \;\geq\; |\hat{I}| \:+\: |I_0| \:+\: p_0 \;\geq\; |\hat{I}| 
        \;\geq\; \sum_{a=0}^k t_a \;\;\; .
        \label{l:5u}
\end{equation}
In the case $h=-1$, (\ref{l:45o}) shows that $|I_a|$, and consequently 
also $t_a$, vanish for $1 \leq a \leq k$. Furthermore, (\ref{l:l3}) 
yields that $|I| \neq 0$. Thus (\ref{l:rn1}) and (\ref{l:bn1}), 
(\ref{l:bn2}) give the bound
\[ r_0 + n_0 \;\geq\; h \:+\: \frac{|I|}{2} \:+\: \frac{1}{2} 
\:\sum_{a=0}^k t_a \;+\; \frac{1}{2} \:t_0
\;\geq\; \frac{1}{2} \:\sum_{a=0}^k t_a 
\;+\; \frac{1}{2} \:t_0 \;\;\; , \]
where we used in the last inequality that $h+|I|/2 \geq -1/2$ and that all the 
other terms are integers. Since $t_0=\sum_{a=0}^k t_a$, we conclude 
that inequality (\ref{l:5u}) also holds in the case $h=-1$.

We finally consider how the bounds (\ref{l:5t}) and (\ref{l:5u}) for $l_0 + n_0$ and
$r_0 + n_0$ must be modified in the subsequent construction steps.
In step {\em{5)}}, the partial derivative $\Pdd_x$ may annihilate a factor 
$(y-x)$, in which case the parameters $t_a$ might decrease. On the 
other hand, the partial derivatives $\Pdd_x$ may produce an additional 
factor $\partial_{z_0}$; in this case, $r_0$ is incremented according 
to (\ref{l:34h}). In step {\em{6)}}, only this additional factor 
$\partial_{z_0}$ may be contracted with $(y-x)^{\hat{I}}$. Step 
{\em{7)}} does not change $l_0$, $r_0$, $n_0$, and $t_0$. Putting 
these transformations together, we conclude that the inequality 
(\ref{l:5t}) for $l_0+n_0$ must be weakened by one, whereas the bound 
(\ref{l:5u}) for $r_0 + n_0$ remains valid as it is.
This gives precisely the inequalities (\ref{l:73f}) for $a=0$.
\QED

\begin{Thm} {\bf{(partial integration of the tangential derivatives)}}
\label{l:thm2}
Every contribution (\ref{l:l1}) to the light cone expansion of Theorem 
\ref{l:thm1} can be written as a finite sum of expressions of the form
\begin{eqnarray}
&& \chi_c\:C \:(y-x)^K \:W^{(0)}(x) \int_x^y [l_1, r_1 \:|\: n_1] \:dz_1 \;
W^{(1)}(z_1) \int_{z_1}^y [l_2, r_2 \:|\: n_2] \:dz_2 \; W^{(2)}(z_2)
\nonumber \\
&&\hspace*{1cm} \cdots \int_{z_{\alpha-1}}^y [l_\alpha, r_\alpha \:|\: 
n_\alpha] \:dz_\alpha \; W^{(\alpha)}(z_\alpha) \;
\gamma^J \; S^{(h)}(x,y) \;\;\;\;\;, \;
\alpha \leq k , \spc
\label{l:70}
\end{eqnarray}
where the factors $W^{(\beta)}$ are composed of the potentials and 
their partial derivatives,
\begin{equation}
        W^{(\beta)} \;=\; (\partial^{K_{a_\beta}} \OBox^{p_{a_\beta}} 
        V^{(a_\beta)}_{J_{a_\beta}, c_{a_\beta}})
        \cdots (\partial^{K_{b_\beta}} \OBox^{p_{b_\beta}} 
        V^{(b_\beta)}_{J_{b_\beta}, c_{b_\beta}})
        \label{l:71}
\end{equation}
with $a_1=1$, $a_{\beta+1}=b_\beta+1$, $b_\beta \geq a_\beta-1$ (in 
the case $b_\beta=a_\beta-1$, $W^{(\beta)}$ is identically one),
and $b_\alpha=k$.
The parameters $l_a$, $r_a$, and $n_a$ are non-negative integers,
$C$ is a complex number, and $c=L\!/\!R$, $c_a=L\!/\!R$ are chiral 
indices. The potentials $V^{(a)}$ are again given by (\ref{l:42e});
their chirality is determined by the rules (i) and (ii) of Theorem \ref{l:thm1}.
The tensor indices of the multi-indices $J$, $K$, $J_a$, and 
$K_a$ are all contracted with each other, according to the rules (a),(b) 
of Theorem \ref{l:thm1} and
\begin{description}
\item[\rm{\em{(c)}}] The tensor indices of $(y-x)^K$ are all 
contracted with the tensor indices of the factors $V^{(a)}_{J_a}$ or $\gamma^J$ 
(but not with the factors $\partial^{K_a}$).
\end{description}
 We have the relation
\begin{equation}
        2h \;=\; k - 1 - |K| + \sum_{a=1}^k (|K_a| + 2 p_a) \;\;\; .
        \label{l:71b}
\end{equation}
\end{Thm}
{\Proof}
The basic method for the proof is to iteratively eliminate those partial
derivatives $\partial_{z_a}^{I_a}$ in (\ref{l:l1}) which are contracted with
a factor $(y-x)$. This is done with the partial integration formula
\begin{eqnarray}
\lefteqn{ (y-x)^j \int_x^y [l, r \:|\: n] \:dz \;\partial_j f(z)
\;\stackrel{(\ref{l:29x})}{=}\; \int_0^1 d\alpha \; \alpha^l 
\:(1-\alpha)^r \:(\alpha-\alpha^2)^n \:\frac{d}{d\alpha} f(\alpha 
y + (1-\alpha) x) } \nonumber \\
&=& \delta_{r+n, 0} \:f(y) \:-\: \delta_{l+n, 0} \:f(x) \nonumber \\
&&-(l+n) \int_x^y [l-1, r \:|\: n] \:dz \; f(z) \;+\;
(r+n) \int_x^y [l, r-1 \:|\: n] \:dz \; f(z) \;\;\; . \spc \label{l:72}
\end{eqnarray}
In order to see the main difficulty, consider the example of two
nested line integrals with two tangential derivatives
\begin{eqnarray}
\lefteqn{ (y-x)^j \:(y-x)^k \int_x^y [0,1 \:|\: 0] \:dz_1 \; 
V^{(1)}(z_1) \int_{z_1}^y [0,1 \:|\: 0] \:dz_2 \; \partial_{jk} 
V^{(2)}(z_2) } \label{l:80} \\
&=& (y-x)^j \int_x^y [0,0 \:|\: 0] \:dz_1 \; 
V^{(1)}(z_1) \: (y-z_1)^k \int_{z_1}^y [0,1 \:|\: 0] \:dz_2 \; \partial_{jk} 
V^{(2)}(z_2) \nonumber \\
&=& -(y-x)^j \int_x^y dz_1 \; V^{(1)}(z_1) \: \partial_j V^{(2)}(z_1)
\label{l:81} \\
&&+(y-x)^j \int_x^y dz_1 \; V^{(1)}(z_1) \int_{z_1}^y dz_2 \; 
\partial_j V^{(2)}(z_2) \;\;\; . \label{l:82}
\end{eqnarray}
Although the line integrals in (\ref{l:80}) satisfy the conditions 
of Theorem \ref{l:thm1}, the expression cannot be transformed into the 
required form (\ref{l:70}). Namely in (\ref{l:81}), we cannot eliminate 
the remaining tangential derivative (because partial integration would
yield a term $(y-x)^j \:\partial_j V^{(1)}(z_1)$).
In (\ref{l:82}), on the other hand, we can successfully perform a second partial 
integration
\[ (\ref{l:82}) \;=\; \int_x^y [0,-1 \:|\: 0] \:dz_1 \;V^{(1)}(z_1) 
\;(V^{(2)}(y) - V^{(2)}(z_1)) \;\;\; , \]
but then the second parameter in the bracket $[.,.\:|\:.]$ becomes 
negative. More generally, we must ensure that the boundary terms contain no 
tangential derivatives, and that the parameters $l_a, r_a$, and $n_a$ 
stay positive in the construction.

Since the chirality of the potentials is not affected by the partial 
integrations, it is obvious that the rules {\em{(i)}} and {\em{(ii)}} 
of Theorem \ref{l:thm1} will remain valid. For simplicity in notation, we will
omit the indices $c_a$ in the following.

First of all, we split up the factor $(y-x)^I$ in (\ref{l:l1}) in the form
$(y-x)^I = (y-x)^K \:(y-x)^L$, where $L$ are those tensor indices which are 
contracted with the partial derivatives $\partial^{I_a}_{z_a}$, $a=1,\ldots, k$.
Setting $b=1$ and $z_0=x$, the first line integral in (\ref{l:l1}) can 
be written as
\begin{equation}
        \cdots \:(y-z_{b-1})^L \int_{z_{b-1}}^y [l_b, r_b \:|\: n_b] \:dz_b 
        \; \partial^{I_b}_{z_b} \:\OBox^{p_b}_{z_b} \:V^{(b)}_{J_b}(z_b) \:\cdots
        \;\;\; .
        \label{l:81a}
\end{equation}
We rewrite the tangential derivatives in this line integral as derivatives in
the integration variable,
\begin{equation}
        \;=\; \cdots \:(y-z_{b-1})^N \int_0^1 d\alpha \: \alpha^l 
        \:(1-\alpha)^r \; \left( \frac{d}{d\alpha} \right)^q
        \partial^{K_b}_{z_b} \:\OBox^{p_b}_{z_b} \:V^{(b)}_{J_b}(z_b) \:\cdots
        \label{l:82a}
\end{equation}
with $|L|=|N|+q$ and $l=l_b+n_b$, $r=r_b+n_b$.
Lemma \ref{l:lemma3} gives the bounds
\begin{equation}
        l \;\geq\; q-1 \;\;\;,\spc r \;\geq\; q + |N| \;\;\; .
        \label{l:83}
\end{equation}
More generally, we use (\ref{l:82a}),(\ref{l:83}) as our induction 
hypothesis, whereby the left factor `$\cdots$' stands for all 
previous line integrals (which contain {\em{no}} tangential 
derivatives), and the right factor `$\cdots$' stands for subsequent 
line integrals. The tensor indices of the factor $(y-z_{a-1})^N$ must all be
contracted with the partial derivatives $\partial^{I_a}_{z_a}$ for $a>b$ and
thus give tangential derivatives in the subsequent line integrals.
The induction step is to show that all the $\alpha$-derivatives 
in (\ref{l:82a}) can be eliminated, and that we can write the 
resulting expressions again in the form (\ref{l:82a}),(\ref{l:83}) with $b$ 
replaced by $b+1$. Under the assumption that this induction step
holds, we can eliminate all tangential derivatives in $k$ steps.
The resulting expressions are very similar to (\ref{l:70}),(\ref{l:71}). 
The only difference is that the derivatives $\partial^{K_a}$ and 
$\OBox^{p_a}$ in the resulting expressions are differential operators 
acting on all the following factors $V^{(a)}$, $V^{(a+1)}$,\ldots; 
in (\ref{l:71}), on the other hand, the partial derivatives act only on the 
adjacent potential $V^{(a)}$. In order to bring the resulting 
expressions into the required form, we finally carry out all the 
derivatives with the Leibniz rule and the chain rule (\ref{l:34h}).

For the proof of the induction step, we integrate in (\ref{l:82a}) $q$ times 
by parts (if $q$ is zero, we can skip the partial integrations; our 
expression is then of the form (\ref{l:84})). Since the powers of 
the factors $\alpha$ and $(1-\alpha)$ are decreased at most by one in 
each partial integration step, (\ref{l:83}) implies that the boundary values
vanish unless in the last step for $\alpha=0$.
We thus obtain a sum of terms of the form
\begin{eqnarray}
&&\cdots (y-z_{b-1})^N \;\partial^{K_b}_{z_b} \:\OBox^{p_b}_{z_b} \:
V^{(b)}_{J_b}(z_b) \:\cdots_{|z_b \equiv z_{b-1}} \spc {\mbox{and}}
\label{l:83b} \\
&&\cdots (y-z_{b-1})^N \int_{z_{b-1}}^y [l, r \:|\: n=0] \:dz_b \;
\partial^{K_b}_{z_b} \:\OBox^{p_b}_{z_b} \:V^{(b)}_{J_b}(z_b) \:
\cdots \;\;\;\;\; {\mbox{with $l \geq 0$, $r \geq |N|$.}}\spc
\label{l:84}
\end{eqnarray}
In (\ref{l:84}), we iteratively use the relation
\begin{equation}
        (y-x)^j \int_x^y [l, r \:|\: n] \:dz \;\cdots \;=\; \int_x^y [l, r-1 
        \:|\: n] \:dz \; (y-z)^j \;\cdots \label{l:74}
\end{equation}
to bring all factors $(y-z_{b-1})$ to the right,
\begin{equation}
        (\ref{l:84}) \;=\; \cdots \int_{z_{b-1}}^y [l, r \:|\: n=0] \:dz_b \;
(y-z_{b})^N \:\partial^{K_b}_{z_b} \:\OBox^{p_b}_{z_b} \:V^{(b)}_{J_b}(z_b) 
\cdots \spc {\mbox{with $l, r \geq 0$.}}
        \label{l:85}
\end{equation}
It might be convenient to reorganize the polynomials in this line 
integral with the identity
\[ \int_x^y [l, r \:|\: n] \:dz \cdots \;=\; \int_x^y [l-1, r-1 \:|\: 
n+1] \:dz \cdots \;\;\; , \]
but this is not relevant for the statement of the theorem.

In both cases (\ref{l:83b}) and (\ref{l:85}), we have an expression of 
the form
\begin{equation}
        \cdots (y-z_b)^N \;\partial^{K_b}_{z_b} \:\OBox^{p_b}_{z_b}\:
        V^{(b)}_{J_b}(z_b) \cdots \;\;\; ,
        \label{l:86}
\end{equation}
where the first factor `$\cdots$' stands for line integrals 
without tangential derivatives, and where none of the factors $(y-z_b)$ 
are contracted with $\partial^{K_b}_{z_b}$. Applying the ``inverse
Leibniz rules''
\begin{eqnarray}
(y-x)^j \:\frac{\partial}{\partial x^k} &=& \frac{\partial}{\partial x^k}
\: (y-x)^j \:+\: \delta^j_k \label{l:75} \\
(y-x)_j \:\OBox_x &=& \OBox_x \: (y-x)_j \:+\:
2 \frac{\partial}{\partial x^j} \;\;\; , \label{l:75b}
\end{eqnarray}
we iteratively commute all factors $(y-z_b)$ in (\ref{l:86}) to the right.
This gives a sum of expressions of the form
\begin{equation}
        \cdots \partial^{K_b}_{z_b} \:\OBox^{p_b}_{z_b}\:
        V^{(b)}_{J_b}(z_b) \; (y-z_b)^L \cdots \;\;\; ,
        \label{l:87}
\end{equation}
where the factors $(y-z_b)$ are all contracted with the partial derivatives
$\partial^{I_a}_{z_a}$, $a=b+1,\ldots, k$. The Leibniz rules may have 
annihilated some factors $(y-z_b)$ (i.e., $|L|$ might be smaller 
than $|N|$); in this case, the parameters $t_a$, $a=b+1,\ldots,k$ 
have decreased. As a consequence, the inequalities of Lemma 
\ref{l:lemma3} are still valid for all expressions (\ref{l:87}).
If we write (\ref{l:87}) in the form (\ref{l:81a}) with $b$ replaced 
by $b+1$, we can thus split up the tangential derivatives in the form
(\ref{l:82a}),(\ref{l:83}), which concludes the proof of the induction step.

It remains to derive equation (\ref{l:71b}): Note that each partial 
integration decreases both the number of factors $(y-z_{a-1})$ and the 
total number of partial derivatives by one. If we carry out the 
remaining derivatives with the Leibniz rule (in the last step of the 
proof), this does not change the total order $\sum_{a=1}^k |K_a|+2 
p_a$ of the derivatives. Therefore relation (\ref{l:l3}) in Theorem
\ref{l:thm1} transforms into (\ref{l:71b}).
\QED

\subsection{Resummation, Reduction to the Phase-Free Contribution}
\label{l:sec_22}
In the previous subsection, we gave a procedure for performing the 
light-cone expansion of every summand of the perturbation series (\ref{l:11b}).
In order to obtain formulas for the light-cone expansion of
$\tilde{s}$, we must sum up the light-cone expansions of all Feynman 
diagrams.
Collecting the contributions of all Feynman diagrams gives, to every order on
the light cone, an infinite number of terms. To get control over all
these terms, we shall reorder the sums and partly carry them out.
In the end, the light-cone expansion for $\tilde{s}(x,y)$ will, to 
any order on the light cone, consist of only a finite number of summands. This
rearrangement and simplification of the sums is called {\em{resummation}}
of the light-cone expansion.

In order to get a first impression of what needs 
to be done, we consider the leading singularity on the light cone 
(more precisely, we neglect all terms of the order 
${\cal{O}}((y-x)^{-2})$). This corresponds to taking only the contributions with
$h=-1$ in Theorem \ref{l:thm1}. According to the bound (\ref{l:45o}), the
multi-indices $I_a$ and the parameters $p_a$ must all vanish. Furthermore,
equation (\ref{l:l3}) yields that $|I|=k+1$. The only possibility to 
satisfy the rules {\em{(a)}} and {\em{(b)}} is to contract one 
factor $(y-x)$ with each potential $V^{(a)}$, $a=1,\ldots,k$,
and the remaining factor $(y-x)$ with a $\gamma$-matrix. Thus the 
potentials $V^{(a)}$ must all be chiral potentials $A_{L\!/\!R}$. 
According to the rules {\em{(a)}} and {\em{(b)}}, the chirality of all 
potentials must coincide with the chirality of the projector $\chi_c$. 
We conclude that the leading order of $((-s B)^k s)(x,y)$ on the light cone
consists of a sum of expressions of the form
\begin{eqnarray*}
&&\chi_c \:C \:(y-x)_{j_1} 
\cdots (y-x)_{j_k} \int_x^y [l_1, r_1 \:|\: n_1] \:dz_1 \; 
A^{j_1}_c(z_1) \\
&& \hspace*{1.5cm} \ldots \int_{z_{k-1}}^y [l_k, r_k \:|\: n_k] \:dz_k \; 
A^{j_k}_c(z_k) \;(y-x)^j \gamma_j \; S^{(-1)}(x,y)
\end{eqnarray*}
with $c=L$ or $c=R$. Thus Theorem \ref{l:thm1} makes a precise statement on
the mathematical structure of all the contributions to the light-cone expansion. 
However, it does not give information about the values 
of the parameters $C$, $l_a$, $r_a$, and $n_a$.
In order to see more precisely how the leading order on the light 
cone looks like, we perform the light-cone expansion directly with 
Lemma \ref{l:lemma2}. To first order in the external potential, we obtain
\begin{eqnarray}
\lefteqn{ (-s \:B\: s)(x,y) \;\stackrel{(\ref{l:11a}),(\ref{l:29z})}{=}\;
\frac{1}{2} \:\Pdd_x \int d^4z \:S^{(0)}(x,z) \left( B(z)\: (y-z)^k 
\gamma_k \right) S^{(-1)}(z,y) } \nonumber \\
&\stackrel{(\ref{l:21})}{\asymp}& \frac{1}{2} \:\Pdd_x \int_x^y [0,-1 \:|\: 
0] \:dz\; B(z) \:(y-z)^k \gamma_k \; S^{(0)}(x,y) \nonumber \\
&\stackrel{(\ref{l:29z})}{\asymp}& \frac{1}{4} \int_x^y dz\; (y-x)^j
\gamma_j \:B(z) \:(y-x)^k \gamma_k \; S^{(-1)}(x,y) \nonumber \\
&\asymp& \frac{1}{2} \int_x^y dz\; (\chi_L \:A^j_L(z) + \chi_R 
\:A^j_R(z))\:(y-x)_j \:(y-x)^k \gamma_k \; S^{(-1)}(x,y) \nonumber \\
&\stackrel{(\ref{l:29z}),(\ref{l:11a})}{=}& -i \:(y-x)_j \int_x^y dz \;
(\chi_L \:A^j_L(z) + \chi_R \:A^j_R(z)) \;s(x,y) \;\;\; ,
\label{l:85a}
\end{eqnarray}
where `$\asymp$' denotes that we only take the leading order on the 
light cone. Since (\ref{l:85a}) is a product of a smooth function with a
single factor $s(x,y)$, this formula can be immediately iterated. We obtain
for the left and right handed component of the
$k^{\mbox{\scriptsize{th}}}$ order Feynman diagram
\begin{eqnarray}
\lefteqn{ \chi_c \:((-s \:B)^k s)(x,y) \;\asymp\; \chi_c \:(-i)^k 
\:(y-x)_{j_1} \cdots (y-x)_{j_k} \int_x^y [0,k-1 \:|\: 0] \:dz_1 \; 
A^{j_1}_c(z_1) } \nonumber \\
&& \times \int_{z_1}^y [0,k-2 \:|\: 0] \:dz_2 \; 
A^{j_2}_c(z_2) \cdots \int_{z_{k-1}}^y [0,0 \:|\: 0] \:dz_k \; A^{j_k}_c(z_k) \; 
s(x,y) \;\;\; . \spc
\label{l:86a}
\end{eqnarray}
(Notice that, according to Lemma \ref{l:lemma1}, the higher order terms 
which were neglected in (\ref{l:85a}) also give contributions of the 
order ${\cal{O}}((y-x)^{-2})$ in the iteration. Therefore it is really 
sufficient to take only the leading contribution on the light cone
in every step.)
The line integrals in (\ref{l:86a}) are particularly simple. Namely, 
they are the $k^{\mbox{\scriptsize{th}}}$ order contributions to the 
familiar Dyson series. From this we conclude that, to 
leading order on the light cone, the sum over all Feynman diagrams converges
absolutely. We can carry out the sum and obtain
\begin{equation}
        \chi_c \:\tilde{s}(x,y) \;\asymp\; \chi_c \:\Pexp \left( -i \int_0^1 
        A^j_{c\:|\alpha y + (1-\alpha) x} \: (y-x)_j \: d\alpha \right) 
        \:s(x,y) \;\;\; ,
        \label{l:87a}
\end{equation}
where $\Pexp$ is the usual ordered exponential. For completeness, we 
give its definition:
\begin{Def}
For a smooth one-parameter family of matrices $F(\alpha)$, $\alpha 
\in \R$, the ordered exponential $\Pexp (\int F(\alpha) \:d\alpha)$
is given by the Dyson series
\begin{eqnarray*}
\Pexp \left( \int_a^b F(\alpha) \:d\alpha \right) 
&=& \1 \:+\: \int_a^b dt_0 \:F(t_0) \:dt_0 \:+\: \int_a^b dt_0 
        \:F(t_0) \int_{t_0}^b dt_1 \: F(t_1) \\
&&+\int_a^b dt_0 
        \:F(t_0) \int_{t_0}^b dt_1 \: F(t_1) \int_{t_1}^b dt_2 \:F(t_2) 
        \:+\: \cdots \;\;\; .
\end{eqnarray*}
\end{Def}
As is verified by a direct calculation, the ordered exponential is a 
solution of the ordinary differential equation
\begin{equation}
        \frac{d}{da} \Pexp \left( \int_a^b F(\alpha) \:d\alpha \right) 
        \;=\; -F(a) \:\Pexp \left( \int_a^b F(\alpha) \:d\alpha \right)
        \label{l:71a}
\end{equation}
with the boundary conditions
\begin{equation}
        \Pexp \left( \int_b^b F(\alpha) \:d\alpha \right) \;=\; \1 \;\;\; .
        \label{l:71aa}
\end{equation}
Because of the uniqueness of the solutions of ordinary differential 
equations, one can alternatively take (\ref{l:71a}),(\ref{l:71aa}) as the 
definition for the ordered exponential.

For the ordered exponential in (\ref{l:87a}), we also use the shorter 
notations
\begin{equation}
\Pexp \left(-i \int_x^y A^j_c \:(y-x)_j \right) \spc {\mbox{and}} \spc
\Pe^{-i \int_x^y A^j_c \:(y-x)_j} \;\;\; .
        \label{l:t30}
\end{equation}
Notice that (\ref{l:t30}) is a unitary
$(f \times f)$-matrix which depends only on the 
chiral potentials along the line segment $\overline{xy}$.
Its inverse is\footnote{We mention for clarity that this is in 
general not the same as
\begin{equation}
\Pexp \left(i \int_x^y A^j_c \:(y-x)_j \right) \;\;\; ,
        \label{l:75c}
\end{equation}
because the exponentials in (\ref{l:75f}) and (\ref{l:75c}) are ordered in 
opposite directions.}
\begin{equation}
        \Pexp \left(-i \int_x^y A^j_c \:(y-x)_j \right)^\dagger \;=\;
        \Pexp \left(-i \int_y^x A^j_c \:(x-y)_j \right) \;\;\; .
        \label{l:75f}
\end{equation}
If the chiral potentials are commutative, i.e.\ $[A_c(x), A_c(y)]=0$, 
then the ordered exponential coincides with the ordinary 
exponential (this is e.g.\ the case if one considers the system $f=1$ 
of only one Dirac sea). For the ordered exponential (\ref{l:t30}) along 
a line segment in Minkowski space, the differential equation (\ref{l:71a}) can
be written with partial derivatives as
\begin{equation}
(y-x)^k \frac{\partial}{\partial x^k} \Pe^{-i \int_x^y A^j_c \:(y-x)_j}
\;=\; i (y-x)_k \:A^k_c(x) \; \Pe^{-i \int_x^y A^j_c \:(y-x)_j} \;\;\; .
        \label{l:71x}
\end{equation}

We conclude that, to leading order on the light cone, the 
special form of the formulas of the light-cone expansion allows us to 
immediately carry out the sum over all Feynman diagrams.
Unfortunately, the situation to higher order on the light cone is more
difficult, because the combinatorics of the partial derivatives and 
of the tensor contractions becomes very complicated. We cannot 
expect that the sum over all Feynman diagrams can then still
be written in a simple, closed form.
Nevertheless, ordered exponentials over the chiral
potentials should be helpful. More precisely, it is promising to write
the light-cone expansion with line integrals which contain 
intermediate ordered exponentials, like e.g.\ the line integral
\begin{equation}
\int_x^y dz \; \Pe^{-i \int_x^z A^j_L \:(z-x)_j} 
\;(\OBox \Aslsh_L(z)) \; \Pe^{-i \int_z^y A^k_L \:(y-z)_k}
        \label{l:ex1}
\end{equation}
(expressions of this form are also suggested in view of the behavior 
of the fermionic projector under local gauge transformations of 
the external potential).
Our basic idea is to arrange the contributions
to the light-cone expansion of Theorem \ref{l:thm2}
to any given order on the light cone in such a way that all infinite sums 
(which arise from the fact that we have an infinite number of Feynman 
diagrams) can be carried out giving ordered exponentials.
We want to end up with a {\em{finite}} number of terms which are 
of the form (\ref{l:70}) with the only exception that 
the nested line integrals contain, similar to (\ref{l:ex1}), additional 
ordered exponentials.

Before we can make this idea mathematically precise, we must clarify 
the connection between the line integrals in (\ref{l:70}) and the 
line integrals with intermediate ordered exponentials. For this, we 
consider the example (\ref{l:ex1}). If we expand the ordered 
exponentials in a Dyson series and reparametrize the integrals, 
(\ref{l:ex1}) goes over into an infinite sum of nested line integrals 
of the form as in (\ref{l:70}), more precisely
\begin{eqnarray}
\lefteqn{ (\ref{l:ex1}) \;=\; \sum_{p,q=0}^\infty
\int_x^y [l_1, r_1 \:|\: n_1] \:dz_1 \:A^{j_1}_L(z_1) \:(y-x)_{j_1}
\int_{z_1}^y [l_2, r_2 \:|\: n_2] \:dz_2 \:A^{j_2}_L(z_2) \:(y-x)_{j_2}
} \nonumber \\
&& \spc \cdots \int_{z_{p-1}}^y [l_p, r_p \:|\: n_p] \:dz_p
\:A^{j_p}_L(z_p)\: (y-x)_{j_p} \int_{z_p}^y [l, r \:|\: n] \:dz \;
\OBox \Aslsh_L(z) \nonumber \\
&&\times \int_z^y [l_{p+1}, r_{p+1} \:|\: n_{p+1}] \:dz_{p+1} \:
A^{j_{p+1}}_L(z_{p+1})  \:(y-x)_{j_{p+1}} \int_{z_{p+1}}^y
A^{j_{p+2}}_L(z_{p+2}) \:(y-x)_{j_{p+2}} \nonumber \\
&& \spc \cdots \int_{z_{p+q-1}}^y [l_{p+1}, r_{p+q} \:|\: 
n_{p+1}] \:dz_{p+q} \:A^{j_{p+q}}_L(z_{p+q}) \:(y-x)_{j_{p+q}} \;\;\; .
\label{l:e1}
\end{eqnarray}
The formulas of our desired light-cone expansion must be such 
that, after expanding the ordered exponentials in this way, we get 
precisely all the contributions to the light-cone expansion of 
Theorem \ref{l:thm2}. We point out that we can view the 
expansion (\ref{l:e1}) as a power series in the functions 
$A^j_{L\!/\!R} \:(y-x)_j$. The leading contribution to this power 
series is simply the line integral
\[ \int_x^y dz \; \OBox \Aslsh_L(z) \;\;\; ; \]
it is obtained from (\ref{l:ex1}) by taking out the ordered exponentials 
there. In view of this example, we can hope to get the contributions to our
desired expansion without the ordered exponentials (i.e.\ with all the 
ordered exponentials removed from the formulas) by picking those
contributions to the light-cone expansion of Theorem \ref{l:thm2}
which contain no factors $A^j_{L\!/\!R}\:(y-x)_j$.
We take these contributions as the starting point for our construction.
\begin{Def}
\label{l:def_pf}
A contribution (\ref{l:70}),(\ref{l:71}) to the light-cone expansion of Theorem
\ref{l:thm2} is called {\bf{phase-free}} if all the tangential potentials
$V^{(a)}_{J_a}$ are differentiated, i.e.
\[ |K_a|+2 p_a > 0 \spc{\mbox{whenever}}\spc {\mbox{$J_a$ is contracted with 
$(y-x)^K$.}} \]
\end{Def}
To leading order on the light cone, the only phase-free contribution is the 
free Green's function $s$ (namely, according to (\ref{l:86a}), the 
contributions with $k \geq 1$ all contain factors 
$A^j_{L\!/\!R}\:(y-x)_j$). The restriction to the phase-free 
contribution also simplifies the situation in the general case. 
Namely, the following proposition shows that the phase-free contributions of 
the higher order Feynman diagrams involve higher mass-derivatives 
of the Green's functions.
\begin{Prp}
\label{l:prp1}
For every phase-free contribution (\ref{l:70}) to the light-cone expansion of
the $k^{\mbox{\scriptsize{th}}}$ order Feynman diagram $((-s \:B)^k 
s)(x,y)$, the parameter $h$ satisfies the bound
\begin{equation}
h \;\geq\; -1 + \left[ \frac{k+1}{2} \right] \;\;\; ,
        \label{l:P1}
\end{equation}
where $[.]$ denotes the Gau{\ss} bracket.
\end{Prp}
{\Proof}
Consider a phase-free contribution (\ref{l:70}) of the 
$k^{\mbox{\scriptsize{th}}}$ order Feynman diagram.
According to the rules for the possible contractions of the 
tensor indices, only one factor $(y-x)$ may be contracted 
with $\gamma^J$; the remaining $|K|-1$ factors $(y-x)$ must be contracted 
with the $V^{(a)}_{J_a, c_a}$. Thus at least $|K|-1$ potentials are 
tangential and must (according to Def.\ \ref{l:def_pf}) be differentiated.
This gives the inequality
\[ |K|-1 \;\leq\; \sum_{a=1}^k (|K_a| + 2 p_a) \;\;\; . \]
We substitute this bound into 
(\ref{l:71b}) and obtain $2h \geq -2+k$; this is equivalent to 
(\ref{l:P1}).
\hspace*{5cm} \QED
According to the explicit formula (\ref{l:12}), the higher 
mass-derivatives of the Green's functions are of higher order on the 
light cone. More precisely, (\ref{l:P1}) and  (\ref{l:24b}) yield that 
the phase-free contribution to the light-cone expansion of the 
$k^{\mbox{\scriptsize{th}}}$ order Feynman diagram is of the order
\begin{equation}
        {\cal{O}}((y-x)^{2g}) \spc {\mbox{with}} \spc
        g=-2+\left[\frac{k+1}{2} \right] \;\;\; .
        \label{l:29n}
\end{equation}
This means that, to every order on the light cone, only a
finite number of Feynman diagrams contribute. As a consequence, there 
are, to every order on the light cone, only a finite number of 
phase-free terms.

The phase-free contributions are useful because our 
desired light-cone expansion is obtained from them by inserting 
ordered exponentials into the line integrals. We do this ``by hand,'' 
according to simple rules.
\begin{Def}
\label{l:def2}
For every phase-free contribution (\ref{l:70}) to the light-cone 
expansion of Theorem \ref{l:thm2}, we introduce a corresponding
{\bf{phase-inserted contribution}}. It is constructed according to the
following rules:
\begin{description}
\item[\rm{\em{(I)}}] We insert one ordered exponential into each line integral 
and one ordered exponential at the very end. More precisely, the 
phase-inserted contribution has the form
\begin{eqnarray}
\lefteqn{ \chi_c\:C \:(y-x)^K \:W^{(0)}(x) \int_x^y [l_1, r_1 \:|\: n_1] \:dz_1
\;\Pe^{-i \int_x^{z_1} A^{j_1}_{c_1} \:(z_1-x)_{j_1}} \: 
W^{(1)}(z_1) } \nonumber \\
&&\times\int_{z_1}^y [l_2, r_2 \:|\: n_2] \:dz_2 \;
\Pe^{-i \int_{z_1}^{z_2} A^{j_2}_{c_2} \:(z_2-z_1)_{j_2}} \:W^{(2)}(z_2) 
\;\cdots \nonumber \\
&&\times \int_{z_{\alpha-1}}^y [l_\alpha, r_\alpha \:|\: 
n_\alpha] \:dz_\alpha \; 
\Pe^{-i \int_{z_{\alpha-1}}^{z_\alpha} A^{j_\alpha}_{c_\alpha} 
\:(z_{\alpha}-z_{\alpha-1})_{j_1}}
\:W^{(\alpha)}(z_\alpha) \nonumber \\
&&\hspace*{2cm} \times
\Pe^{-i \int_{z_\alpha}^y A^{j_{\alpha+1}}_{c_{\alpha+1}} 
\:(y-z_\alpha)_{j_{\alpha+1}}} \:\;\gamma^J \; S^{(h)}(x,y) \;\;\; .
\label{l:p30}
\end{eqnarray}
\item[\rm{\em{(II)}}] The chirality $c_\beta$, $\beta=1,\ldots, 
\alpha+1$ of the potentials in the ordered exponentials is determined 
by the number of dynamic mass matrices in the factors $W^{(\beta)}$; 
namely
\[ {\mbox{$c_{\beta-1}$ and $c_\beta$}} \left\{ \!\!\!\begin{array}{c} 
{\mbox{coincide}} \\ {\mbox{are opposite}} \end{array} \!\!\right\}
{\mbox{if $W^{(\beta-1)}$ contains an}} \left\{ \!\!\!\begin{array}{c} 
{\mbox{even}} \\ {\mbox{odd}} \end{array} \!\!\right\}
{\mbox{number of factors $Y$,}} \]
where $c_0:=c$ is the chirality of the projector $\chi_c$ in 
(\ref{l:p30}).
\end{description}
\end{Def}
To illustrate these insertion rules, we consider the example 
of two nested line integrals
\[ \chi_L \int_x^y dz_1 \;(\OBox \Aslsh_L)(z_1)
\int_{z_1}^y dz_2 \; m Y_L(z_2) \;S^{(1)}(x,y) \;\;\; . \]
The corresponding phase-inserted contribution is
\begin{eqnarray*}
\lefteqn{ \chi_L \int_x^y dz_1 \;\Pe^{-i \int_x^{z_1} A_L^j 
\:(z_1-x)_j} \:(\OBox \Aslsh_L)(z_1) } \\
&&\times \int_{z_1}^y dz_2 \;\Pe^{-i \int_{z_1}^{z_2} A_L^k \:(z_2-z_1)_k}
\: m Y_L(z_2) \:\Pe^{-i \int_{z_2}^y A_R^l \:(y-z_2)_l}
\;S^{(1)}(x,y) \;\;\; .
\end{eqnarray*}

\begin{Thm}
\label{l:thm3}
The light-cone expansion of the Green's function $\tilde{s}(x,y)$ 
coincides with the sum of all phase-inserted contributions.
\end{Thm}
{\Proof}
A possible method for the proof would be to rearrange all the
contributions to the light-cone expansion of Theorem \ref{l:thm2}
until recovering the Dyson 
series of the ordered exponentials in (\ref{l:p30}). However, this method
has the disadvantage of being technically complicated. It is more elegant
to use a particular form of ``local gauge invariance'' of the Green's function
for the proof: First we will, for given $x$ and $y$, locally transform the
spinors. 
The transformation will be such that the light-cone expansion for the 
transformed Green's function $\hat{s}(x,y)$ consists precisely of all 
phase-free contributions. Using the transformation law of 
the Green's function, we then show that the light-cone expansion of 
$\tilde{s}(x,y)$ is obtained from that of $\hat{s}(x,y)$ by inserting 
unitary matrices into the line integrals. Finally, we prove that 
these unitary matrices coincide with the ordered exponentials in Def.\ 
\ref{l:def2}.

In preparation, we consider the transformation law of the Dirac 
operator and the Green's function under generalized local phase 
transformations of the spinors: We let $U_L$ and $U_R$ be two unitary matrix 
fields acting on the Dirac sea index,
\[ U_L \;=\; (U_L(x)^l_m)_{l,m=1,\ldots,f} \;,\;\;\;
U_R \;=\; (U_R(x)^l_m)_{l,m=1,\ldots,f} 
\;\;\;\;\;{\mbox{with}}\;\;\;\;\; U_L\:U_L^*=\1=U_R\:U_R^*, \]
and transform the wave functions
$\Psi=(\Psi^\alpha_l(x))^{\alpha=1,\ldots,4}_{l=1,\ldots,f}$ according to
\[      \Psi(x) \;\rightarrow\; \hat{\Psi}(x) \:=\: U(x)\:\Psi(x) \spc
        {\mbox{with}} \spc
        U(x) \;=\; \chi_L \:U_L(x) \:+\: \chi_R \:U_R(x) \;\;\; . \]
Thus $U_L$ and $U_R$ independently transform the left and right handed 
component of the wave functions, respectively.
Notice that the transformation $U$ is {\em{not}} unitary
with respect to our scalar product (\ref{l:3s}), because
\begin{eqnarray*}
V &:=& U^{-1} \;=\; \chi_L \:U_L^{-1} \:+\: \chi_R \:U_R^{-1} \spc 
{\mbox{but}} \\
U^* &=& \gamma^0 \:U^\dagger \:\gamma^0 \;=\;
\chi_R \:U_L^{-1} \:+\: \chi_L \:U_R^{-1} \;\;\; .
\end{eqnarray*}
Therefore we must in the following carefully distinguish between $U$, 
$U^*$ and their inverses $V$, $V^*$. As an immediate consequence of 
the Dirac equation $(i \Pdd + {\cal{B}} - m) \Psi =0$, the transformed
wave functions $\hat{\Psi}$ satisfy the equation
\[ V^* (i \Pdd + B) V \:\hat{\Psi} \;=\; 0 \;\;\; . \]
A short computation yields for the transformed Dirac operator
\[ V^* (i \Pdd + B) V \;=\; i \Pdd + \hat{B} \]
with
\begin{equation}
\hat{B} \;=\; \chi_L \:(\hat{A}_R \!\!\!\!\!\!\!\slash \;\:- m \:\hat{Y}_R) \:+\:
\chi_R \:(\hat{A}_L \!\!\!\!\!\!\!\slash \;\;- m \:\hat{Y}_L) \;\;\; ,
\end{equation}
whereby $\hat{A}_{L\!/\!R}$ and $\hat{Y}_{L\!/\!R}$ are the potentials
\begin{eqnarray}
\hat{A}_{L\!/\!R}^j &=& U_{L\!/\!R} \:A_{L\!/\!R}^j \:U_{L\!/\!R}^{-1} 
\:+\: i U_{L\!/\!R} (\partial^j U_{L\!/\!R}^{-1}) \label{l:p8} \\
\hat{Y}_{L\!/\!R} &=& U_{L\!/\!R} \:Y_{L\!/\!R} \:U_{R\!/\!L}^{-1} 
\;\;\; . \label{l:p9}
\end{eqnarray}
We denote the advanced and retarded Green's functions of the transformed
Dirac operator $i \Pdd + \hat{B}$ by $\hat{s}$. They satisfy the
equation
\begin{equation}
(i \Pdd_x \:+\: \hat{B}(x)) \:\hat{s}(x,y) \;=\; \delta^4(x-y) \;\;\; .
        \label{l:p0}
\end{equation}
Since we can view $\hat{B}$ as the perturbation of the Dirac 
operator, the Green's function $\hat{s}$ is, in analogy to 
(\ref{l:11b}), given by the perturbation series
\begin{equation}
\hat{s} \;=\; \sum_{n=0}^\infty (-s \:\hat{B})^n \:s \;\;\; .
        \label{l:p10}
\end{equation}
The important point for the following is that the Green's functions 
$\tilde{s}$ and $\hat{s}$ are related to each other by the local 
transformation
\begin{equation}
\hat{s}(x,y) \;=\; U(x) \:\tilde{s}(x,y) \:U(y)^* \;\;\; .
        \label{l:p1}
\end{equation}
This is verified as follows: The right side of (\ref{l:p1}) also satisfies the
defining equation (\ref{l:p0}) of the Green's functions; namely
\begin{eqnarray*}
\lefteqn{ (i \Pdd_x + \hat{B}(x)) \; U(x) 
\:\tilde{s}(x,y)\:U(y)^* \;=\; V(x)^* \:(i \Pdd_x + B(x)) \:V(x) \;U(x) 
\:\tilde{s}(x,y)\:U(y)^* } \\
&=& V(x)^* \:(i \Pdd_x + B(x)) \:\tilde{s}(x,y)\:U(y)^* \;=\;
V(x)^* \:\delta^4(x-y) \:U(y)^* \spc\spc \\
&=&  V(x)^* \:U(x)^* \:\delta^4(x-y) \;=\; \delta^4(x-y) \;\;\; .
\end{eqnarray*}
Furthermore, the supports of both sides of (\ref{l:p1}) lie (depending 
on whether we consider the advanced or retarded Green's functions) 
either in the upper or in the lower light cone. A uniqueness argument 
for the solutions of hyperbolic differential equations yields that 
both sides of (\ref{l:p1}) must coincide.

Next, we specify the unitary transformations $U_L$ and $U_R$: We fix the
points $x$ and $y$. For any point $z$ on the line segment $\overline{xy}$,
we chose $U_{L\!/\!R}(z)$ as
\begin{equation}
U_{L\!/\!R}(z) \;=\; \Pexp \left(-i \int_x^z A^j_{L\!/\!R} \;
(z-x)_j \right) \;\;\; . \label{l:p3}
\end{equation}
The differential equation (\ref{l:71x}) yields that
\begin{eqnarray*}
(y-x)^j \;U_c(z) \:(\partial_j U_c(z)^{-1}) &=& \Pe^{-i \int_x^z 
A^k_c \: (z-x)_k} \:(y-x)^j \frac{\partial}{\partial z^j}
\Pe^{-i \int_z^x A^k_c \: (x-z)_k} \\
&=& \Pe^{-i \int_x^z A^k_c \: (z-x)_k} \:i (y-x)_j\:A^j_c(z) \:
\Pe^{-i \int_z^x A^k_c \: (x-z)_k} \\
&=& i (y-x)_j \:U_c(z) \:A^j_c(z) \: U_c(z)^{-1} \;\;\; .
\end{eqnarray*}
Substituting into (\ref{l:p8}) gives
\begin{equation}
\hat{A}^j_{L\!/\!R}(z) \:(y-x)_j \;=\; 0 \spc
{\mbox{for $z \in \overline{xy}$.}}
        \label{l:p11}
\end{equation}
Thus our choice of $U_L$ and $U_R$ makes the potentials $\hat{A}_L(z)$ 
and $\hat{A}_R(z)$, $z \in \overline{xy}$, orthogonal to $(y-x)$. 
Notice, however, that since $Y_{L\!/\!R}$ in (\ref{l:p9}) is arbitrary and
independent of $U_{L\!/\!R}$, (\ref{l:p3}) gives no constraints for the 
dynamic mass matrices $\hat{Y}_{L\!/\!R}$.

We point out that we did not specify $U_{L\!/\!R}(z)$ outside the 
line segment $z \in \overline{xy}$; the unitary transformation $U_{L\!/\!R}$ 
may be arbitrary there. This means that also $\hat{A}_{L\!/\!R}$ is 
undetermined outside the line segment $\overline{xy}$. Especially, all the 
non-tangential derivatives of $\hat{A}_{L\!/\!R}(z)$, $z \in \overline{xy}$,
are undetermined. However, equation (\ref{l:p3}) does give
constraints for the tangential derivatives. For example, 
differentiating (\ref{l:p11}) in the direction $(y-x)$ yields
\begin{equation}
        (y-x)^j \:(y-x)_k \:\partial_j \hat{A}^k_{L\!/\!R}(z) =0 \spc 
        {\mbox{for}} \spc z \in \overline{xy} \;\;\; .
        \label{l:21v}
\end{equation}

We now consider the perturbation expansion (\ref{l:p10}). The light-cone 
expansion of all Feynman diagrams according to Theorem \ref{l:thm2} 
gives a sum of terms of the form
\begin{eqnarray}
\lefteqn{ \chi_c\:C \:(y-x)^K \:\hat{W}^{(0)}(x) \int_x^y [l_1, r_1 \:|\: n_1]
\:dz_1 \; \hat{W}^{(1)}(z_1) \int_{z_1}^y [l_2, r_2 \:|\: n_2] \:dz_2 \;
\hat{W}^{(2)}(z_2) } \nonumber \\
&&\hspace*{1cm} \cdots \int_{z_{\alpha-1}}^y [l_\alpha, r_\alpha \:|\: 
n_\alpha] \:dz_\alpha \; \hat{W}^{(\alpha)}(z_\alpha) \;
\gamma^J \; S^{(h)}(x,y) \;\;\; , \spc \spc
\label{l:t1}
\end{eqnarray}
where the factors $\hat{W}^{(\beta)}$ are given by
\begin{equation}
        \hat{W}^{(\beta)} \;=\; (\partial^{K_{a_\beta}} \OBox^{p_{a_\beta}} 
        \hat{V}^{(a_\beta)}_{J_{a_\beta}, c_{a_\beta}})
        \cdots (\partial^{K_{b_\beta}} \OBox^{p_{b_\beta}} 
        \hat{V}^{(b_\beta)}_{J_{b_\beta}, c_{b_\beta}}) \;\;\; .
        \label{l:t2}
\end{equation}
Because of condition (\ref{l:p11}), all the contributions which are not 
phase-free vanish. Furthermore, according to Theorem \ref{l:thm2}, the 
contributions (\ref{l:t1}),(\ref{l:t2}) contain no tangential derivatives. 
Clearly, the derivatives in these formulas may have a component in 
direction of $(y-x)$; but the contribution of the derivatives transversal 
to $(y-x)$ uniquely determines the form of each derivative term. 
Therefore all the phase-free contributions of the form 
(\ref{l:t1}),(\ref{l:t2}) are independent in the sense that we have no 
algebraic relations between them. We conclude that, as long as the 
potentials $\hat{A}_{L\!/\!R}$ and $\hat{Y}_{L\!/\!R}$ are only specified 
by (\ref{l:p8}), (\ref{l:p9}), and (\ref{l:p3}), the light-cone expansion 
(\ref{l:t1}),(\ref{l:t2}) consists precisely of all phase-free 
contributions.

Next, we exploit the local transformation law (\ref{l:p1}) of the Green's 
functions: We solve this equation for $\tilde{s}$,
\begin{equation}
        \tilde{s}(x,y) \;=\; V(x) \:\hat{s}(x,y) \:V(y)^* \;\;\; .
        \label{l:t0}
\end{equation}
The transformation $U_{L\!/\!R}$ does not enter on the 
left side of this equation. Thus the right side of (\ref{l:t0}) 
is also independent of $U_{L\!/\!R}$. Especially, we conclude that the 
light-cone expansion of $\hat{s}(x,y)$ must be independent of the 
derivatives of $U_{L\!/\!R}$ along the line segment $\overline{xy}$.
At first sight, this might seem inconsistent because the individual 
contributions (\ref{l:t1}),(\ref{l:t2}) do depend on the derivatives of 
$U_{L\!/\!R}$ (this is obvious if one substitutes (\ref{l:p8}) and 
(\ref{l:p9}) into (\ref{l:t2}) and carries out the derivatives with the 
Leibniz rule). The right way to understand the independence of 
$\hat{s}(x,y)$ on the derivatives of $U_{L\!/\!R}$ is that
all derivative terms of $U_{L\!/\!R}$ cancel each other to every 
order on the light cone if the (finite) sum over all contributions 
(\ref{l:t1}) to the light-cone expansion of $\hat{s}(x,y)$ is carried out.
Since we will form the sum over all contributions to the light-cone 
expansion in the end, it suffices to consider only those contributions to the
light-cone expansion which contain no derivatives of $U_{L\!/\!R}$.
This means that we can substitute (\ref{l:p8}) and 
(\ref{l:p9}) into (\ref{l:t2}), forget about the derivative term $i 
U_{L\!/\!R} (\partial^j U_{L\!/\!R}^{-1})$ in 
(\ref{l:p8}), and pull the unitary transformations $U_{L\!/\!R}, U_{L\!/\!R}^{-1}$
out of the derivatives. In other words, we can replace $\hat{W}^{(\beta)}$, 
(\ref{l:t2}), by
\begin{equation}
        \hat{W}^{(\beta)} \;=\; U_{c_{a_\beta}}
        (\partial^{K_{a_\beta}} \OBox^{p_{a_\beta}} 
        V^{(a_\beta)}_{J_{a_\beta}, c_{a_\beta}}) U_{d_{a_\beta}}^{-1} \:
        \cdots U_{c_{b_\beta}} (\partial^{K_{b_\beta}} \OBox^{p_{b_\beta}} 
        V^{(b_\beta)}_{J_{b_\beta}, c_{b_\beta}}) U_{d_{b_\beta}}^{-1}
        \label{l:t3}
\end{equation}
with chiral indices $c_a, d_a = L\!/\!R$.
The light-cone expansion for $\hat{s}(x,y)$ consists precisely of the sum of 
all phase-free contributions of the form (\ref{l:t1}), (\ref{l:t3}).

The chiralities $c_a$, $d_a$ of the unitary transformations
$U_{L\!/\!R}$, $U_{L\!/\!R}^{-1}$ in (\ref{l:t3}) are determined by the rules 
{\em{(i)}} and {\em{(ii)}} (in Theorem \ref{l:thm1})
and by (\ref{l:p8}) and (\ref{l:p9}).
According to the rule {\em{(ii)}}, the indices $c_a$ and $c_{a+1}$ 
coincide iff $V^{(a)}$ is a chiral potential. According to (\ref{l:p8}) 
and (\ref{l:p9}), on the other hand, the indices $c_a$ and $d_a$ 
coincide iff $V^{(a)}=A_{L\!/\!R}$. We conclude that the indices $d_a$ and 
$c_{a+1}$ always coincide. Thus all the intermediate factors
$U_{d_a} \:U_{c_{a+1}}$ give the identity, and (\ref{l:t3}) simplifies to
\begin{equation}
\hat{W}^{(\beta)} \;=\; U_{c_\beta} \:W^{(\beta)} 
\:U_{d_\beta}^{-1} \;\;\; .
        \label{l:t4}
\end{equation}
Furthermore, the chiralities $c_\beta$ and $d_\beta$ coincide if and only if 
$W^{(\beta)}$ contains an even number of dynamic mass matrices.

Finally, we substitute the light-cone expansion (\ref{l:t1}), 
(\ref{l:t4}) for $\hat{s}(x,y)$ into (\ref{l:t0}). This gives for the light-cone 
expansion of $\tilde{s}(x,y)$ a sum of expressions of the form
\begin{eqnarray}
&&\chi_c\:C \:(y-x)^K \:U_c^{-1}(x)
\:(U_{c_0} W^{(0)} U_{d_0}^{-1})(x) \int_x^y [l_1, r_1 \:|\: n_1] \:dz_1 \;
(U_{c_1} W^{(1)} U_{d_1}^{-1})(z_1) \nonumber \\
&&\hspace*{1cm} \cdots \int_{z_{\alpha-1}}^y [l_\alpha, r_\alpha \:|\: 
n_\alpha] \:dz_\alpha \; (U_{c_\alpha} W^{(0)} U_{d_\alpha}^{-1})(z_\alpha) 
\;U_{c_{\alpha+1}}(y) \;
\gamma^J \; S^{(h)}(x,y) \;\;\; ,\spc\;\;\;
\label{l:t5}
\end{eqnarray}
where the sum runs over all phase-free contributions of this type. 
Similar to the considerations before (\ref{l:t4}), one sees that
adjacent unitary transformations always have the same 
chirality. Thus (\ref{l:t5}) can be simplified to
\begin{eqnarray*}
\lefteqn{ \chi_c\:C \:(y-x)^K \:W^{(0)}(x) \int_x^y [l_1, r_1 \:|\: n_1]
\:dz_1 \; U_{c_1}(x)^{-1} \: U_{c_1}(z_1) \: W^{(1)} } \\
&&\cdots \int_{z_{\alpha-1}}^y [l_\alpha, r_\alpha \:|\: 
n_\alpha] \:dz_\alpha \; U_{c_\alpha}(z_{\alpha-1})^{-1} \: 
U_{c_\alpha}(z_\alpha) \:W^{(0)}(z_\alpha) \;U_{c_{\alpha+1}}(z_\alpha)^{-1} \\
&& \times \: U_{c_{\alpha+1}}(y) \; \gamma^J \; S^{(h)}(x,y) \;\;\; ,
\end{eqnarray*}
whereby the indices $c_a$ satisfy the rule {\em{(II)}}
of Def.\ \ref{l:def2}. 
According to (\ref{l:p3}), the factors $U_c^{-1}(.) \:U_c(.)$ coincide 
with the ordered exponentials in (\ref{l:p30}), which concludes the proof.
\hspace*{5cm} \QED
For clarity, we point out that all the constructions following Def.\ 
\ref{l:def2} are based on the light-cone expansion of Theorem 
\ref{l:thm2}. It is essential for the statement of Theorem \ref{l:thm3} 
that the phase-free contributions contain no tangential derivatives. 
If we had worked with the light-cone expansion of Theorem \ref{l:thm1} 
(instead of Theorem \ref{l:thm2}), the light-cone expansion of 
$\hat{s}(x,y)$ would not have consisted of all the phase-free 
contributions to the light-cone expansion. For example, a line 
integral containing the tangential derivative (\ref{l:21v}) would 
vanish, although it is phase-free. As a consequence of this problem,
the whole construction would break down.

The introduction of the phase-free and phase-inserted contributions 
has simplified the light-cone expansion of the Green's functions 
considerably: Assume that we want to perform the light-cone expansion 
to some given order on the light cone. Then we first calculate the 
phase-free contribution to the light-cone expansion; according to 
Proposition \ref{l:prp1}, this gives only a finite number of terms. Using 
the rules of Def.\ \ref{l:def2}, we can easily construct the 
corresponding phase-inserted contributions. According to Theorem 
\ref{l:thm3}, this finite number of phase-inserted contributions gives
precisely the light-cone expansion of the Green's functions to the 
desired order on the light cone. This procedure is called the
{\em{reduction to the phase-free contribution.}}

\subsection{Calculation of the Phase-Free Contribution}
\label{l:sec2_3}
According to the reduction to the phase-free contribution, it remains 
to calculate the phase-free contribution to any given order on the 
light cone. Although this is still a very complicated problem, we know 
from Proposition \ref{l:prp1} that we only get a finite number of 
terms. This makes it possible to use a computer algebra 
program for the calculation. The author has developed the C++ 
program ``class\_commute'' specifically for this problem. It generates 
explicit formulas for the phase-free contribution to any order on 
the light cone. We now outline how this program works, without entering
implementation details\footnote{The commented source code of the 
program ``class\_commute'' is available from the author on request. It 
is an extension of the program used in \cite{F2} for the light-cone 
expansion to first order in the external potential.}.

All the objects occurring in the calculation (like integrals, partial 
derivatives, Laplacians, potentials, Dirac matrices, etc.) are described 
by different data structures (classes in C++). Formulas are built 
up as sums of lists of these data structures.
The calculation is performed by manipulating the lists. More precisely, this
works as follows: Each data structure carries an ordering number.
At the beginning of the computation, the lists are disordered in the 
sense that their elements do not occur with increasing ordering numbers.
For ordering the lists, the program iteratively commutes adjacent elements
of a list. Each commutation is performed by a function of the program 
which is specific to the particular pair of data structures; i.e., there 
is a function for commuting a partial derivative with a potential, a 
function for commuting two Dirac matrices, etc. (this is easily implemented in
C++ using virtual class functions). The data structures and 
commutation rules are designed in such a way that, after the ordering process
has come to an end, the lists consist of the desired formulas of the
light-cone expansion.

The main advantage of this implementation with commutation rules is that
the programmer must only think of the calculation on a ``local'' scale
by telling the computer the rules for commuting a
given pair of data structures. Furthermore, this gives a convenient 
segmentation of the computer program into small, independent parts, which can 
be written and debugged separately. As soon as all commutation 
rules are specified correctly, the program can perform the whole
calculation ``globally'' by recursively applying the commutation 
rules. Compared to standard computer algebra packages like e.g.\ 
Mathematica or Maple, this concept is very flexible and efficient.

For the more complex manipulations like partial integrations and
the handling of tensor indices, the program uses a so-called ``message 
pipe,'' through which the
objects in the formulas can pass information to other objects
(this is again implemented with virtual class functions). In this way, the
elements of the lists can exchange data and give commands to each 
other. This allows a convenient coordination of the formula manipulations.

The calculation rules of the program ``class\_commute'' are very 
similar to the construction steps described in the proofs of Theorem 
\ref{l:thm1} and Theorem \ref{l:thm2}. The only difference is that, according 
to our implementation as commutation rules, the calculation does
not follow the same strict and clear order as in Subsection 
\ref{l:sec_21}. Basically, one may think of the construction steps of 
Theorem \ref{l:thm1} and Theorem \ref{l:thm2} as being performed 
simultaneously in a disordered way, whereby the program ensures that 
all rules are applied consistently.

Some formulas generated by the program ``class\_commute'' are compiled 
in the appendix; they give a picture of the leading singularities
of $\tilde{s}(x,y)$ on the light cone. We remark that ``class\_commute'' 
was also a valuable tool for finding and checking the combinatorial 
results of Theorem \ref{l:thm1}, Lemma \ref{l:lemma3}, and Theorem \ref{l:thm2}.

\section{The Light-Cone Expansion of the Dirac Sea}
\setcounter{equation}{0}
\label{l:sec_dirac}
In this section, we shall perform the light-cone expansion of the 
fermionic projector as defined in \cite{F1}. In Subsection \ref{l:sec_31}, we 
establish a formal analogy between the light-cone expansions of the 
Dirac sea and of the Green's functions. This {\em{residual argument}} 
allows us to use the results of the previous section also for the 
fermionic projector. However, the analogy between the Dirac sea and the Green's 
functions cannot be extended beyond a purely formal level. The basic 
reason is that, in contrast to the Green's functions, the Dirac sea is  
a {\em{non-causal}} object. This is developed in detail in Subsections 
\ref{l:sec_32} and \ref{l:sec_33}.

We point out that, in this section, we do not work with the 
dynamic mass matrices $Y_{L\!/\!R}(x)$, (\ref{l:n20}). The reason is 
that, for the Dirac sea, the 
regularity conditions of Lemma \ref{l:lemma0} are necessary for 
the contributions to the perturbation expansion to be well-defined.
Working with the dynamic 
mass matrices, however, implies that we consider the potential $B$, 
(\ref{l:18a}), as the perturbation of the Dirac operator; but $B$ does 
in general not go to zero at infinity. For our notation, the reader is
referred to \cite{F1}.

\subsection{The Residual Argument}
\label{l:sec_31}
We begin by describing how the light-cone expansion of the Green's 
functions can be understood in momentum space. Apart from giving a 
different point of view, this allows us to get a connection
to the light-cone expansion of the Dirac sea. For clarity, we 
begin with the special case $m Y=0$ of zero fermion mass. This case is 
particularly simple because then $B={\cal{B}}$, so that the perturbation 
expansions (\ref{l:9}) and (\ref{l:11b}) coincide. This is sufficient to 
explain the basic construction; the extension to $m Y \neq 0$ will later be
accomplished by a general argument.
Furthermore, we only consider the advanced Green's function; for the
retarded Green's function, the calculation is analogous.

Suppose that we want to perform the light-cone expansion of the
$k^{\mbox{\scriptsize{th}}}$ order contribution to the perturbation 
series (\ref{l:9})=(\ref{l:11b}). We write the contribution as a multiple
Fourier integral,
\begin{eqnarray}
\lefteqn{ ((-s^\vee \:B)^k \:s^\vee)(x,y) } \nonumber \\
&=& \int \frac{d^4p}{(2 \pi)^4}
        \int \frac{d^4q_1}{(2 \pi)^4} \cdots \int \frac{d^4q_k}{(2 \pi)^4}
        \: \Delta s^\vee(p;q_1,\ldots,q_k) \:e^{-i(p+q_1+\cdots+q_k)x \:+\: i p y} 
        \;\;\; ,
        \label{l:31}
\end{eqnarray}
where the distribution $\Delta s^\vee(p;q_1,\ldots,q_k)$ is the Feynman 
diagram in momentum space,
\begin{eqnarray}
\Delta s^\vee(p;q_1,\ldots,q_k) &=& (-1)^k \:s^\vee(p+q_1+\cdots+q_k) 
\:\tilde{B}(q_k)\: s^\vee(p+q_1+\cdots+q_{k-1}) \:\tilde{B}(q_{k-1}) 
\nonumber \\
&& \hspace*{1cm} \;\cdots\; \tilde{B}(q_2) \:s^\vee(p+q_1) \:\tilde{B}(q_1)\:
s^\vee(p) \label{l:A2}
\end{eqnarray}
($\tilde{B}$ denotes the Fourier transform of the potential $B$, and 
$s^\vee(p)$ is the multiplication operator in momentum space).
For the arguments of the Green's functions, we introduce the 
abbreviation
\begin{equation}
         p_0 \;:=\; p \spc {\mbox{and}} \spc p_l\;:=\;p+q_1+\cdots+q_l 
        \;,\;\;\;\;\; 1 \leq l \leq k .
        \label{l:32x}
\end{equation}
Substituting the explicit formulas (\ref{l:10}) and 
(\ref{l:21x}) into (\ref{l:A2}), we obtain
\begin{eqnarray}
\lefteqn{ \Delta s^\vee(p; q_1,\ldots,q_k) \;=\; (-1)^k \;
p \slsh_k \:\tilde{B}(q_k) \:p \slsh_{k-1}
\;\cdots\; p \slsh_1 \:\tilde{B}(q_1)\: p \slsh_0 } \nonumber \\
&&\times \lim_{0<\varepsilon \rightarrow 0} 
\frac{1}{(p_k)^2 - i \varepsilon p^0_k} \:
\frac{1}{(p_{k-1})^2- i \varepsilon p^0_{k-1}} \:\cdots\:
\frac{1}{(p_0)^2- i \varepsilon p^0_0} \;\;\; . \label{l:A22}
\end{eqnarray}
We now expand the Klein-Gordon Green's functions in (\ref{l:A22})
with respect to the 
momenta $p_l - p$. If we expand the terms $i \varepsilon p^0_l$
with a geometric series,
\[ \frac{1}{(p_l)^2 - i \varepsilon p^0_l} \;=\; \sum_{n=0}^\infty 
\frac{(i \varepsilon \:(p^0_l - p^0))^n}{((p_l)^2 - i \varepsilon 
p^0)^{1+n}} \;\;\; , \]
all contributions with $n \geq 1$ contain factors $\varepsilon$ and 
vanish in the limit $\varepsilon \rightarrow 0$. Therefore we must only
expand with respect to the parameters $((p_l)^2 - p^2)$. This gives, 
again with geometric series,
\begin{eqnarray*}
\lefteqn{ \Delta s^\vee(p; q_1,\ldots,q_k) \;=\; (-1)^k \;
p \slsh_k \:\tilde{B}(q_k) \:p \slsh_{k-1}
\;\cdots\; p \slsh_1 \:\tilde{B}(q_1)\: p \slsh_0 } \\
&&\times \sum_{n_1,\ldots,n_k=0}^\infty (p^2 - p_k^2)^{n_k}
\:\cdots\: (p^2 - p_1^2)^{n_1} \; \lim_{0<\varepsilon \rightarrow 0} 
\frac{1}{(p^2 - i \varepsilon p^0)^{1+k+n_1+\cdots+n_k}} \;\;\; .
\end{eqnarray*}
Rewriting the negative power of $(p^2-i \varepsilon p^0)$ as a mass-derivative,
\begin{eqnarray}
\lefteqn{ \frac{1}{(p^2 - i \varepsilon p^0)^{1+k+n_1+\cdots+n_k}} } 
\nonumber \\
&=& \frac{1}{(k+n_1+\cdots+n_k)!} \left( \frac{d}{da} 
\right)^{k+n_1+\cdots+n_k} \frac{1}{p^2 - a - i \varepsilon 
p^0}_{|a=0} \;\;\; , \label{l:34z}
\end{eqnarray}
we obtain a formula containing only one Green's function. Namely,
using the notation (\ref{l:23b}),
\begin{eqnarray}
\lefteqn{ \Delta s^\vee(p; q_1,\ldots,q_k) \;=\; (-1)^k \;
p \slsh_k \:\tilde{B}(q_k) \:p \slsh_{k-1}
\;\cdots\; p \slsh_1 \:\tilde{B}(q_1)\: p \slsh_0 } \nonumber \\
&&\times \sum_{n_1,\ldots,n_k=0}^\infty \frac{1}{(k+n_1+\cdots+n_k)!}
\;(p^2 - p_k^2)^{n_k} \:\cdots\: (p^2 - p_1^2)^{n_1} \;
S^{\vee (k+n_1+\cdots+n_k)}(p) \; . \;\;\;\;\;\;\;\;\;\; \label{l:A3}
\end{eqnarray}
This is the basic equation for the light-cone expansion of the Green's 
functions in momentum space. Similar to the light-cone expansion of the 
previous section, (\ref{l:A3}) involves the differentiated Green's 
functions $S^{\vee (.)}$. It remains to transform the 
polynomials in the momenta $p_0,\ldots,p_k$ until 
getting a connection to the nested line integrals of, say, Theorem 
\ref{l:thm1}: Substituting (\ref{l:32x}), we rewrite (\ref{l:A3}) in terms 
of the momenta $p$, $q_1,\ldots,q_k$ and multiply out. 
Furthermore, we simplify the Dirac matrices with the anti-commutation 
rules (\ref{l:ar1}). This gives for (\ref{l:A3}) a sum of terms of the form
\begin{equation}
\chi_c \:C\: \gamma^I \:q_k^{I_k} \cdots q_1^{I_1} 
\;\tilde{V}^{(k)}_{J_k, c_k}(q_k) \cdots \tilde{V}^{(1)}_{J_1, c_1}(q_1)
\:p^L \;S^{\vee (h)}(p) \spc (h \geq \left[ |L|/2 \right]),
        \label{l:A4}
\end{equation}
where the tensor indices of the multi-indices
$I$, $I_l$, $J_l$, and $L$ are contracted 
with each other (similar to the notation of Theorem \ref{l:thm1}, the 
factors $\tilde{V}^{(l)}_{J_l, c_l}$ stand for the individual 
potentials of $\tilde{B}$). If tensor indices of the power $p^L$ are 
contracted with each other, we can iteratively eliminate the 
corresponding factors $p^2$ with the rule (\ref{l:5}), more precisely
\begin{equation}
p^2 \:S^{\vee (h)}(p) \;=\; h \:S^{\vee (h-1)}(p) \spc (h \geq 1).
        \label{l:37x}
\end{equation}
Thus we can arrange that the tensor indices of $p^L$ in 
(\ref{l:A4}) are all contracted with tensor indices of the 
factors $\gamma^I$, $q_l^{I_l}$, or $\tilde{V}^{(l)}_{J_l, c_l}$.
By iteratively applying the differentiation rule (\ref{l:21a}), we can
now rewrite the power $p^L$ in (\ref{l:A4}) with $p$-derivatives, e.g.
\begin{eqnarray*}
p_j\: p_k\: S^{\vee (2)}(p) &=& -\frac{1}{2} 
\:p_j\:\frac{\partial}{\partial p^k} S^{\vee (1)}(p) \;=\; -\frac{1}{2} 
\:\frac{\partial}{\partial p^k} (p_j \:S^{\vee (1)}(p)) \:+\: \frac{1}{2} 
\:g_{jk}\: S^{\vee (1)}(p) \\
&=& \frac{1}{4} \:\frac{\partial^2}{\partial p^j \:\partial p^k} 
S^{(0)}(p) \:+\: \frac{1}{2} \:g_{jk} \:S^{(1)}(p) \;\;\; .
\end{eqnarray*}
In this way, we obtain for $\Delta s^\vee(p; q_1,\ldots,q_k)$ a sum of terms
of the form
\begin{equation}
\chi_c \:C\: \gamma^I \:q_k^{I_k} \cdots q_1^{I_1} 
\;\tilde{V}^{(k)}_{J_k, c_k}(q_k) \cdots \tilde{V}^{(1)}_{J_1, c_1}(q_1)
\:\partial_p^K \;S^{\vee (h)}(p) \;\;\; ,
        \label{l:A5}
\end{equation}
whereby no tensor indices of the derivatives $\partial_p^K$ are 
contracted with each other.
We substitute these terms into (\ref{l:31}) and transform them to
position space. Integrating the derivatives 
$\partial_p^K$ by parts gives factors $(y-x)^K$. The factors 
$q_l^{I_l}$, on the other hand, can be written as partial 
derivatives $\partial^{I_l}$ acting on the potentials $V^{(l)}$.
More precisely, the term (\ref{l:A5}) gives after substitution into (\ref{l:31}) 
the contribution
\begin{equation}
\chi_c \:C\:i^{|I_1|+\cdots+|I_k|} \:(-i)^{|K|} \;\gamma^I
\;(\partial^{I_k} V^{(k)}_{J_k, c_k}(x)) \cdots
(\partial^{I_1} V^{(1)}_{J_1, c_1}(x))
\:(y-x)^K \;S^{\vee (h)}(x,y) \;\;\; ,
        \label{l:A6}
\end{equation}
where the tensor indices of the factor $(y-x)^K$ are all contracted 
with tensor indices of the multi-indices $I$, $I_l$, or $J_l$.
The Feynman diagram $((-s B)^k s)(x,y)$ coincides with the sum of
all these contributions.

This expansion has much similarity with the light-cone expansion of 
Theorem \ref{l:thm1}. Namely, if one expands the nested line 
integrals in (\ref{l:l1}) in a Taylor series around $x$, one gets 
precisely the expansion into terms of the form (\ref{l:A6}). Clearly, 
the light-cone expansion of Theorem \ref{l:thm1} goes far beyond 
the expansion (\ref{l:A6}), because the dependence on the external potential is 
described by non-local line integrals. Nevertheless, the 
expansion in momentum space (\ref{l:A3}) and subsequent Fourier 
transformation give an easy way of understanding in principle how the 
formulas of the light-cone expansion come about. We remark that, after going through the
details of the combinatorics and rearranging the 
contributions (\ref{l:A6}), one can recover the Taylor series of the 
line integrals in (\ref{l:l1}). This gives an alternative method for 
proving Theorem \ref{l:thm1}. However, it is obvious that this becomes 
complicated and does not yield the most elegant approach (the reader 
interested in the details of this method is referred to \cite{F2}, 
where a very similar technique is used for the light-cone expansion to first 
order in the external potential).

Our next aim is to generalize the previous construction. Since we 
must, similar to (\ref{l:34z}), rewrite a product of Green's functions as 
the mass-derivative of a single Green's function, we can only expect 
the construction to work if all Green's functions in the product 
(\ref{l:A2}) are of the same type (e.g.\ the construction breaks down 
for a ``mixed'' operator product containing both advanced and retarded 
Green's functions). But we need not necessarily
work with the advanced or retarded Green's functions. Instead, we can use
Green's functions with a different position of the poles in the complex
$p^0$-plane: We consider the Green's functions
\begin{equation}
s^\pm(p) \;=\; p \slsh \: S^\pm_{a \:|\: a=0}(p) \spc {\mbox{with}} 
        \spc S^\pm(p) \;=\; \lim_{0<\varepsilon \rightarrow 0} \frac{1}{p^2-a 
        \mp i \varepsilon}
        \label{l:G}
\end{equation}
and again use the notation (\ref{l:23b}),
\[ S^{\pm \:(l)} \;=\; \left( \frac{d}{da} \right)^l S^\pm_{a \:|\: a=0}
\;\;\; . \]
The perturbation expansion for these Dirac Green's functions is,
similar to (\ref{l:9}) or (\ref{l:11b}), given by the formal series
\begin{equation}
        \tilde{s}^+ \;:=\; \sum_{n=0}^\infty (-s^+ \:B)^n 
        s^+ \;\;\;,\spc \tilde{s}^- \;:=\; \sum_{n=0}^\infty
        (-s^- \:B)^n s^- \;\;\; .
        \label{l:F}
\end{equation}
The light-cone expansion in momentum space is performed exactly as for 
the advanced and retarded Green's functions; we obtain in analogy to 
(\ref{l:31}) and (\ref{l:A3}) the formula
\begin{eqnarray*}
\lefteqn{ ((-s^\pm \:B)^k \:s^\pm)(x,y) } \nonumber \\
&=& \int \frac{d^4p}{(2 \pi)^4}
        \int \frac{d^4q_1}{(2 \pi)^4} \cdots \int \frac{d^4q_k}{(2 \pi)^4}
        \: \Delta s^\pm(p;q_1,\ldots,q_k) \:e^{-i(p+q_1+\cdots+q_k)x \:+\: i p y}
\end{eqnarray*}
with
\begin{eqnarray*}
\lefteqn{ \Delta s^\pm(p; q_1,\ldots,q_k) \;=\; (-1)^k \;
p \slsh_k \:\tilde{B}(q_k) \:p \slsh_{k-1}
\;\cdots\; p \slsh_1 \:\tilde{B}(q_1)\: p \slsh_0 } \nonumber \\
&&\times \sum_{n_1,\ldots,n_k=0}^\infty \frac{1}{(k+n_1+\cdots+n_k)!}
\;(p^2 - p_k^2)^{n_k} \:\cdots\: (p^2 - p_1^2)^{n_1} \;
S^{\pm\: (k+n_1+\cdots+n_k)} \;\;\; .
\end{eqnarray*}
Since $S^\pm$ are Green's functions of the Klein-Gordon equation, 
they clearly also satisfy the identity (\ref{l:37x}).
Furthermore, the differentiation rule (\ref{l:21a}) is also valid 
for $S^\pm$; namely
\begin{eqnarray*}
\frac{\partial}{\partial p^j} S^{\pm \:(l)}(p) &=& \left( 
\frac{d}{da} \right)^l \:\lim_{0<\varepsilon \rightarrow 0} 
\frac{\partial}{\partial p^j} \left( \frac{1}{p^2 - a \mp i 
\varepsilon} \right)_{|a=0} \\
&=& \left( \frac{d}{da} \right)^l \:\lim_{0<\varepsilon \rightarrow 0} 
\frac{-2 p_j}{(p^2 - a \mp i \varepsilon)^2}_{|a=0} \;=\; -2 p_j 
\:S^{\pm\: (l+1)}(p) \;\;\; .
\end{eqnarray*}
Therefore we can, exactly as in (\ref{l:A5}), rewrite the power $p^L$ 
with $p$-derivatives. Thus the expansion (\ref{l:A6}) is valid in 
the same way for the Green's functions $s^\pm$ if one only replaces 
the index ``$^\vee$'' in (\ref{l:A6}) by ``$^\pm$''. As explained 
before, the expansion (\ref{l:A6}) is obtained from the light-cone 
expansion of Theorem \ref{l:thm1} by expanding the potentials around 
the space-time point $x$. Since the formulas of the light-cone 
expansion are uniquely determined by this Taylor expansion, we 
immediately conclude that the statement of Theorem \ref{l:thm1} is also 
valid for the $k^{\mbox{\scriptsize{th}}}$ order contribution to the 
perturbation expansion (\ref{l:F}) if the factor $S^{(h)}$ in (\ref{l:l1}) 
stands more generally for $S^{+\:(h)}$ or $S^{-\:(h)}$, respectively. 
This simple analogy between the formulas of the light-cone expansion 
of the Feynman diagrams $((-s^{\vee / \wedge} \:B)^k \:s^{\vee / 
\wedge})$ and $((-s^\pm \:B)^k \:s^\pm)$, which is obtained by changing the 
position of the poles of the free Green's functions in momentum space, 
is called the {\em{residual argument}}.

After these preparations, we come to the fermionic projector in the general 
case $m Y \neq 0$. We want to extend 
the light-cone expansion to an object 
$\tilde{P}^{\mbox{\scriptsize{res}}}$ being a perturbation of the free 
fermionic projector. Our method is to define $\tilde{P}^{\mbox{\scriptsize{res}}}$ 
in such a way that it can be easily expressed in terms of the Green's 
functions $\tilde{s}^\vee$, $\tilde{s}^\wedge$, $\tilde{s}^+$, and $\tilde{s}^-$.
The light-cone expansion of the Green's functions then immediately 
carries over to $\tilde{P}^{\mbox{\scriptsize{res}}}$.
We denote the lower mass shell by $T_a$, i.e. in momentum space
\begin{equation}
        T_a(q) \;=\; \Theta(-q^0)\: \delta(q^2-a) \;\;\; ,
        \label{l:31s}
\end{equation}
and set
\begin{equation}
        T^{(l)} \;=\; \left( \frac{d}{da} \right)^l T_{a \:|\: a=0} \;\;\; .
        \label{l:F2}
\end{equation}
Furthermore, we introduce, exactly as in \cite{F1}, the series of operator 
products
\[ b^< \;=\; \sum_{k=0}^\infty (-s \:{\cal{B}})^k \;\;\;,\spc b \;=\; 
\sum_{k=0}^\infty (-{\cal{B}} \:s)^k \:{\cal{B}} \;\;\;,\spc b^> \;=\; 
\sum_{k=0}^\infty (-{\cal{B}} \:s)^k \;\;\; . \]
\begin{Def}
\label{l:def_res}
The {\bf{residual fermionic projector}}
$\tilde{P}^{\mbox{\scriptsize{res}}}(x,y)$ is defined by
\begin{equation}
\tilde{P}^{\mbox{\scriptsize{res}}}(x,y) \;=\; \frac{1}{2} \:X\:
(\tilde{p}^{\mbox{\scriptsize{res}}} - \tilde{k})(x,y) \;\;\; ,
        \label{l:E0}
\end{equation}
where the operators $\tilde{p}^{\mbox{\scriptsize{res}}}$ and 
$\tilde{k}$ are given by the perturbation series
\begin{eqnarray}
\tilde{p}^{\mbox{\scriptsize{res}}} &=& \sum_{\beta=0}^\infty (-i 
\pi)^{2 \beta} \;b^< \:p\: (b \:p)^{2 \beta} \:b^> \label{l:E1} \\
\tilde{k} &=& \sum_{\beta=0}^\infty (-i 
\pi)^{2 \beta} \;b^< \:k\: (b \:k)^{2 \beta} \:b^> \;\;\; . \label{l:E2}
\end{eqnarray}
\end{Def}

\begin{Prp} {\bf{(formal light-cone expansion of the residual 
fermionic projector)}}
\label{l:prp4}
The results of Section \ref{l:sec_2} also apply to the residual fermionic
projector. More precisely, the light-cone expansion of Theorem \ref{l:thm1} 
holds for $\tilde{P}^{\mbox{\scriptsize{res}}}(x,y)$ if we replace $S^{(h)}$
by $T^{(h)}$ and multiply the formulas of the light-cone 
expansion from the left with the chiral asymmetry
matrix $X$. According to Theorem \ref{l:thm2}, all tangential 
derivatives can be integrated by parts. With Def.\ \ref{l:def_pf}, 
Def.\ \ref{l:def2}, and Theorem \ref{l:thm3}, the light-cone expansion can 
be reduced to the phase-free contribution. According to Proposition 
\ref{l:prp1}, the phase-free contribution consists, to every order $\sim 
T^{(h)}$, of only a finite number of terms.
\end{Prp}
{\Proof}
First of all, we must generalize the residual argument to the case 
$m Y \neq 0$ of massive fermions. According to (\ref{l:9}) and (\ref{l:11b}),
there are two equivalent perturbation series for 
$\tilde{s}^\vee$,
\begin{eqnarray}
        \tilde{s}^\vee &=& \sum_{k=0}^\infty (-s^\vee\: {\cal{B}})^k\: 
        s^\vee \label{l:n8a} \\
        &=& \sum_{k=0}^\infty (-s^\vee_{m=0}\: B)^k\: s^\vee_{m=0}
        \;\;\; . \label{l:n8b}
\end{eqnarray}
In both perturbation series, each summand is a well-defined tempered 
distribution (this follows from the smoothness of $B$, ${\cal{B}}$ and
from the causality of the perturbation expansion). In 
Section \ref{l:sec_2}, we developed the light-cone expansion from the 
series in (\ref{l:n8b}). But by arranging the contributions to this
light-cone expansion in powers of the potential ${\cal{B}}$, one also 
obtains formulas for the light-cone expansion of every Feynman 
diagram of the perturbation series (\ref{l:n8a}). For the 
Green's functions $s^\pm$, we have similar perturbation expansions
\begin{eqnarray}
\tilde{s}^\pm &=& \sum_{k=0}^\infty (-s^\pm\: {\cal{B}})^k\: s^\pm 
\label{l:n8c} \\
&=& \sum_{k=0}^\infty (-s^\pm_{m=0}\: B)^k\: s^\pm_{m=0} \;\;\; .
        \label{l:n8d}
\end{eqnarray}
Since the support of the distributions $s^\pm(x,y)$ does {\em{not}} 
vanish outside the light cone, we now need the conditions of Lemma \ref{l:lemma0} 
on the decay of the potentials at infinity.
According to our assumptions on ${\cal{B}}$, each summand of the 
perturbation expansion (\ref{l:n8c}) is a well-defined distribution.
The potential $B$, however, does in general not decay at infinity; 
thus the Feynman diagrams of the perturbation expansion
(\ref{l:n8d}) are ill-defined. This is a problem, especially because 
in our above consideration, the residual argument was derived for the Feynman 
diagrams of the expansions (\ref{l:n8b}) and (\ref{l:n8d}).
The solution to this problem is an approximation argument using the
``causality'' of the formulas 
of the light-cone expansion: We consider a smooth function $\eta_R(x)$ 
which is equal to one inside the ball of radius $R$ around the origin
and vanishes outside the ball of radius $2R$ (in $\R^4$ equipped with
the standard Euclidean metric). Then the
potential $\eta_R B$ has compact support and, according to 
Lemma \ref{l:lemma0}, the Feynman diagrams
\begin{equation}
\left((-s^\pm_{m=0} \:\eta_R B)^k \:s^\pm_{m=0} \right)(x,y)
        \label{l:n9}
\end{equation}
are well-defined. We can apply the above residual argument for $m Y =0$;
this yields formulas of the light-cone expansion in terms of 
the potential $(\eta_R B)$ and its partial derivatives. Since the 
potential enters into the formulas of the light-cone expansion
only along the convex line
$\overline{xy}$ , we can, by taking the limit $R \rightarrow \infty$,
remove the cutoff function $\eta_R$ from these formulas.
This limiting process shows that the summands 
of the perturbation series in (\ref{l:n8d}) make mathematical 
sense in terms of the light-cone expansion. By reordering the 
contributions, we immediately also get formulas for the light-cone 
expansion of the Feynman diagrams of the perturbation series 
(\ref{l:n8c}). The analogy between the light-cone expansions of the 
Feynman diagrams of the perturbation series (\ref{l:n8a}) and (\ref{l:n8c}) 
finally yields the extension of the residual argument to a 
general mass matrix $m Y$.

Evaluating the poles in (\ref{l:21x}) and (\ref{l:F}) in the complex 
$p^0$-plane gives (using the formula $\lim_{0<\varepsilon \rightarrow 
0} ((x-i \varepsilon)^{-1} - (x+i \varepsilon)^{-1}) = 2 \pi i 
\delta(x)$) the relations
\begin{eqnarray}
s^\vee &=& s \:+\: i \pi \:k \;\;\;,\spc s^\wedge \;=\; s \:-\: i \pi \:k
\label{l:fl1} \\
s^+ &=& s \:+\: i \pi \:p \;\;\;,\spc s^- \;=\; s \:-\: i \pi \:p \;\;\; ,
\label{l:fl2}
\end{eqnarray}
where $s$ denotes as in \cite{F1} the arithmetic mean of the advanced and 
retarded Green's functions,
\[ s \;=\; \frac{1}{2} \:(s^\vee + s^\wedge) \;\;\; . \]
We substitute (\ref{l:fl1}) and (\ref{l:fl2}) into the perturbation series
(\ref{l:9}), (\ref{l:n8c}) and multiply out. After rearranging the sums, one sees
that the series (\ref{l:E1}) and (\ref{l:E2}) can be written as
\begin{equation}
        \tilde{p}^{\mbox{\scriptsize{res}}} \;=\; \frac{1}{2 \pi i} 
        \:(\tilde{s}^+ - \tilde{s}^-) \spc {\mbox{and}} \spc
        \tilde{k} \;=\; \frac{1}{2 \pi i} \:(\tilde{s}^\vee - \tilde{s}^\wedge) 
        \;\;\; ,
        \label{l:fl3}
\end{equation}
respectively (see \cite[proof of Theorem 3.2]{F1} for the details of the
combinatorics).
According to the residual argument, all Green's functions have a
light-cone expansion according to Theorem \ref{l:thm1}. By substituting into 
(\ref{l:fl3}), this light-cone expansion immediately generalizes to
$\tilde{p}^{\mbox{\scriptsize{res}}}$ and $\tilde{k}$.
Using (\ref{l:E0}), we conclude that Theorem \ref{l:thm1} is 
also valid for $\tilde{P}^{\mbox{\scriptsize{res}}}$ after the 
replacement $S^{(h)} \rightarrow T^{(h)}$ and multiplication with the 
chiral asymmetry matrix.
Since the results of Theorem \ref{l:thm2}, 
Proposition \ref{l:prp1}, and Theorem \ref{l:thm3} are obtained merely by 
manipulating and rearranging the formulas of the light-cone expansion, 
they also hold for the residual fermionic projector.
\QED
We point out that the argumentation in this subsection was only 
formal in the sense that we did not care about the convergence of the 
infinite sums. Also, the approximation argument in the proof of Proposition 
\ref{l:prp4} requires a mathematical justification.
Furthermore, the derivative (\ref{l:F2}) is ill-defined 
because, for $a=0$, the mass shell degenerates to the cone 
$\{q^2=0,\; q^0<0\}$, which is not differentiable at $q=0$.
We postpone the mathematical analysis of these problems to 
Subsection \ref{l:sec_33}.

\subsection{The Non-Causal High Energy Contribution}
\label{l:sec_32}
Unfortunately, the residual fermionic projector 
$\tilde{P}^{\mbox{\scriptsize{res}}}$ of the previous subsection does 
not coincide with the fermionic projector $\tilde{P}$ of \cite{F1},
\begin{equation}
\tilde{P}(x,y) \;=\; \frac{1}{2} \:X\: (\tilde{p}-\tilde{k})(x,y) \;\;\; .
        \label{l:E4}
\end{equation}
The difference is that, instead of the operator
$\tilde{p}^{\mbox{\scriptsize{res}}}$ in the residual fermionic 
projector (\ref{l:E0}), the fermionic projector (\ref{l:E4}) involves the operator 
$\tilde{p}$, which is formally given by
\begin{equation}
\tilde{p} \;\stackrel{\mbox{\scriptsize{formally}}}{=}\; \sqrt{ 
\tilde{k}^2 } \;\;\; .
        \label{l:E5}
\end{equation}
Using an operator calculus method, this formal definition is made 
mathematically precise in \cite{F1} in terms of a perturbation series 
for $\tilde{p}$. Similar to (\ref{l:E1}), this perturbation expansion 
consists of a sum of operator products. But the operator products are 
more complicated; they also contain operators $k$ with some 
combinatorial factors (see \cite{F1} for details).

Before entering the mathematical analysis of the operator products, 
we point out that it is not just a matter of taste to take (\ref{l:E4}), 
and not (\ref{l:E0}), as the definition of the fermionic projector; only the 
definition (\ref{l:E4}) makes physical sense. This comes
as follows: As explained in \cite{F1}, the operator $\tilde{k}$ 
generalizes the splitting of the solutions of the Dirac equation
into solutions of positive 
and negative frequency to the case with interaction. The ``generalized 
positive and negative frequency solutions'' are given by the 
eigenstates of $\tilde{k}$ with positive and negative eigenvalue, 
respectively. The construction (\ref{l:E5}),(\ref{l:E4}) projects out 
all positive eigenstates of $\tilde{k}$; the operator
$\frac{1}{2}(\tilde{p}-\tilde{k})$ consists 
precisely of all eigenstates of $\tilde{k}$ with negative eigenvalue. 
The residual fermionic projector (\ref{l:E0}), however, consists of a mixture of 
positive and negative eigenstates of $\tilde{k}$, which is not a 
reasonable physical concept.

We begin by giving the difference between the fermionic projector and the 
residual fermionic projector a name.
\begin{Def}
\label{l:def_he}
The {\bf{non-causal high energy contribution}}
$\tilde{P}^{\mbox{\scriptsize{he}}}(x,y)$ to the fermionic projector is given by
\[ \tilde{P}^{\mbox{\scriptsize{he}}}(x,y) \;=\; \tilde{P}(x,y) \:-\:
\tilde{P}^{\mbox{\scriptsize{res}}}(x,y) \;\;\; . \]
\end{Def}
Clearly, this definition is only helpful if
$\tilde{P}^{\mbox{\scriptsize{he}}}$ has some nice properties. The 
reason why the definition makes sense is that every contribution to 
the perturbation expansion of 
$\tilde{P}^{\mbox{\scriptsize{he}}}(x,y)$ is a smooth function. Thus 
the singular behavior of the fermionic projector on the light cone is completely 
described by the residual fermionic projector and its light-cone expansion,
Proposition \ref{l:prp4}.
\begin{Thm}
\label{l:thm4}
The non-causal high energy contribution 
$\tilde{P}^{\mbox{\scriptsize{he}}}(x,y)$ is, to every order in
perturbation theory, a smooth function in $x$ and $y$.
\end{Thm}
{\Proof}
The perturbation series \cite[Theorem 4.1]{F1} defines $\tilde{p}$ as a 
sum of operator products of the form
\begin{equation}
C_n \:{\cal{B}}\:C_{n-1} \:{\cal{B}}\: \cdots \: {\cal{B}}\:C_0 \;\;\; ,
        \label{l:n7}
\end{equation}
where the factors $C_l$ coincide with either $k$, $p$, or $s$. The number of 
factors $k$ in these operator products is always even. If one 
replaces all factors $k$ by $p$, one gets precisely the perturbation 
series for $\tilde{p}^{\mbox{\scriptsize{res}}}$, (\ref{l:E1}) (this is 
verified using the details of the combinatorics in \cite{F1}). 
Therefore we can convert the perturbation series for $\tilde{p}$ into that for
$\tilde{p}^{\mbox{\scriptsize{res}}}$ by iteratively replacing pairs 
of factors $k$ in the operator products by factors $p$. Thus the 
difference $(\tilde{p} - \tilde{p}^{\mbox{\scriptsize{res}}})$ can, 
to every order in perturbation theory, be written as a finite sum of 
expressions of the form
\begin{eqnarray}
&&C_n \:{\cal{B}} \cdots C_{b+1} \:{\cal{B}} \:( p \:{\cal{B}}\: 
C_{b-1} \cdots C_{a+1} \:{\cal{B}}\: p \nonumber \\
&& \hspace*{2.3cm}\:-\: k \:{\cal{B}}\: C_{b-1} \cdots C_{a+1} \:{\cal{B}}\: k) 
\:{\cal{B}}\:C_{a-1} \cdots {\cal{B}}\:C_0 \;\;\; ,
\label{l:n6}
\end{eqnarray}
where the factors $C_l$ again stand for $k$, $p$, or $s$. Since 
$\tilde{P}^{\mbox{\scriptsize{he}}} = 
\frac{1}{2}\:X (\tilde{p}-\tilde{p}^{\mbox{\scriptsize{res}}})$, it 
suffices to show that (\ref{l:n6}) is a smooth function in position 
space.

We first simplify our problem: Once we have shown that 
the bracket in (\ref{l:n6}) is smooth and bounded in position space, 
the additional multiplications to the very left and right can be carried out 
by iteratively multiplying with ${\cal{B}}$ and forming the 
convolution with $C_l$, which again gives a smooth and bounded 
function in each step (notice that, according to the assumptions of 
Lemma \ref{l:lemma0}, ${\cal{B}}$ decays sufficiently fast at infinity). 
Thus we must only consider the bracket in (\ref{l:n6}). We rewrite this 
bracket with the projectors $\frac{1}{2}(p-k)$ and $\frac{1}{2}(p+k)$ 
on the lower and upper mass shells,
\begin{eqnarray*}
\lefteqn{p \:{\cal{B}}\: C_{n-1} \cdots C_1 \:{\cal{B}}\: p
\:-\: k \:{\cal{B}}\: C_{n-1} \cdots C_1 \:{\cal{B}}\: k } \\
&=& \frac{1}{2}\:(p+k) \:{\cal{B}}\: C_{n-1} \cdots C_1 
\:{\cal{B}}\: (p-k) \:+\:\frac{1}{2}\:(p-k) \:{\cal{B}}\: C_{n-1} \cdots C_1 
\:{\cal{B}}\: (p+k) \;\;\; .
\end{eqnarray*}
For symmetry reasons, it suffices to show that the first summand of this 
decomposition,
\begin{equation}
        ((p+k) \:{\cal{B}}\: C_{n-1} \cdots C_1 \:{\cal{B}}\: (p-k))(x,y) 
        \label{l:p-k} \;\;\; ,
\end{equation}
is smooth and bounded.

We proceed in momentum space. We say that a function $f(q)$ has 
{\em{rapid decay for positive frequency}} if it is $C^1$, bounded 
together with its first derivatives (i.e.\ $\sup |f|, \sup|\partial_l 
f| < \infty$), and satisfies for every $\alpha>0$ the bounds
\begin{equation}
\sup_{\omega >0,\; \vec{k} \in \sR^3} |\omega^\alpha \:f(\omega, \vec{k})|,\;
\sup_{\omega >0,\; \vec{k} \in \sR^3} |\omega^\alpha \:\partial_l
f(\omega, \vec{k})| \;<\; \infty \;\;\; .
        \label{l:n12}
\end{equation}
After setting $C_0=p-k$ and $C_n = p+k$,
the operator product (\ref{l:p-k}) is of the form (\ref{l:2}). We choose 
a function $g$ with rapid decay for positive frequency and decompose 
the operator product in the form (\ref{l:4z}),(\ref{l:4a}). It follows by 
induction that the functions $F_j$ all have rapid decay for positive 
frequency: The induction hypothesis is obvious by setting $F_0=g$. The 
induction step is to show that for a function $F_{j-1}$ with rapid 
decay for positive frequency, the convolution
\begin{equation}
F_j(\omega, \vec{k}) \;=\; \int \frac{d\omega^\prime}{2 \pi}
\int\frac{ d\vec{k}^\prime}{(2 \pi)^3} \; 
\tilde{B}(\omega - \omega^\prime, \vec{k}-\vec{k}^\prime) 
\:C_{j-1}(\omega^\prime, \vec{k}^\prime) \:F_{j-1}(\omega^\prime, 
\vec{k}^\prime)
        \label{l:n13}
\end{equation}
also has rapid decay for positive frequency. In Lemma \ref{l:lemma0}, 
it was shown that $F_j$ is $C^1$ and bounded together with its first 
derivatives. As a consequence, we must only establish the bounds (\ref{l:n12}) 
for $\omega>1$. Because of monotonicity $\omega^\alpha < 
\omega^\beta$ for $\alpha<\beta$ (and $\omega >1$), it furthermore suffices
to show that there are arbitrarily large numbers $\alpha$ satisfying 
the bounds (\ref{l:n12}); we only consider $\alpha=2n$ with $n \in \N$.
For $\omega>1$ and $\omega^\prime \in \R$, we have the inequality
\[ \omega^{2n} \;\leq\; (2 \omega^\prime)^{2n} 
\:\Theta(\omega^\prime) \:+\: (2 (\omega-\omega^\prime))^{2n} \;\;\; , \]
as is immediately verified by checking the three regions 
$\omega^\prime \leq 0$, $0 < \omega^\prime \leq \omega/2$, and 
$\omega^\prime>\omega/2$. We combine this inequality with (\ref{l:n13}) and
obtain for $\omega>1$ the estimate
\begin{eqnarray}
\lefteqn{ |\omega^{2n} \:F_j(\omega, \vec{k})| } \nonumber \\
&\leq& \left| \int \frac{d\omega^\prime}{(2 \pi)} 
\int \frac{d\vec{k}^\prime}{(2 \pi)^3} \: \tilde{B}(\omega-\omega^\prime, 
\vec{k}-\vec{k}^\prime) \:C_{j-1}(\omega^\prime, \vec{k}^\prime) 
\left[ (2 \omega^\prime)^{2n} \:\Theta(\omega^\prime) 
\:F_{j-1}(\omega^\prime, \vec{k}) \right] \right| \label{l:n14} \\
&&\!\!\!\!\!+\left| \int \frac{d\omega^\prime}{2 \pi} \int 
\frac{d\vec{k}^\prime}{(2 \pi)^3} \:
\left[ (2(\omega-\omega^\prime))^{2n} 
\:\tilde{B}(\omega-\omega^\prime, \vec{k}-\vec{k}^\prime) \right]
\:C_{j-1}(\omega^\prime, \vec{k}^\prime) 
\:F_{j-1}(\omega^\prime, \vec{k}) \right| . \spc \label{l:n15}
\end{eqnarray}
According to the induction hypothesis, the square bracket in 
(\ref{l:n14}) is bounded together with its first derivatives. Since 
$\tilde{B}$ has rapid decay at infinity, the square bracket in (\ref{l:n15}) also 
has rapid decay at infinity. Thus both integrals in (\ref{l:n14}) and (\ref{l:n15}) 
satisfy the hypothesis considered in Lemma \ref{l:lemma0} for (\ref{l:3a}),
and are therefore bounded.
For estimating $|\omega^{2n}\: \partial_l F_j|$, we
differentiate (\ref{l:n13}) and obtain similar to 
(\ref{l:n14}),(\ref{l:n15}) the inequality
\begin{eqnarray*}
\lefteqn{ |\omega^{2n} \:\partial_l F_j(\omega, \vec{k})| } \\
&\leq& \left| \int \frac{d\omega^\prime}{2 \pi} 
\int \frac{d\vec{k}^\prime}{(2 \pi)^3} \;
\partial_l \tilde{B}(\omega-\omega^\prime, 
\vec{k}-\vec{k}^\prime) \:C_{j-1}(\omega^\prime, \vec{k}^\prime) 
\left[ (2 \omega^\prime)^{2n} \:\Theta(\omega^\prime) 
\:F_{j-1}(\omega^\prime, \vec{k}) \right] \right| \\
&&+\left| \int \frac{d\omega^\prime}{d \omega} \int 
\frac{d\vec{k}^\prime}{(2 \pi)^3} \:
\left[ (2(\omega-\omega^\prime))^{2n} 
\:\partial_l \tilde{B}(\omega-\omega^\prime, \vec{k}-\vec{k}^\prime) \right]
\:C_{j-1}(\omega^\prime, \vec{k}^\prime) 
\:F_{j-1}(\omega^\prime, \vec{k}) \right| \;\;\; .
\end{eqnarray*}
This concludes the proof of the induction step.

We have just shown that for a function $g$ with rapid decay for 
positive frequency, the function
\begin{equation}
F_n(q) \;=\; \int \frac{d^4q_1}{(2 \pi)^4} \left( {\cal{B}}\:C_{n-1}\:{\cal{B}}
\cdots {\cal{B}}\:C_1\:{\cal{B}}\:C_0 \right)(q,q_1) \:g(q_1)
        \label{l:n15a}
\end{equation}
has rapid decay for positive frequency. We now consider what this 
means for our operator product (\ref{l:p-k}) in position space. For a 
given four-vector $y=(y^0, \vec{y})$, we choose
\[ g(\omega, \vec{k}) \;=\; \eta(\omega)\; e^{-i 
(\omega y^0 - \vec{k} \vec{y})} \;\;\; , \]
where $\eta$ is a smooth function with $\eta(\omega)=1$ for 
$\omega \leq 0$ and $\eta(\omega)=0$ for $\omega>1$ (this choice of $g$ clearly 
has rapid decay for positive frequency). Since the support of the 
factor $C_0=(p-k)$ is the lower mass cone $\{q^2 \geq 0,\: q^0 \leq 
0\}$, $g(\omega, \vec{k})$ enters into the integral (\ref{l:n15a}) only 
for negative $\omega$. But for $\omega \leq 0$, the cutoff function $\eta$ is 
identically one. Thus the integral (\ref{l:n15a}) is simply a Fourier 
integral; i.e., with a mixed notation in momentum and position space,
\[ F_n(q) \;=\; \left( {\cal{B}}\:C_{n-1}\:{\cal{B}} \cdots 
{\cal{B}}\:C_1\:{\cal{B}}\:(p-k) \right)(q,y) \;\;\; . \]
Next, we multiply from the left with the operator $(p+k)$,
\begin{equation}
\left( (p+k)\:{\cal{B}}\:C_{n-1}\:{\cal{B}} \cdots 
{\cal{B}}\:C_1\:{\cal{B}}\:(p-k) \right)(q,y) \;=\;
(p+k)(q) \:F_n(q) \;\;\; .
        \label{l:n16}
\end{equation}
Since $F_n$ has rapid decay for positive frequency and $(p+k)$ has 
its support in the upper mass cone $\{q^2 \geq 0,\:q^0>0\}$, their product 
decays fast at infinity. More precisely,
\[ \left| q^I \:(p+k)(q) \:F_n(q) \right| \;\leq\; {\mbox{const}}(I) 
\: (p+k)(q) \]
for any multi-index $I$. As a consequence, the Fourier transform of 
(\ref{l:n16}) is even finite after multiplying with an arbitrary number 
of factors $q$, i.e.
\[ \left| \int \frac{d^4q}{(2 \pi)^4} \:q^I
\;(p+k)(q) \:F_n(q) \; e^{-i q x} \right| \;\leq\; 
{\mbox{const}}(I) \;<\; \infty \]
for all $x$ and $I$.
This shows that our operator product in position space (\ref{l:p-k}) is 
bounded and, for fixed $y$, a smooth function in $x$ (with derivative 
bounds which are uniform in $y$). Similarly, one 
obtains that (\ref{l:p-k}) is, for fixed $x$, a smooth function in $y$. We 
conclude that the distribution (\ref{l:p-k}) is a smooth and bounded function.
\QED
We point out that $\tilde{P}^{\mbox{\scriptsize{he}}}(x,y)$ is a 
non-causal object in the sense that it does not only depend on the 
potential ${\cal{B}}$ in the ``diamond'' $(L^\vee_x \cap L^\wedge_y) \cup 
(L^\wedge_x \cap L^\vee_y)$ (with $L^\vee_x$ and $L^\wedge_x$ 
according to (\ref{l:n18})), but on the external potential in the entire space-time. 
This becomes clear from the fact that the support of the operators 
$p(z_1, z_2)$ in the perturbation expansion for
$\tilde{P}^{\mbox{\scriptsize{he}}}$ is the whole space $(z_1, z_2) 
\in \R^4 \times \R^4$. Especially, it is not possible to express
$\tilde{P}^{\mbox{\scriptsize{he}}}(x,y)$ similar to the formulas of 
the light-cone expansion in terms of the potential and its partial 
derivatives along the convex line $\overline{xy}$.

The non-causal high energy contribution is an effect of 
higher order perturbation theory; it vanishes to first order in the 
external potential \cite{F2}.
According to the decomposition into terms of the form (\ref{l:p-k}), it
comes about because states on the 
upper and lower mass shell are mixed through multiplication with the 
potential. Qualitatively speaking, this mixing only becomes an 
important effect if the energy (i.e.\ frequency) of the external 
potential is high enough to overcome the energy difference between the 
states on the upper and lower mass shell. This gives the justification 
for the name ``high energy'' contribution.

\subsection{The Non-Causal Low Energy Contribution}
\label{l:sec_33}
In this subsection, we will put the residual argument and the 
formal light-cone expansion of Proposition \ref{l:prp4} on a satisfying 
mathematical basis.
In order to explain what we precisely need to do, we first
recall how the light-cone expansion of the Green's functions makes 
mathematical sense: Theorem \ref{l:thm1} gives a representation of every 
Feynman diagram of the perturbation series (\ref{l:11b}) as an infinite 
sum of contributions of the form (\ref{l:l1}). According to the bound 
(\ref{l:45o}), there are, for any given $h$, only a finite number of
possibilities to choose $I_a$ and $p_a$; as a consequence, we get, for
fixed $h$, only a finite number of contributions (\ref{l:l1}).
Thus we can write the light-cone expansion in the symbolic form
\begin{equation}
((-s \:B)^k \:s)(x,y) \;=\; \sum_{h=-1}^\infty 
\:\sum_{\mbox{\scriptsize{finite}}} \:\cdots\: S^{(h)}(x,y) \;\;\; ,
        \label{l:m1}
\end{equation}
where `$\cdots$' stands for a configuration of the $\gamma$-matrices 
and nested line integrals in (\ref{l:l1}). According to the
explicit formula (\ref{l:12}), the higher $a$-derivatives of $S_a(x,y)$ 
contain more factors $(y-x)^2$ and are thus of higher order on the 
light cone. This makes it possible to understand the infinite 
sum in (\ref{l:m1}) in terms of Def.\ \ref{l:def1}; we can give it a
mathematical meaning via the approximation by the finite 
partial sums (\ref{l:6b}). In Subsection \ref{l:sec_22}, it is shown that
understanding the light-cone expansion via these partial sums even 
makes it possible to explicitly carry out the sum over all Feynman 
diagrams.

According to Proposition \ref{l:prp4}, all the results of Section 
\ref{l:sec_2} are, on a formal level, also valid for the residual Dirac 
sea. We begin by considering the light-cone expansion of the individual
Feynman diagrams in more detail. Similar to 
(\ref{l:m1}), the $k^{\mbox{\scriptsize{th}}}$ order contribution 
$\Delta P^{\mbox{\scriptsize{res}}}$ to the residual Dirac sea has an 
expansion of the form
\begin{equation}
\Delta P^{\mbox{\scriptsize{res}}}(x,y) \;=\; \sum_{h=-1}^\infty 
\:\sum_{\mbox{\scriptsize{finite}}} \:\cdots\: T^{(h)}(x,y) \;\;\; ,
        \label{l:m2}
\end{equation}
where $T^{(h)}$ is the $a$-derivative (\ref{l:F2}) of the lower mass 
shell $T_a$, (\ref{l:31s}). In position space, $T_a$ has the explicit form
\begin{eqnarray}
        T_a(x,y) & = & -\frac{1}{8 \pi^3} \:\lim_{0<\varepsilon \rightarrow 0}
        \: \frac{1}{\xi^2-i \varepsilon \xi^0} \nonumber \\
        &&+\: \frac{a}{32 \pi^3} \:\lim_{0<\varepsilon \rightarrow 0}
        \left( \log(a \xi^2-i \varepsilon \xi^0) + i \pi + c \right)
        \sum_{j=0}^\infty \frac{(-1)^j}{j! \: (j+1)!} \: \frac{(a 
        \xi^2)^j}{4^j} \nonumber \\
         &  & -\frac{a}{32 \pi^3} \sum_{j=0}^\infty \frac{(-1)^j}{j! \: 
         (j+1)!} \: \frac{(a \xi^2)^j}{4^j} \: (\Phi(j+1) + \Phi(j))
        \label{l:3.1}
\end{eqnarray}
with $\xi \equiv (y-x)$, $c=2C-\log 2$ with Euler's constant $C$, and the
function
\[ \Phi(0)=0 \;\;\;\;,\spc \Phi(n)=\sum_{k=1}^n \frac{1}{k}
\;\;\;\;{\mbox{for}} \;\;\; n \geq 1 \;\;\; . \]
The logarithm in (\ref{l:3.1}) is understood in 
the complex plane which is cut along the positive real axis
(so that $\lim_{0<\varepsilon \rightarrow 0} \log(x+i \varepsilon) = 
\log |x|$ is real for $x>0$).
Alternatively, one can avoid the complex logarithm using the formula
\[ \lim_{0<\varepsilon \rightarrow 0} \log(a \xi^2 - i \varepsilon 
\xi^0) \:+\: i \pi \;=\; \log |a \xi^2| \:+\: i \pi \:\Theta(\xi^2) 
\:\epsilon(\xi^0) \]
($\epsilon$ is the step function $\epsilon(x)=1$ for $x \geq 0$ and 
$\epsilon(x)=-1$ otherwise);
thus the complex logarithm describes both a logarithmic pole 
on the light cone and a constant contribution in the interior of the 
light cone. The basic difference between the light-cone expansions 
(\ref{l:m1}) and (\ref{l:m2}) is related to the logarithmic pole $\log |a|$
in (\ref{l:3.1}). Namely, as a consequence of this logarithm, the higher 
$a$-derivatives of $T_a$ are {\em{not}} of higher order on the light 
cone. To the order ${\cal{O}}((y-x)^2)$, for example, one has
\begin{equation}
\left( \frac{d}{da} \right)^n T_a(x,y) \;=\; \frac{1}{32 \pi^3} \left( 
\frac{d}{da} \right)^n (a \:\log |a|) \:+\: {\cal{O}}((y-x)^2)
\spc (,n \geq 2) .
        \label{l:m3}
\end{equation}
This means that the infinite sum in (\ref{l:m2}) cannot be understood 
in terms of Def.\ \ref{l:def1}; the number of summands is already infinite 
to a given order on the light cone. In our context of an expansion 
around $a=0$, the situation is even worse, because the 
$a$-derivatives of $T_a$ are singular for $a \rightarrow 0$ (as one 
sees e.g.\ in (\ref{l:m3})). Thus not even the individual contributions 
to the light-cone expansion make mathematical sense. These 
difficulties arising from the logarithm in (\ref{l:3.1}) are called the 
{\em{logarithmic mass problem}} (see \cite{F2} for a more detailed 
discussion in a slightly different setting). Since we know from Lemma 
\ref{l:lemma0} that the Feynman diagrams are all well-defined, the 
logarithmic mass problem is not a problem of the perturbation 
expansion, but shows that something is wrong with the light-cone 
expansion of Proposition \ref{l:prp4}.

In order to resolve the logarithmic mass problem, we first 
``regularize'' the formal light-cone expansion by taking out the 
problematic $\log |a|$ term. By resumming the formal light-cone 
expansion, we then show that the difference between the residual 
Dirac sea and the ``regularized'' Dirac sea is a smooth function in 
position space. We introduce the notation
\begin{eqnarray}
T_a^{\mbox{\scriptsize{reg}}}(x,y) &=& T_a(x,y) \:-\:
\frac{a}{32 \pi^3} \:\log|a| \: \sum_{j=0}^\infty
\frac{(-1)^j}{j! \: (j+1)!} \: \frac{(a \xi^2)^j}{4^j} \\
T^{\mbox{\scriptsize{reg}}\:(l)} &=& \left( \frac{d}{da} \right)^l
T_{a \:|\: a=0}^{\mbox{\scriptsize{reg}}} \;\;\; . \label{l:3zz}
\end{eqnarray}

\begin{Def}
\label{l:def_le}
The {\bf{causal contribution}} 
$\tilde{P}^{\mbox{\scriptsize{causal}}}$ to the fermionic projector is obtained 
from the residual Dirac sea $\tilde{P}^{\mbox{\scriptsize{res}}}$ by 
replacing all factors $T^{(h)}$ in the formal light-cone expansion by
$T^{\mbox{\scriptsize{reg}}\:(h)}$.
The {\bf{non-causal low energy contribution}}
$\tilde{P}^{\mbox{\scriptsize{le}}}$ to the fermionic projector is given by
\[ \tilde{P}^{\mbox{\scriptsize{le}}}(x,y) \;=\;
\tilde{P}^{\mbox{\scriptsize{res}}}(x,y) \:-\:
\tilde{P}^{\mbox{\scriptsize{causal}}}(x,y) \;\;\; . \]
\end{Def}
By the replacement $T^{(h)} \rightarrow 
T^{\mbox{\scriptsize{reg}}\:(h)}$, the formal light-cone expansion of 
Proposition \ref{l:prp4} becomes mathematically meaningful in terms of 
Def.\ \ref{l:def1}. Thus we can restate this result as a theorem,
leaving out the word ``formal.''
\begin{Thm} {\bf{(light-cone expansion of the causal contribution)}}
\label{l:thm5}
The results of Section \ref{l:sec_2} also apply to the causal 
contribution to the fermionic projector. More precisely, the
light-cone expansion of Theorem \ref{l:thm1} holds for
$\tilde{P}^{\mbox{\scriptsize{causal}}}$ if we replace
$S^{(h)}$ by $T^{\mbox{\scriptsize{reg}}\:(h)}$, (\ref{l:3zz}),
and multiply the formulas of the light-cone 
expansion from the left with the chiral asymmetry
matrix $X$. According to Theorem \ref{l:thm2}, all tangential 
derivatives can be integrated by parts. With Def.\ \ref{l:def_pf}, 
Def.\ \ref{l:def2}, and Theorem \ref{l:thm3}, the light-cone expansion can 
be reduced to the phase-free contribution. According to Proposition 
\ref{l:prp1}, the phase-free contribution consists, to every order 
${\cal{O}}((y-x)^{2g})$ on the light cone, of only a finite number of terms.
\end{Thm}
We come to the analysis of the low energy contribution. In the 
following lemma, we reformulate the light-cone expansion of Theorem 
\ref{l:thm1} in a way where the infinite sums are handled more explicitly.
\begin{Lemma}
\label{l:lemma_new}
The light-cone expansion of the $k^{\mbox{\scriptsize{th}}}$ order 
contribution $((-s \:B)^k \:s)(x,y)$ to the perturbation series 
(\ref{l:11b}) can be written as a finite sum of expressions of the form
\begin{eqnarray}
\lefteqn{\sum_{n_1,\ldots,n_k=0}^\infty \frac{1}{n_1! \cdots n_k!}\:
\chi_c \:C \:(y-x)^I \int_x^y [a_1, b_1+n_2+\cdots+n_k \:|\: n_1] \:dz_1 \; 
\partial_{z_1}^{I_1}\: \OBox_{z_1}^{n_1+q_1} \:V^{(1)}_{J_1, c_1}(z_1) }
\nonumber \\
&& \times \int_{z_1}^y [a_2, b_2+n_3+\cdots+n_k \:|\: n_2] \:dz_2 \;
\partial_{z_2}^{I_2}\:
\OBox_{z_2}^{n_2+q_2}\: V^{(2)}_{J_2, c_2}(z_2) \nonumber \\
&& \hspace*{1cm} \cdots \int_{z_{k-1}}^y [a_k, b_k \:|\: n_k] \:dz_k \; 
\partial_{z_k}^{I_k}\: \OBox_{z_k}^{n_k+q_k}\: V^{(k)}_{J_k, c_k}(z_k) \;
\gamma^J \;S^{(r+n_1+\cdots+n_k)}(x,y) \;\;\; . \spc\; \label{l:Al1}
\end{eqnarray}
In this formula, we use for the chiral and tensor indices the same
notation as in Theorem \ref{l:thm1}; 
$a_l$, $b_l$, and $q_l$ are non-negative integers.
The parameters $r$ and $b_l$ satisfy the bounds
\begin{eqnarray}
r &\leq& k - |I| \label{l:Al2} \\
b_l &\geq& r - l + |I| \;\;\;,\spc 1 \leq l \leq k. \label{l:Al3}
\end{eqnarray}
\end{Lemma}
{\Proof}
The form of the expression (\ref{l:Al1}) is straightforward if one 
keeps track of the infinite sums in the inductive construction of 
Theorem \ref{l:thm1}; it is also obvious that we only get a finite 
number of such expressions. The only point which needs an explanation 
is how one can arrange that all infinite sums are of the form
\begin{equation}
\sum_{n=0}^\infty \frac{1}{n!} 
\:[.,. \:|\: n] \;\OBox^n \;\cdots \;\;\; .
        \label{l:sn}
\end{equation}
For this, we must manipulate the sums
when then Laplacian $\OBox^n$ is carried 
out after (\ref{l:33n}). Whenever a Laplacian acts on a factor $(y-x)^I$, 
we shift the summation index by one.
More precisely, in the case of one factor $(y-x)$, we use the
transformation
\begin{eqnarray}
\lefteqn{ \sum_{n=0}^\infty \frac{1}{n!} \:[a, b \:|\: n] \: 
\OBox^n_x \: (y-x)^i \:f_{(n)}(x) \;=\;
\sum_{n=0}^\infty \frac{1}{n!} \:[a, b \:|\: n] \left(
(y-x)^i \:\OBox^n_x \:f_{(n)} - 2n \:\partial^i \OBox^{n-1} f_{(n)}
\right) } \nonumber \\
&=& \!\!(y-x)^i \sum_{n=0}^\infty \frac{1}{n!} \:[a, b \:|\: n] \: 
\OBox^n_x \:f_{(n)}(x) \:-\: 2 \sum_{n=0}^\infty \frac{1}{n!}
\:[a+1, b+1 \:|\: n] \: \OBox^n_x \partial^i f_{(n+1)}(x) \spc \label{l:Al4}
\end{eqnarray}
(where $f_{(n)}$ denotes a function depending on $n$);
in the general case of several factors $(y-x)$, we inductively 
apply (\ref{l:Al4}).

It remains to show that the parameters $a_l$, $b_l$, and $q_l$ are
non-negative, and that the inequalities (\ref{l:Al2}) and (\ref{l:Al3}) 
hold. For this, it suffices to consider the leading summand $n_1=\cdots=n_k=0$
of (\ref{l:Al1}). Since this is a (special) contribution of the form
(\ref{l:l1}), we can apply Theorem \ref{l:thm1} 
with $a_i=l_i+n_i$, $b_i=r_i+n_i$, and $p_i=q_i$. It follows that $a_l$, $b_l$,
and $q_l$ are non-negative.
For the proof of the inequalities (\ref{l:Al2}) and (\ref{l:Al3}), we 
proceed inductively in the order $k$ of the perturbation theory. For 
$k=0$, we have $r=-1$ and $|I|=1$, so that the inequalities are 
satisfied. Assume that (\ref{l:Al2}),(\ref{l:Al3}) hold for a given 
$k$. We go through the construction steps of Theorem \ref{l:thm1} using 
the index shift (\ref{l:Al4}) and verify that (\ref{l:Al2}) and 
(\ref{l:Al3}) are also valid to $(k+1)^{\mbox{\scriptsize{st}}}$ order:

For the proof of (\ref{l:Al2}), we note that additional factors $(y-x)$
are generated at most 
once in the construction; namely, if the derivative $\Pdd_x$ acts 
on $S^{(\hat{h})}$ in step {\rm{\em{5)}}. The parameter $r$ is only 
increased if either a Laplacian acts on the factor $(y-x)^I$ in step
{\rm{\em{3)}} (leading 
to the index shift (\ref{l:Al4})) or if the derivative $\Pdd_x$ does 
not act on $S^{(\hat{h})}$ in step {\rm{\em{5)}}. In both cases, one 
loses at least one factor $(y-x)$. This gives (\ref{l:Al2}).

For the proof of (\ref{l:Al3}), we take the 
contribution (\ref{l:Al1}) with $n_1=\cdots=n_k=0$ and apply the 
construction of Theorem \ref{l:thm1}. When 
computing the Laplacian $\OBox^n_z$ in step {\rm{\em{3)}}, we shift 
the index according to (\ref{l:Al4}) whenever a Laplacian acts on
a factor $(y-x)$. Denoting the number of index shifts by $s$, we get a 
finite number of terms of the form
\begin{eqnarray*}
\lefteqn{ \sum_{n=0}^\infty \frac{1}{n!}\:
\chi_c \:C \:i \Pdd_x \:S^{(\hat{r}+n)}(x,y)
\int_x^y [s,r+s \:|\: 0] \:dz\; (y-x)^{\hat{I}} \:
\partial_z^{I_0} \:\OBox^n V^{(0)}_{J_0}(z) } \\
&& \times
\int_z^y [a_1, b_1 \:|\: 0] \:dz_1 \; 
\partial_{z_1}^{I_1}\: \OBox_{z_1}^{q_1} \:V^{(1)}_{J_1, c_1}(z_1)
\cdots \int_{z_{k-1}}^y [a_k, b_k \:|\: 0] \:dz_k \; 
\partial_{z_k}^{I_k}\: \OBox_{z_k}^{q_k}\: V^{(k)}_{J_k, c_k}(z_k) \;
\gamma^J \;\;\; .
\end{eqnarray*}
Each index shift decreases the number of factors $(y-z)$
and increments the order of the mass-derivative of the Green's function, thus
\begin{equation}
        |\hat{I}| \;\leq\; |I| - s \;\;\;,\spc \hat{r} \;=\; r + 1 + s 
        \;\;\;\; .
        \label{l:li0}
\end{equation}
It again suffices to consider the leading summand $n=0$; this is a 
contribution of the form (\ref{l:l1}).
After extracting the factors $(y-x)$ in step {\rm{\em{4)}}, the 
parameter $b_0=r_0+n_0$ satisfies
\[ b_0 \;=\; r + |\hat{I}| + s \;
\stackrel{(\ref{l:li0})}{\geq}\; \hat{r} + |\hat{I}| - 1 \;\;\; . \]
The parameters $b_l$, $1 \leq l \leq k$, remain unchanged in the construction;
they are still the same as in (\ref{l:Al3}),
\[ b_l \;\geq\; r - l + |I|
\;\geq\; \hat{r} - (l+1) + |\hat{I}| \;\;\;,\spc l=1,\ldots,k. \]
When the derivative $\Pdd_x$ is carried out in step {\rm{\em{5)}}, 
either $r$ is decremented and $|\hat{I}|$ increased by one, or 
$|\hat{I}|$ is decreased. In steps {\rm{\em{6)}} and {\rm{\em{7)}}, the 
transformations (\ref{l:25t}) and (\ref{l:25u}) may only decrease the 
sum $\hat{r} + |\hat{I}|$. We conclude that, after performing all the
construction steps,
\[ b_l \;\geq\; \hat{r} - (l+1) + |\hat{I}| \;\;\;,\spc l=0,\ldots,k. \]
The index shift $l \rightarrow l+1$ finally gives
the inequalities (\ref{l:Al3}) for $l=1,\ldots,k+1$.
\QED
  
\begin{Thm}
\label{l:thm6}
The non-causal low energy contribution 
$\tilde{P}^{\mbox{\scriptsize{le}}}(x,y)$ is, to every order in
perturbation theory, a smooth function in $x$ and $y$.
\end{Thm}
{\Proof}
We first outline our strategy: According to Def.~\ref{l:def_res},
Proposition~\ref{l:prp4}, and Def.~\ref{l:def_le}, the
$k^{\mbox{\scriptsize{th}}}$ order contribution to
$\tilde{P}^{\mbox{\scriptsize{le}}}(x,y)$ is obtained from (\ref{l:m1}) 
by the replacement $S^{(h)} \rightarrow (T^{(h)} - 
T^{\mbox{\scriptsize{reg}} \:(h)})$,
\begin{equation}
\Delta P^{\mbox{\scriptsize{le}}}(x,y) \;=\; X \sum_{h=-1}^\infty 
\:\sum_{\mbox{\scriptsize{finite}}} \:\cdots\:
(T^{(h)} - T^{\mbox{\scriptsize{reg}} \:(h)})(x,y) \;\;\; .
        \label{l:m6}
\end{equation}
Because of the logarithmic mass problem, the infinite sum in 
(\ref{l:m6}) is ill-defined. In order to give (\ref{l:m6}) a 
mathematical meaning, we manipulate the infinite sum 
until recovering it as a formal Taylor series, which can be carried 
out explicitly. Finally, we show that the mathematical object
$\Delta P^{\mbox{\scriptsize{le}}}(x,y)$ obtained in this way
is a smooth function in $x$ and $y$.

Consider the light-cone expansion of the Green's functions of 
Lemma \ref{l:lemma_new}. Since there are only a finite number of 
contributions of the form (\ref{l:Al1}), we can restrict ourselves to
one of them. In order to get the corresponding contribution to 
$\tilde{P}^{\mbox{\scriptsize{le}}}$, we replace the factor $S^{(h)}$ 
in (\ref{l:Al1}) according to (\ref{l:m6}) by the operator
$L^{(h)}:=T^{(h)}-T^{\mbox{\scriptsize{reg}} \:(h)}$, i.e. more 
explicitly
\[ L^{(h)}(x,y) \;=\; \left( \frac{d}{da} \right)^h L_{a \:|\: a=0} \]
with
\begin{equation}
        L_a(x,y) \;=\; \frac{a}{32 \pi^3} \:\log |a| \:
                \sum_{j=0}^\infty \frac{(-1)^j}{j! \: (j+1)!} \: \frac{(a 
                \:(y-x)^2)^j}{4^j} \;\;\; .
        \label{l:m9a}
\end{equation}
We can leave out the factor $(y-x)^I$ in (\ref{l:Al1}) and disregard the 
chiral asymmetry matrix $X$ in (\ref{l:m6}), because they are irrelevant 
as smooth functions. Furthermore, we can carry out the partial derivatives
$\partial^{I_l}_{z_l}$ in (\ref{l:Al1}) with the Leibniz rule.
According to the chain rule (\ref{l:34h}),
this may increase the parameters $b_l$; nevertheless, the inequalities
(\ref{l:Al2}) and (\ref{l:Al3}) remain valid. We conclude that
it suffices to consider the formal series
\begin{eqnarray}
\Delta P^{\mbox{\scriptsize{le}}}(x,y) &=&
\sum_{n_1,\ldots,n_k=0}^\infty \:\frac{1}{n_1! \cdots n_k!}\:
\int_x^y [a_1, b_1+n_2+\cdots+n_k \:|\: n_1] \:dz_1 \; 
\OBox^{n_1}_{z_{1}} \:W_1(z_1) \nonumber \\
&&\:\cdots\: \int_{z_{k-1}}^y [a_k, b_k \:|\: n_k] \:dz_k \; 
\OBox^{n_k}_{z_k} \:W_k(z_k) \; L^{(r+n_1+\cdots+n_k)}(x,y)
\label{l:m10}
\end{eqnarray}
together with the bounds (\ref{l:Al2}) and (\ref{l:Al3}), where $W_l$ 
stand for partial derivatives of the potential $V^{(l)}_{J_l, c_l}$.
Our task is to give this series a mathematical meaning 
and to show that it is a smooth function in $x$ and $y$.

We first consider the case that the potentials $W_l$ 
are plane waves,
\[ W_l(x) \;=\; e^{-i q_l x} \;\;\; . \]
This allows us to explicitly carry out the Laplacians in (\ref{l:m10}), and we
obtain
\begin{eqnarray}
\lefteqn{ \Delta P^{\mbox{\scriptsize{le}}}(x,y) \;=\; \int_x^y 
[a_1,b_1 \:|\: 0] \:dz_1\: e^{-i q_1 z_1} \:\cdots\:
\int_{z_{k-1}}^y [a_k, b_k \:|\: 0] 
\:dz_k \: e^{-i q_k z_k} } \nonumber \\
&& \times \sum_{n_1,\ldots,n_k=0}^\infty \:\frac{1}{n_1! \cdots n_k!}
\left( (\alpha_1^2 - \alpha_1) \:p_1^2 \right)^{n_1}
\left( (1-\alpha_1) (\alpha_2^2 - \alpha_2) \:p_2^2 
\right)^{n_2}
\nonumber \\
&& \hspace*{0.7cm} \cdots\:
\left( (1-\alpha_1) (1-\alpha_2) \cdots (1-\alpha_{k-1})
(\alpha_k^2 - \alpha_k) \:p_k^2 \right)^{n_k} \; 
L^{(r+n_1+\cdots+n_k)}(x,y) \;\;\; , \spc
\label{l:m11}
\end{eqnarray}
where $\alpha_l$ denote the integration variables of the line integrals 
(all running from zero to one), and where $p_l$ are the momenta
\begin{equation}
p_l \;=\; q_l + (1-\alpha_l) \:q_{l+1} \:+\: \cdots \:+\:
(1-\alpha_l)(1-\alpha_{l+1}) \cdots (1-\alpha_{k-1}) \:q_k \;\;\; .
        \label{l:m12}
\end{equation}
For fixed values of the parameters $\alpha_1,\ldots,\alpha_k$, 
(\ref{l:m11}) is a product of
$k$ formal Taylor series. For example by using the continuation of the
logarithm into the complex plane
\[ \log |a| \;=\; \frac{1}{2} \:\lim_{0<\varepsilon \rightarrow 0} 
\left( \log(a+i \varepsilon) \:+\: \log(a-i \varepsilon) \:+\: 2 i 
\pi \right) \]
(which is, as in (\ref{l:3.1}), cut along the positive real axis), we 
can carry out these formal Taylor series and obtain
\begin{equation}
\Delta P^{\mbox{\scriptsize{le}}}(x,y) \;=\; \int_x^y [a_1, b_1 \:|\: 0] 
\:dz_1\: e^{-i q_1 z_1} \:\cdots\: \int_{z_{k-1}}^y [a_k, b_k \:|\: 0] 
\:dz_k \: e^{-i q_k z_k} \:\left( \frac{d}{da} \right)^r L_a(x,y)
\label{l:m13}
\end{equation}
with
\begin{eqnarray}
a &=& (\alpha_1^2 - \alpha_1) \:p_1^2 \:+\:
(1-\alpha_1) (\alpha_2^2 - \alpha_2) \:p_2^2 \nonumber \\
&&\hspace*{1cm}
+\:\cdots\:+\: (1-\alpha_1) (1-\alpha_2) \cdots (1-\alpha_{k-1})
(\alpha_k^2 - \alpha_k) \:p_k^2
        \label{l:m14}
\end{eqnarray}
and the momenta $p_l$ according to (\ref{l:m12}).
This construction is helpful in two ways: all infinite sums have 
disappeared, and the mass parameter $a$ is now in general non-zero, so 
that the $a$-derivatives of $L_a$ are no longer singular.

After this preparation, we consider the case that the potentials 
$V^{(l)}$ in (\ref{l:Al1}) are smooth and satisfy the conditions of Lemma 
\ref{l:lemma0} on the decay at infinity. Then the Fourier transform 
$\tilde{V}^{(l)}$ is $C^2$ and has rapid decay at infinity. Since the 
potentials $W^{(l)}$ are partial derivatives of $V^{(l)}$, their 
Fourier transforms $\tilde{W}_l$ are also $C^2$ and have rapid decay 
at infinity. The low energy contribution is obtained
from (\ref{l:m13}) by integrating over the momenta $q_l$. More precisely,
\begin{eqnarray}
\lefteqn{ \Delta P^{\mbox{\scriptsize{le}}}(x,y) \;=\; \left( \frac{d}{db} 
\right)^r \int \frac{d^4q_1}{(2 \pi)^4} \:\cdots\: \int \frac{d^4q_k}{(2 
\pi)^4} \;\tilde{W}_1(q_1) \cdots \tilde{W}_k(q_k) } \nonumber \\
&& \times \; \int_x^y [a_1,b_1 \:|\: 0] 
\:dz_1\: e^{-i q_1 z_1} \:\cdots\: \int_{z_{k-1}}^y [a_k, b_k \:|\: 0] 
\:dz_k \: e^{-i q_k z_k} \;
L_{a+b}(x,y)_{\;|\: b=0} \;\;\;,\spc \label{l:m15}
\end{eqnarray}
where the parameter $a$ depends on the momenta $q_l$ via (\ref{l:m14}) 
and (\ref{l:m12}).
We must show that (\ref{l:m15}) is well-defined and depends smoothly 
on $x$ and $y$. Qualitatively speaking, we can view the 
$q_l$-integrations as a multiple convolution in the parameter $a$. 
These convolutions mollify $L_{a+b}$ in such a way that the 
$b$-derivatives can be carried out giving a smooth function in $x$ 
and $y$. Unfortunately, this ``mollifying argument'' is quite delicate. 
Complications arise from the fact that the $q_l$-integrals are 
multi-dimensional and that we must handle the additional line 
integrals over $\alpha_l$. The main problem is that the dependence 
of $a$ on the momenta $q_l$ becomes singular when the parameters 
$\alpha_l$ approach zero or one (see (\ref{l:m14})).
Because of these difficulties, we give the mollifying argument in detail. 
Equivalently to analyzing the regularity in the parameter $a$, one can
take its Fourier transform,
\begin{equation}
        L_a(x,y) \;=\; \int_{-\infty}^\infty \frac{d \tau}{2 \pi} 
        \:\tilde{L}_\tau(x,y) \; e^{-i \tau a} \;\;\; ,
        \label{l:ma0}
\end{equation}
and study the decay properties in $\tau$ for $\tau \rightarrow \pm \infty$.
We prefer working with the parameter $\tau$, because this is a bit easier
and makes our argument clearer.

In the first step of the mollifying argument, we transform the 
$q_l$-integrals into integrals over the momenta $p_l$. Since the 
transformation (\ref{l:m12}) is volume preserving, we get
\begin{eqnarray}
\lefteqn{ \Delta P^{\mbox{\scriptsize{le}}}(x,y) \;=\; \left( \frac{d}{db} 
\right)^r \int \frac{d^4p_1}{(2 \pi)^4} \:\cdots\: \int \frac{d^4p_k}{(2 
\pi)^4} } \nonumber \\
&& \times \; \int_x^y [a_1,b_1 \:|\: 0] 
\:dz_1 \:\cdots\: \int_{z_{k-1}}^y [a_k, b_k \:|\: 0] 
\:dz_k \; f^{\{\alpha\}}(p_1,\ldots,p_k) \;L_{a+b}(x,y)_{\;|\: b=0}
\spc \label{l:ma1}
\end{eqnarray}
with
\[ f^{\{\alpha\}}(p_1,\ldots,p_k) \;:=\;
\tilde{W}_1(q_1) \cdots \tilde{W}_k(q_k) \;e^{-i (q_1 z_1+\cdots+q_k 
z_k)} \;\;\;, \]
where $p_l$ and $q_l$ are related to each other via (\ref{l:m12}). The 
explicit inverse of (\ref{l:m12}),
\[ q_l \;=\; p_l - (1-\alpha_l) \:p_{l+1} \;,\;\;\; 1 \leq l \leq k
\;\;\;\;\; {\mbox{and}} \;\;\;\;\; q_k=p_k \;\;\; , \]
shows that this transformation is regular for all values of the 
parameters $\alpha_l$. As a consequence, the function
$f^{\{\alpha\}}(p_1,\ldots,p_k)$ is $C^2$ and has 
rapid decay at infinity, both uniformly in $\alpha_l$ (more precisely,
the Schwartz norms $\| f^{\{\alpha\}} \|_{p,2}$, $p \geq 0$,
are all uniformly bounded in $\alpha_l$). Because of 
this uniformity, we need not care about the dependence of 
$f^{\{\alpha\}}$ on the parameters $\alpha_l$ in the following.

According to the bound (\ref{l:Al2}), $r \leq k$. Thus we can insert 
into (\ref{l:ma1}) the identity
\[ 1 \;=\; \int_{-\infty}^\infty da_1 \:\cdots \int_{-\infty}^\infty 
da_r \; \delta(a_1 - p_1^2) \:\cdots\: \delta(a_r - p_r^2) \;\;\; , \]
pull the $a_l$-integrals outside and carry out the integrations over 
$p_1,\ldots,p_r$. This gives
\begin{eqnarray}
\lefteqn{ \Delta P^{\mbox{\scriptsize{le}}}(x,y) \;=\; \left( \frac{d}{db} 
\right)^r \int_{-\infty}^\infty da_1 \cdots \int_{-\infty}^\infty da_r
\int \frac{d^4p_{r+1}}{(2 \pi)^4} \:\cdots\: \int \frac{d^4p_k}{(2 
\pi)^4} } \nonumber \\
&& \hspace*{-.5cm} \times \!\int_x^y [a_1,b_1 \:|\: 0] 
\:dz_1 \:\cdots\: \int_{z_{k-1}}^y [a_k, b_k \:|\: 0] 
\:dz_k \; U(a_1,\ldots,a_r,p_{r+1},\ldots,p_k) \;L_{a+b}(x,y)_{\;|\: b=0}
\;\;\;\;\;\;\;\;\;
\label{l:ma2}
\end{eqnarray}
with
\begin{equation}
U \;=\;\int \frac{d^4 p_1}{(2 \pi)^4} \cdots \int \frac{d^4 p_r}{(2 \pi)^4}
\; \delta(a_1 - p_1^2) \:\cdots\: \delta(a_r - p_r^2) \;
f^{\{\alpha\}}(p_1,\ldots,p_k) \;\;\;\; .
        \label{l:ma3}
\end{equation}
In (\ref{l:ma2}), the parameter $a$ depends on $a_l$, $p_l$, and $\alpha_l$ via
\begin{eqnarray*}
a &=& (\alpha_1^2 - \alpha_1) \:a_1 \:+\: \cdots \:+\:
(1-\alpha_1) \cdots (1-\alpha_{r-1})(\alpha_r^2 - \alpha_r) \:a_r \nonumber \\
&&+(1-\alpha_1) \cdots (1-\alpha_r) (\alpha_{r+1}^2 - \alpha_{r+1})
\:p_{r+1}^2 \\
&& \hspace*{2cm} \:+\:\cdots\:+\:
(1-\alpha_1) (1-\alpha_2) \cdots (1-\alpha_{k-1})
(\alpha_k^2 - \alpha_k) \:p_k^2 \;\;\; .
\end{eqnarray*}
Since $f^{\{\alpha\}}$ is $C^2$ and has rapid decay at infinity, it 
follows by evaluating the integrals over the mass shells in 
(\ref{l:ma3}) that $U$ is $C^1$ in the variables $a_l$ and has rapid 
decay at infinity.

Next, we take the Fourier transform in the variables $a$ and 
$a_1,\ldots,a_r$ by substituting (\ref{l:ma0}) and
\begin{eqnarray*}
\lefteqn{ U(a_1,\ldots,a_r, p_{r+1},\ldots,p_l) } \\
&=& \int_{-\infty}^\infty 
\frac{d\tau_1}{2 \pi} \cdots \int_{-\infty}^\infty \frac{d\tau_r}{2 
\pi} \;\tilde{U}(\tau_1,\ldots,\tau_r, p_{r+1},\ldots,p_k) \;
e^{-i(\tau_1 a_1 + \cdots + \tau_r a_r)}
\end{eqnarray*}
into (\ref{l:ma2}). The $a_l$-dependence of the resulting expression for
$\Delta P^{\mbox{\scriptsize{le}}}$ has the form of plane waves; thus
the $a_l$-integrals give $\delta$-distributions. We can then also
carry out the $\tau_l$-integrations. Finally, the $b$-derivatives in
(\ref{l:ma2}) give a factor $(-i \tau)^r$, and we 
obtain
\begin{eqnarray}
\Delta P^{\mbox{\scriptsize{le}}}(x,y) &=&
\int_x^y [a_1,b_1 \:|\: 0] \:dz_1 \:\cdots\: \int_{z_{k-1}}^y
[a_k, b_k \:|\: 0] \:dz_k
\int \frac{d^4p_{r+1}}{(2 \pi)^4} \:\cdots\: \int \frac{d^4p_k}{(2 
\pi)^4} \nonumber \\
&& \times \; \int_{-\infty}^\infty \frac{d \tau}{2 \pi}
\; \tilde{U}(\tau_1,\ldots,\tau_r,p_{r+1},\ldots,p_k) \:
(-i \tau)^r \;\tilde{L}_\tau(x,y) \;\;\;\;\; ,
\label{l:ma5}
\end{eqnarray}
where the parameters $\tau_l$ are given in terms of $\tau$ and 
$\alpha_l$ by
\[ \tau_l \;=\; (1-\alpha_1) \cdots (1-\alpha_{l-1}) \:(\alpha_l - 
\alpha_l^2) \:\tau \;\;\; . \]
The Fourier transform $\tilde{L}_\tau(x,y)$ is an integrable function 
in $\tau$ which depends smoothly on $x$ and $y$ (this can e.g.\ be 
verified by writing the infinite sum (\ref{l:m9a}) with the Bessel 
functions $J_1$ and $K_1$ which decay at infinity). Since $U$ is $C^2$ 
in the parameters $a_l$ and has rapid decay, its Fourier transform 
$\tilde{U}$ is a function in $\tau_l$ which decays at infinity at 
least like ${\cal{O}}(\tau_l^{-1})$.

We now estimate the $\alpha_l$-integrals in (\ref{l:ma5})
for $l=1,\ldots,r$, from the right to the left.
Since $\tilde{U}$ decays in $\tau_r$ at 
infinity like ${\cal{O}}(\tau_r^{-1})$, we have the bound
\[ \int_0^1 d\alpha_r \;\cdots \tilde{U}_{| \tau_r = (1-\alpha_1)\cdots
(1-\alpha_{r-1}) (\alpha_r - \alpha_r^2) \tau}
\;\leq\; \sup_{\alpha_r \in [0,1]} \;\cdots
\left( (1-\alpha_1)\cdots (1-\alpha_{r-1}) \tau \right)^{-1} \; g 
\;\;\; , \]
where we have for clarity only written out the $\alpha_r$-integral;
$g$ is a function depending on the variables 
$\tau_1,\ldots,\tau_{r-1}$ and $p_{r+1},\ldots,p_k$.
The inequality (\ref{l:Al3}) implies that all 
factors $(1-\alpha_l)^{-1}$ cancel against corresponding factors 
$(1-\alpha_l)$ in the nested line integrals. The decay properties in 
the remaining parameters $\tau_1,\ldots,\tau_{r-1}$ remain unchanged 
by our estimate of the $\alpha_r$-integral. Therefore we can proceed 
inductively in the same way for the integrals over
$\alpha_{r-1}, \alpha_{r-2}, \ldots,
\alpha_1$. The bounds (\ref{l:Al3}) ensure that all 
factors $(1-\alpha_l)^{-1}$ drop out. Since we get a 
factors $\tau^{-1}$ in each step, the factor $(-i \tau)^r$ in (\ref{l:ma5})
disappears.
Using that $\tilde{L}_\tau$ is integrable and that we have fast decay 
at infinity in the variables $p_{r+1},\ldots,p_k$, all the remaining 
integrals are finite. We conclude that $\Delta 
P^{\mbox{\scriptsize{le}}}(x,y)$ is well-defined.

If we take partial derivatives of (\ref{l:m15}) with respect to $x$ and 
$y$, the derivatives act either on the exponentials, yielding 
additional factors $\alpha_l$, $(1-\alpha_l)$, and $q_l$, or they act
on $L_a(x,y)$. Since $L_a$ and its Fourier transform $\tilde{L}_\tau$ 
depend smoothly on $x$ and $y$, we can repeat the above mollifying argument 
and conclude that $\Delta P^{\mbox{\scriptsize{le}}}(x,y)$ is smooth.
\QED
For a very rigid mathematician, it might not seem quite satisfying that 
the light-cone expansion of the residual fermionic projector was first 
performed on a formal level and later made rigorous by resumming the 
formal expansion. We remark that one could avoid all formal series in 
intermediate steps of the construction by already ``regularizing'' the 
logarithmic mass terms in the Green's functions $s^\pm$ after (\ref{l:F}). 
However, this has the disadvantage of becoming quite technical. Our 
procedure is easier to understand because we could introduce the 
residual argument in Subsection \ref{l:sec_31} without entering the
mathematical details right away.

Similar to the high energy contribution, the low energy 
contribution $\tilde{P}^{\mbox{\scriptsize{le}}}(x,y)$ is non-causal 
in the sense that it depends on the external potential in the entire 
Minkowski space. This can be understood from the fact that the 
operator products in the perturbation expansion (\ref{l:E1}) contain 
factors $p(z_1, z_2)$, whose support is the whole space $(z_1, z_2) 
\in \R^4 \times \R^4$. We point out that, although the statements of 
Theorem \ref{l:thm4} and Theorem \ref{l:thm6} are very similar, their 
proofs are completely different. This shows and illustrates that the 
high and low energy contributions describe two different physical 
effects. In contrast to the high energy contribution, the potential 
in the low energy contribution need not overcome an ``energy gap;'' as 
a consequence, the low energy contribution plays an important role
even if the energy of the external potential is small.

This concludes our analysis of the Dirac sea. We briefly summarize our 
main results: According to Def.\ \ref{l:def_res}, Def.\ \ref{l:def_he}, 
and Def.\ \ref{l:def_le}, we decompose the fermionic projector in the form
\begin{equation}
\tilde{P}(x,y) \;=\; \tilde{P}^{\mbox{\scriptsize{causal}}}(x,y) \:+\:
\tilde{P}^{\mbox{\scriptsize{le}}}(x,y) \:+\:
\tilde{P}^{\mbox{\scriptsize{he}}}(x,y) \;\;\; .
        \label{l:m4}
\end{equation}
The causal contribution $\tilde{P}^{\mbox{\scriptsize{causal}}}(x,y)$ 
has singularities on the light cone, which are completely described by 
the light-cone expansion of Theorem \ref{l:thm5}. The non-causal low 
and high energy contributions 
$\tilde{P}^{\mbox{\scriptsize{le}}}(x,y)$ and
$\tilde{P}^{\mbox{\scriptsize{he}}}(x,y)$, on the other hand, are, to 
every order in perturbation theory, smooth functions in $x$ and $y$ (see 
Theorem \ref{l:thm4} and Theorem \ref{l:thm6}).

We finally point out that, in contrast to
$\tilde{P}^{\mbox{\scriptsize{causal}}}$, both non-causal contributions
$\tilde{P}^{\mbox{\scriptsize{le}}}$ and $\tilde{P}^{\mbox{\scriptsize{he}}}$
were only studied to every order in perturbation theory, but we did 
not consider the convergence of their perturbation series. For
$\tilde{P}^{\mbox{\scriptsize{le}}}$, this convergence problem could 
be studied by resumming all phase-free contributions 
to the formal light-cone expansion of Proposition \ref{l:prp4}.
However, there seems to be no easy method at the moment to control the 
convergence of the perturbation expansion of the high energy 
contribution $\tilde{P}^{\mbox{\scriptsize{he}}}(x,y)$.
Nevertheless, Theorem \ref{l:thm4} and Theorem \ref{l:thm6} give a strong 
indication that the non-causal contribution is a smooth function. 
Even if singularities or divergences occurred when carrying out the
sum over the perturbation series (which seems unlikely), these 
singularities would be of very different nature than the 
singularities of $\tilde{P}^{\mbox{\scriptsize{causal}}}(x,y)$. More 
precisely, they could not be expressed in terms of the external 
potential and its derivatives along the convex line $\overline{xy}$, 
because singularities of this type necessarily show up in finite order
perturbation theory.
Thus we can say that, although the non-causal contributions
$\tilde{P}^{\mbox{\scriptsize{le}}}$ and $\tilde{P}^{\mbox{\scriptsize{he}}}$
require further study in order to get a complete understanding, the 
decomposition (\ref{l:m4}) is well-established. The light-cone 
expansion gives a method to explicitly calculate the causal 
contribution $\tilde{P}^{\mbox{\scriptsize{causal}}}(x,y)$.

\appendix
\section{Some Formulas of the Light-Cone Expansion}
\setcounter{equation}{0}
In this appendix, we give a compilation of explicit formulas of the 
light-cone expansion. More precisely, we list the phase-free 
contribution to the light-cone expansion of the Green's functions 
(cf.\ Def.\ \ref{l:def_pf}). According to the reduction to the 
phase-free contribution, the light-cone expansion of the Green's 
functions is immediately obtained by inserting ordered exponentials 
into the line integrals, see Def.\ \ref{l:def2} and Theorem \ref{l:thm3}.
Furthermore, using Theorem \ref{l:thm5}, the formulas can be directly applied
to the fermionic projector; they then describe the singularities of 
$\tilde{P}(x,y)$ on the light cone.

All the following formulas were generated by the C++ program 
``class\_commute'' (see Subsection \ref{l:sec2_3}). Our listings are 
not intended to be in any sense complete; we made a selection in 
order to give the reader a first impression of the form of the 
singularities. The more detailed formulas to higher order on the 
light cone can be easily obtained with ``class\_commute.'' Without 
loss of generality, we restrict ourselves to the left handed 
component of the Green's functions; for the right handed component, 
the formulas are analogous.

We begin with the perturbation by a chiral perturbation to first order. 
The phase-free contribution to the order ${\cal{O}}((y-x)^2)$ on the 
light cone is
\begin{eqnarray}
\lefteqn{ \chi_L \:(-s \:(\chi_L \Aslsh_R + \chi_R \Aslsh_L) 
\:s)(x,y) \;\stackrel{\mbox{\footnotesize{phase-free}}}{\asymp}\; 
{\cal{O}}((y-x)^2) } \nonumber \\
&&  +\chi_L \: S^{(0)}(x,y) \: \xi^i \int_x^y dz\:[0,1\:|\: 0]\:
(\Pdd A_{L i}) \label{l:A00} \\
&&  -\chi_L \: S^{(0)}(x,y) \: \int_x^y dz\:[0,0\:|\: 0]\: \Aslsh_L
\label{l:A01} \\ 
&&  +\chi_L \: S^{(0)}(x,y) \: \Aslsh_L(x) \label{l:A02} \\ 
&&  +\frac{1}{2} \: \chi_L \: S^{(0)}(x,y) \: \xi \slsh
\int_x^y dz\: [0,0\:|\: 0]\: (\Pdd \Aslsh_L) \\ 
&&  -\chi_L \: S^{(0)}(x,y) \: \xi \slsh \: \int_x^y dz\: [1,0\:|\: 0]\:
(\partial^i A_{Li}) \\ 
&&  +\frac{1}{2} \chi_L \: S^{(0)}(x,y) \: \xi \slsh  \: \xi^i \:
\int_x^y dz\: [0,0\:|\: 1]\: (\OBox A_{Li}) \\ 
&&  +\chi_L \: S^{(1)}(x,y) \: \xi^i \: \int_x^y dz\:
[0,1\:|\: 1]\: (\Pdd \OBox A_{Li}) \\ 
&&  +\chi_L \: S^{(1)}(x,y) \: \int_x^y dz\: [0,2\:|\: 0]\: (\OBox 
\Aslsh_L) \\ 
&&  -2 \chi_L \: S^{(1)}(x,y) \: \int_x^y dz\: [0,0\:|\: 1]\:
(\Pdd \partial^i A_{Li}) \\ 
&&  +\frac{1}{2} \: \chi_L \: S^{(1)}(x,y) \: \xi \slsh  \:
\int_x^y dz\: [0,0\:|\: 1]\: (\Pdd \OBox \Aslsh_L) \\ 
&&  -\chi_L \: S^{(1)}(x,y) \: \xi \slsh  \: \int_x^y dz\: [1,0\:|\: 1]\:
(\partial^i \OBox A_{Li}) \\ 
&&  +\frac{1}{4}\: \chi_L \: S^{(1)}(x,y) \: \xi \slsh  \: \xi^i \:
\int_x^y dz\: [0,0\:|\: 2]\: (\OBox^2 A_{Li}) \;\;\; , \label{l:A0e}
\end{eqnarray}
where we used the abbreviation $\xi \equiv (y-x)$.
This formula has the disadvantage that it contains ordinary partial 
derivatives of the chiral potential; it would be better for physical 
applications to work instead with the Yang-Mills field tensor and 
the Yang-Mills current. Therefore, we introduce left and right handed 
gauge-covariant derivatives $D^{L\!/\!R}$,
\[ D^L_j \;=\; \frac{\partial}{\partial x^j} \:-\: i A_{L j} 
\;\;\;,\spc D^R_j \;=\; \frac{\partial}{\partial x^j} \:-\: i A_{R j} 
\;\;\; , \]
and define the corresponding field tensor and current as usual by 
the commutators
\begin{equation}
F^c_{jk} \;=\; i \left[ D^c_j,\:D^c_k \right] \;\;\;,\spc
j^c_l \;=\; \left[ D^{c\:k},\: F^c_{lk} \right] \spc {\mbox{($c=L$ or $R$)}}.
        \label{l:A1}
\end{equation}
In the Abelian case of a single Dirac sea (i.e.\ $f=1$), (\ref{l:A1}) 
reduces to the familiar formulas for the electromagnetic field tensor and
current,
\[ F^c_{jk} \;=\; \partial_j A_{c\:k} - \partial_k A_{c\:j} \;\;\;,\spc
j^c_l \;=\; \partial_{lk} A^k_c - \OBox A_{c\:l} \;\;\; . \]
Notice, however, that in the general case of a system of Dirac seas, 
(\ref{l:A1}) involves quadratic and cubic terms in the potential.

By substituting (\ref{l:A1}) into (\ref{l:A00})--(\ref{l:A0e}) and 
manipulating the line integrals with partial integrations, one can 
rewrite the phase-free contribution in a way where the linear terms in the
potential are gauge invariant. For example, we can combine (\ref{l:A00}), 
(\ref{l:A01}), and (\ref{l:A02}) by transforming the line integrals as
\begin{eqnarray}
\lefteqn{ \xi^k \int_x^y dz \:[0,1 \:|\: 0] \: (\Pdd A_{Lk}) \;=\;
\xi^k \int_x^y dz \:[0,1 \:|\: 0] \: (\gamma^j F^L_{jk} \:+\: 
\partial_k \Aslsh_L) \:+\: {\cal{O}}(A_L^2) } \nonumber \\
&=& \xi^k \int_x^y dz \:[0,1 \:|\: 0] \: \gamma^j F^L_{jk} \:-\: \Aslsh_L(x) 
\:+\: \int_x^y dz \:[0,0 \:|\: 0] \:\Aslsh_L \:+\: 
{\cal{O}}(A_L^2) \;\;\; .
        \label{l:A2a}
\end{eqnarray}
However, this procedure yields (in the non-Abelian case) quadratic 
and cubic terms in the potential which are {\em{not}} gauge 
invariant. Fortunately, these gauge-dependent terms are all 
compensated by corresponding contributions of the higher order 
Feynman diagrams. More generally, it turns out that, after summing up 
the perturbation series for the chiral perturbation, we can arrange a 
gauge invariant phase-free contribution for which the insertion rules 
of Def.\ \ref{l:def2} and the statement of Theorem \ref{l:thm3} hold. This 
is not astonishing in view of the behavior (\ref{l:p1}) 
of the Green's functions under local gauge transformations; we 
verified it explicitly term by term for all following formulas.
We now list the gauge invariant phase-free contribution to the order 
${\cal{O}}((y-x)^2)$ on the light cone. For simplicity, we omit all 
third order terms in the potential which are of the order 
${\cal{O}}((y-x)^0)$ on the light cone and which have a prefactor $\xi 
\slsh$ (the combinatorics of the tensor contractions leads to many 
such terms, but they are not very instructive here).
\begin{eqnarray*}
\lefteqn{ \chi_L \sum_{k=0}^\infty ((-s \:(\chi_L \Aslsh_R + \chi_R 
\Aslsh_L))^k \:s)(x,y) } \\
&&\!\!\!\!\!\!\!\!\!\!\!\!
\stackrel{\mbox{\footnotesize{phase-free}}}{\asymp}\; 
{\cal{O}}((y-x)^2) \:+\: \xi\slsh \:A^i_L A^j_L A^k_L  \:{\cal{O}}((y-x)^0) \\
&&  +\chi_L \:S^{(0)}(x,y)\: \xi^i \int_x^y dz\: [0,1\:|\: 0]\: \gamma^l F^L_{li} \\ 
&&  +\frac{1}{4}\:\chi_L \:S^{(0)}(x,y)\: \xi \slsh\: \int_x^y dz\: [0,0\:|\: 0]\:
\gamma^j \gamma^k\:F^L_{jk} \\ 
&&  -\frac{1}{2}\:\chi_L \:S^{(0)}(x,y)\: \xi \slsh\: \xi^i \int_x^y dz\:
[0,0\:|\: 1]\: j^L_i \\ 
&&  +\chi_L \:S^{(1)}(x,y)\: \xi^i \int_x^y dz\: [0,1\:|\: 1]\: (\Pdd j^L_i) \\ 
&&  +\chi_L \:S^{(1)}(x,y)\: \int_x^y dz\: [0,2\:|\: 0]\: j^L_k \:\gamma^k \\ 
&&  -\frac{1}{2}\:\chi_L \:S^{(1)}(x,y)\: \xi \slsh\: \int_x^y dz\:
[0,0\:|\: 1]\: (\Pdd j^L_k) \:\gamma^k \\ 
&&  -\frac{1}{4}\:\chi_L \:S^{(1)}(x,y)\: \xi \slsh\: \xi^i \int_x^y dz\:
[0,0\:|\: 2]\: (\OBox j^L_i) \\ 
&&  -i\chi_L \:S^{(0)}(x,y)\: \xi \slsh\: \xi_i \xi^j \int_x^y dz_1\: 
[0,1\:|\: 1]\:F^L_{kj} \int_{z_1}^y dz_2\: [0,1\:|\: 0]\: F_L^{ki} \\ 
&&  +i\chi_L \:S^{(1)}(x,y)\: \xi^i \xi^j \int_x^y dz_1\: [0,3\:|\: 0]\: \gamma^k 
\:F^L_{kj} \int_{z_1}^y dz_2\: [0,0\:|\: 1]\: j^L_i \\ 
&&  +i\chi_L \:S^{(1)}(x,y)\: \xi^i \xi^j \int_x^y dz_1\: [0,2\:|\: 1]\: j^L_j
\int_{z_1}^y dz_2\: [0,1\:|\: 0]\: \gamma^l \:F^L_{li} \\ 
&&  -2i\chi_L \:S^{(1)}(x,y)\: \xi_i \xi^j \int_x^y dz_1\: [0,2\:|\: 1]\: F^L_{mj}
\int_{z_1}^y dz_2\: [0,2\:|\: 0]\: (\Pdd F_L^{mi}) \\ 
&&  -2i\chi_L \:S^{(1)}(x,y)\: \xi_i \xi^j \int_x^y dz_1\: [0,2\:|\: 
1]\: (\Pdd F^L_{kj}) \int_{z_1}^y dz_2\: [0,1\:|\: 0]\: F_L^{ki} \\ 
&&  +i\chi_L \:S^{(1)}(x,y)\: \xi^i \xi^j \int_x^y dz_1\: [0,2\:|\: 1]\: \gamma^k F^L_{kj}
\int_{z_1}^y dz_2\: [0,2\:|\: 0]\: j^L_i \\ 
&&  -\frac{i}{2}\:\chi_L \:S^{(1)}(x,y)\: \xi^i \int_x^y dz_1\: [0,2\:|\: 0]\:
\gamma^j F^L_{ji} \int_{z_1}^y dz_2\: [0,0\:|\: 0]\: \gamma^k \gamma^l F^L_{kl} \\ 
&&  -\frac{i}{2}\:\chi_L \:S^{(1)}(x,y)\: \xi^i \int_x^y dz_1\: [0,2\:|\: 0]\:
\gamma^j \gamma^k F^L_{jk} \int_{z_1}^y dz_2\: [0,1\:|\: 0]\: \gamma^l F^L_{li} \\ 
&&  +2i\chi_L \:S^{(1)}(x,y)\: \xi_i \int_x^y dz_1\: [0,3\:|\: 0]\: \gamma^j F^L_{jk}
\int_{z_1}^y dz_2\: [0,1\:|\: 0]\: F_L^{ki} \\ 
&&  -2i\chi_L \:S^{(1)}(x,y)\: \xi^j \int_x^y dz_1\: [0,1\:|\: 1]\: F^L_{ij}
\int_{z_1}^y dz_2\: [0,1\:|\: 0]\: \gamma_k F_L^{ki} \\ 
&&  -\frac{i}{2}\:\chi_L \:S^{(1)}(x,y)\: \xi \slsh\: \xi^j \xi^k \int_x^y dz_1\:
[0,2\:|\: 1]\: j^L_j \int_{z_1}^y dz_2\: [0,0\:|\: 1]\: j^L_k \\ 
&&  -\frac{i}{2}\:\chi_L \:S^{(1)}(x,y)\: \xi \slsh\: \xi^j \xi^k
\int_x^y dz_1\: [0,1\:|\: 2]\: j^L_j \int_{z_1}^y dz_2\: [0,2\:|\: 0]\: j^L_k \\ 
&&  +i\chi_L \:S^{(1)}(x,y)\: \xi \slsh\: \xi^i \xi^j \int_x^y dz_1\:
[0,2\:|\: 1]\: F^L_{kj} \int_{z_1}^y dz_2\: [0,1\:|\: 1]\: (\partial^k  
j^L_i) \\
&&  +i\chi_L \:S^{(1)}(x,y)\: \xi \slsh\: \xi^i \xi^j \int_x^y dz_1\:
[0,1\:|\: 2]\: F^L_{kj} \int_{z_1}^y dz_2\: [0,3\:|\: 0]\: (\partial^k  j^L_i) \\ 
&&  -i\chi_L \:S^{(1)}(x,y)\: \xi \slsh\: \xi_i \xi^j \int_x^y dz_1\:
[0,1\:|\: 2]\: (\partial_k  F^L_{lj}) \int_{z_1}^y dz_2\:
[0,2\:|\: 0]\: (\partial^k  F_L^{li}) \\ 
&&  +\frac{i}{4}\:\chi_L \:S^{(1)}(x,y)\: \xi \slsh\: \xi^i \int_x^y dz_1\:
[0,2\:|\: 0]\: \gamma^j \gamma^k F^L_{jk} \int_{z_1}^y dz_2\:
[0,0\:|\: 1]\: j^L_i \\ 
&&  +\frac{i}{4}\:\chi_L \:S^{(1)}(x,y)\: \xi \slsh\: \xi^i \int_x^y dz_1\:
[0,1\:|\: 1]\: \gamma^j \gamma^k F^L_{jk} \int_{z_1}^y dz_2\:
[0,2\:|\: 0]\: j^L_i \\ 
&&  +\frac{i}{4}\:\chi_L \:S^{(1)}(x,y)\: \xi \slsh\: \xi^i \int_x^y dz_1\:
[0,1\:|\: 1]\: j^L_i \int_{z_1}^y dz_2\:
[0,0\:|\: 0]\: \gamma^j \gamma^k F^L_{jk} \\ 
&&  -\frac{i}{2}\:\chi_L \:S^{(1)}(x,y)\: \xi \slsh\: \xi^i \int_x^y dz_1\:
[0,1\:|\: 1]\: \gamma^j \gamma^k (\partial^l  F^L_{jk}) \int_{z_1}^y dz_2\:
[0,1\:|\: 0]\: F^L_{li} \\ 
&&  -\frac{i}{2}\:\chi_L \:S^{(1)}(x,y)\: \xi \slsh\: \xi^i \int_x^y dz_1\:
[0,1\:|\: 1]\: F^L_{ji} \int_{z_1}^y dz_2\:
[0,1\:|\: 0]\: \gamma^k \gamma^l (\partial^j  F^L_{kl}) \\ 
&&  +i\chi_L \:S^{(1)}(x,y)\: \xi \slsh\: \xi^i \xi^j \int_x^y dz_1\:
[0,1\:|\: 2]\: (\partial^k  j^L_j) \int_{z_1}^y dz_2\:
[0,1\:|\: 0]\: F^L_{ki} \\ 
&&  -i\chi_L \:S^{(1)}(x,y)\: \xi \slsh\: \xi^j \int_x^y dz_1\:
[0,1\:|\: 1]\: F^L_{ij} \int_{z_1}^y dz_2\:
[0,0\:|\: 1]\: j_L^i \\ 
&&  -i\chi_L \:S^{(1)}(x,y)\: \xi \slsh\: \xi^j \int_x^y dz_1\:
[0,0\:|\: 2]\: F^L_{ij} \int_{z_1}^y dz_2\:
[0,2\:|\: 0]\: j_L^i \\ 
&&  +i\chi_L \:S^{(1)}(x,y)\: \xi \slsh\: \xi^i \int_x^y dz_1\:
[0,1\:|\: 1]\: j_L^j \int_{z_1}^y dz_2\:
[0,1\:|\: 0]\: F^L_{ji} \\ 
&&  +2i\chi_L \:S^{(1)}(x,y)\: \xi \slsh\: \xi_i \int_x^y dz_1\:
[0,2\:|\: 1]\: F^L_{kl} \int_{z_1}^y dz_2\:
[0,2\:|\: 0]\: (\partial^k  F_L^{li}) \\ 
&&  -2i\chi_L \:S^{(1)}(x,y)\: \xi \slsh\: \xi^j \int_x^y dz_1\:
[0,0\:|\: 2]\: (\partial_k  F^L_{lj}) \int_{z_1}^y dz_2\:
[0,1\:|\: 0]\: F_L^{kl} \\ 
&&  -\frac{i}{8}\:\chi_L \:S^{(1)}(x,y)\: \xi \slsh\: \int_x^y dz_1\:
[0,1\:|\: 0]\: \gamma^j \gamma^k F^L_{jk} 
\int_{z_1}^y dz_2\: [0,0\:|\: 0]\: \gamma^l \gamma^m F^L_{lm} \\ 
&&  +3i\chi_L \:S^{(1)}(x,y)\: \xi \slsh\: \int_x^y dz_1\:
[0,1\:|\: 1]\: F^L_{kl} \int_{z_1}^y dz_2\: [0,1\:|\: 0]\: F_L^{kl} \\ 
&&  -2 \chi_L \:S^{(1)}(x,y)\: \xi_i \xi^j \xi^k \int_x^y dz_1\:
[0,4\:|\: 0]\: \gamma^l F^L_{lk} \int_{z_1}^y dz_2\:
[0,1\:|\: 1]\: F^L_{mj} \int_{z_2}^y dz_3\:[0,1\:|\: 0]\: F_L^{mi} \\ 
&&  -2\chi_L \:S^{(1)}(x,y)\: \xi_i \xi^j \xi^k \int_x^y dz_1\:
[0,3\:|\: 1]\: \gamma^l F^L_{lk} \int_{z_1}^y dz_2\:
[0,3\:|\: 0]\: F^L_{mj} \int_{z_2}^y dz_3\:[0,1\:|\: 0]\: F_L^{mi} \\ 
&&  -2\chi_L \:S^{(1)}(x,y)\: \xi_i \xi^j \xi^k \int_x^y dz_1\:
[0,3\:|\: 1]\: F^L_{mk} \int_{z_1}^y dz_2\:
[0,3\:|\: 0]\: \gamma^l F^L_{lj} \int_{z_2}^y dz_3\:[0,1\:|\: 0]\: F_L^{mi} \\ 
&&  -2\chi_L \:S^{(1)}(x,y)\: \xi^i \xi_j \xi^k \int_x^y dz_1\:
[0,3\:|\: 1]\: F^L_{mk} \int_{z_1}^y dz_2\:
[0,3\:|\: 0]\: F_L^{mj} \int_{z_2}^y dz_3\:
[0,1\:|\: 0]\: \gamma^l F^L_{li}
\end{eqnarray*}
We call this formulation of the phase-free contributions purely in 
terms of the Yang-Mills field tensor and the Yang-Mills current 
the {\em{gauge invariant form}} of the light-cone expansion.

For clarity, we mention a subtlety in the transformation to the 
gauge invariant form: The gauge invariant phase-free 
contribution implicitly contains tangential derivatives of the chiral 
potential (as one sees by writing out $F^c$ and $j^c$ in 
in terms of the chiral potentials). Thus it is not a phase-free contribution
in correspondence with Def.\ \ref{l:def_pf}; as a consequence,
it is not obvious that the 
statement of Theorem \ref{l:thm3} holds. In other words, one must be very 
careful when transforming the line integrals in order to ensure that 
the insertion rules of Def.\ \ref{l:def2} and Theorem \ref{l:thm3} remain 
valid. A safe method is to insert all ordered exponentials into 
the line integrals before performing the partial integrations. According to
Theorem \ref{l:thm3}, the phase-inserted formulas coincide with the light-cone 
expansion of the Green's functions. Therefore the partial 
integrations become identical transformations of the Green's functions; 
we need not worry about the insertion rules. In the final step, we
again take out the ordered exponentials from the line integrals, 
verifying that they are still in accordance with Def.\ \ref{l:def2}. In 
the example of the line integral (\ref{l:A2a}), the partial integration 
of the phase-inserted contribution gives
\begin{eqnarray}
\lefteqn{ \xi^k \int_x^y dz \:[0,1 \:|\: 0] \: \Pe^{-i \int_x^z A^l_L 
\:(z-x)_l} \:\Pdd A_{L k}(z) \:\Pe^{-i \int_z^y A^m_L \:(y-z)_m} } 
\nonumber \\
&=& \xi^k \int_x^y dz \:[0,1 \:|\: 0] \: \Pe^{-i \int_x^z A^l_L 
\:(z-x)_l} \left( \gamma^j F^L_{jk} + i \left[ \Aslsh_L, \: A_{Lk} 
\right] + \partial_k \Aslsh_L \right)_{|z}
\Pe^{-i \int_z^y A^m_L \:(y-z)_m} \nonumber \\
&=& \xi^k \int_x^y dz \:[0,1 \:|\: 0] \: \Pe^{-i \int_x^z A^l_L 
\:(z-x)_l} \left( \gamma^j F^L_{jk} \:+\: i \left[ \Aslsh_L, \: A_{Lk} 
\right]\right)_{|z} \:\Pe^{-i \int_z^y A^m_L \:(y-z)_m} \\
&&-\Aslsh_L(x) \:\Pe^{-i \int_x^y A^m_L \:(y-z)_x} \\
&&+\int_x^y dz \:[0,0 \:|\: 0] \: \Pe^{-i \int_x^z A^l_L 
\:(z-x)_l} \Aslsh_L(z) \:\Pe^{-i \int_z^y A^m_L \:(y-z)_m} \\
&&+\xi^k \int_x^y dz \:[0,1 \:|\: 0] \: \Pe^{-i \int_x^z A^l_L 
\:(z-x)_l}  \:i \left[ A_{Lk}, \: \Aslsh_L 
\right](z) \:\Pe^{-i \int_z^y A^m_L \:(y-z)_m} \;\;\;\; .\label{l:A8} 
\end{eqnarray}
Thus the correct transformation of the phase-free contribution is
\[ \xi^k \int_x^y dz \:[0,1 \:|\: 0] \: (\Pdd A_{Lk}) 
\;\stackrel{\mbox{\footnotesize{phase-free}}}{\asymp}\;
\xi^k \int_x^y dz \:[0,1 \:|\: 0] \: \gamma^j F^L_{jk} \:+\: 
\int_x^y dz \:[0,0 \:|\: 0] \:\Aslsh_L \:-\: \Aslsh_L(x) \;\;\; . \]
Notice that (\ref{l:A8}) are the derivative terms 
of the ordered exponentials; they get lost if the phase-free 
contribution is transformed in a naive way.

It remains to consider the scalar/pseudoscalar 
perturbation; i.e., we must study how the dynamic mass matrices 
$Y_{L\!/\!R}(x)$ show up in the light-cone expansion. We begin with 
the case of a single mass matrix. To first order in the external 
potential, the corresponding Feynman diagram has the light-cone 
expansion
\begin{eqnarray}
\lefteqn{ \chi_L \:m\:(-s \:(-\chi_L Y_R - \chi_R 
Y_L) \:s)(x,y) \;\stackrel{\mbox{\footnotesize{phase-free}}}{\asymp}\; 
{\cal{O}}((y-x)^2) } \nonumber \\
&& +\frac{1}{2}\:\chi_L\: m  \:S^{(0)}(x,y)\: \xi \slsh\: \int_x^y dz\:
[0,0\:|\: 0]\: (\Pdd Y_L) \nonumber \\ 
&& +\chi_L\: m  \:S^{(0)}(x,y)\: Y_L(x) \nonumber \\ 
&& +\chi_L\: m  \:S^{(1)}(x,y)\: \int_x^y dz\: [0,1\:|\: 0]\: (\OBox 
Y_L) \nonumber \\ 
&& +\frac{1}{2}\:\chi_L\: m  \:S^{(1)}(x,y)\: \xi \slsh\: \int_x^y dz\:
[0,0\:|\: 1]\: (\Pdd \OBox Y_L) \;\;\; . \label{l:A4a}
\end{eqnarray}
Similar to (\ref{l:A00})--(\ref{l:A0e}), this formula involves partial 
derivatives of the potential, which is not a gauge invariant 
formulation. Since the chirality changes at every mass matrix (see 
e.g.\ Def.\ \ref{l:def2}), the correct way to make the light-cone 
expansion gauge invariant is to work with the {\em{gauge-covariant 
mass derivatives}} $D$, $\Delta$ given by
\begin{eqnarray*}
(D_i Y_L) &=& D^L_i \:Y_L \:-\: Y_L \:D^R_i \;=\; (\partial_i Y_L) 
\:-\: i A_{L i}\:Y_L \:+\: i Y_L \:A_{R i} \\
(\Delta Y_L) &=& D_i (D^i Y_L) \;=\; D^L_i \:(D^i Y_L) \:-\: (D^i 
Y_L)\: D^R_i \;\;\; ,
\end{eqnarray*}
and similarly for $Y_R$ and higher derivatives. 
Rewriting (\ref{l:A4a}) with gauge-covariant mass derivatives yields 
additional terms which are linear or quadratic in the chiral 
potentials and which are {\em{not}} gauge invariant. But, similar as 
described for the chiral perturbation above, all these terms cancel if 
the sum over the perturbation series for the chiral potentials is 
carried out. To the order ${\cal{O}}((y-x)^2)$ on the light cone, we 
obtain in this way the gauge invariant phase-free contribution
\begin{eqnarray*}
\lefteqn{ \chi_L \:m\: \sum_{n_1, n_2=0}^\infty ((-s \:(\chi_L \Aslsh_R + 
\chi_R \Aslsh_L))^{n_1} \:s\: (\chi_L Y_R + \chi_R Y_L) \:s\:
((-\chi_L \Aslsh_R - \chi_R \Aslsh_L) \:s)^{n_2})(x,y) } \\
&=& \chi_L \sum_{n_1, n_2=0}^\infty ((-s \:\Aslsh_L)^{n_1} \:s\:
Y_L \:s\: (-\Aslsh_R \:s)^{n_2})(x,y)
\;\stackrel{\mbox{\footnotesize{phase-free}}}{\asymp}\; 
{\cal{O}}((y-x)^2) \\
&& +\frac{1}{2}\:\chi_L\: m  \:S^{(0)}(x,y)\: \xi \slsh\: \int_x^y dz\:
[0,0\:|\: 0]\: \gamma^j \:(D_j Y_L) \\ 
&& +\chi_L\: m  \:S^{(0)}(x,y)\: Y_L(x) \\ 
&& +\chi_L\: m  \:S^{(1)}(x,y)\: \int_x^y dz\: [0,1\:|\: 0]\: (\Delta Y_L) \\ 
&& +\frac{1}{2}\:\chi_L\: m  \:S^{(1)}(x,y)\: \xi \slsh\: \int_x^y dz\:
[0,0\:|\: 1]\: \gamma^j \:(D_j \Delta Y_L) \\
&&  +\frac{i}{2} \:\chi_L\: m  \:S^{(1)}(x,y)\: \xi \slsh\: \xi^i \int_x^y dz_1\:
[0,1\:|\: 1]\: j^L_i \int_{z_1}^y dz_2\:
[0,0\:|\: 0]\: \gamma^j \:(D_j Y_L) \\ 
&&  -i\chi_L\: m  \:S^{(1)}(x,y)\: \xi \slsh\: \xi^i \int_x^y dz_1\:
[0,1\:|\: 1]\: F^L_{ji} \int_{z_1}^y dz_2\:
[0,1\:|\: 0]\: \gamma_k \:(D^{kj} Y_L) \\ 
&&  -\frac{i}{4}\:\chi_L\: m  \:S^{(1)}(x,y)\: \xi \slsh\: \int_x^y dz_1\:
[0,1\:|\: 0]\: \gamma^k \gamma^l F^L_{kl} 
\int_{z_1}^y dz_2\: [0,0\:|\: 0]\: \gamma^j \:(D_j Y_L) \\ 
&&  -i\chi_L\: m  \:S^{(1)}(x,y)\: \xi^i \int_x^y dz_1\:
[0,2\:|\: 0]\: \gamma^k F^L_{ki} \int_{z_1}^y dz_2\:
[0,0\:|\: 0]\: \gamma^j \:(D_j Y_L) \\ 
&&  -\frac{i}{2}\:\chi_L\: m  \:S^{(1)}(x,y)\: \int_x^y dz\:
[0,1\:|\: 0]\: \gamma^i \gamma^j F^L_{ij} \:Y_L \\ 
&&  +\frac{i}{2}\:\chi_L\: m  \:S^{(1)}(x,y)\: \xi \slsh\: \int_x^y dz\:
[0,0\:|\: 1]\: j^L_k \:\gamma^k\: Y_L \\ 
&&  -i\chi_L\: m  \:S^{(1)}(x,y)\: \xi \slsh\: \int_x^y dz\:
[0,0\:|\: 1]\: \gamma^j F^L_{ij} (D^i  Y_L) \\ 
&&  +\frac{i}{2}\:\chi_L\: m  \:S^{(1)}(x,y)\: \xi \slsh\: \xi^i \int_x^y dz_1\:
[0,2\:|\: 0]\: \gamma^j \:(D_j Y_L) \int_{z_1}^y dz_2\:
[0,0\:|\: 1]\: j^R_i \\ 
&&  +\frac{i}{2}\:\chi_L\: m  \:S^{(1)}(x,y)\: \xi \slsh\: \xi^i \int_x^y dz_1\:
[0,1\:|\: 1]\: \gamma^j \:(D_j Y_L) \int_{z_1}^y dz_2\:
[0,2\:|\: 0]\: j^R_i \\ 
&&  -i\chi_L\: m  \:S^{(1)}(x,y)\: \xi \slsh\: \xi^i \int_x^y dz_1\:
[0,1\:|\: 1]\: \gamma_k \:(D^{kj}  Y_L) \int_{z_1}^y dz_2\:
[0,1\:|\: 0]\: F^R_{ji} \\ 
&&  -1\frac{i}{4}\:\chi_L\: m  \:S^{(1)}(x,y)\: \xi \slsh\: \int_x^y dz_1\:
[0,1\:|\: 0]\: \gamma^j \:(D_j Y_L) \int_{z_1}^y dz_2\:
[0,0\:|\: 0]\: \gamma^k \gamma^l F^R_{kl} \\ 
&&  +i\chi_L\: m  \:S^{(1)}(x,y)\: \xi^i \int_x^y dz_1\:
[0,2\:|\: 0]\: \gamma^j \:(D_j Y_L) \int_{z_1}^y dz_2\:
[0,1\:|\: 0]\: \gamma^k F^R_{ki} \\ 
&&  -2i\chi_L\: m  \:S^{(1)}(x,y)\: \xi^i \int_x^y dz_1\:
[0,2\:|\: 0]\: (D^j  Y_L) \int_{z_1}^y dz_2\:
[0,1\:|\: 0]\: F^R_{ji} \\ 
&&  +i\chi_L\: m  \:S^{(1)}(x,y)\: \xi \slsh\: \int_x^y dz\:
[0,0\:|\: 1]\: (D^j  Y_L) \:\gamma^i F^R_{ji} \\ 
&&  -\frac{i}{2}\:\chi_L\: m  \:S^{(1)}(x,y)\: \xi \slsh\: \int_x^y dz\:
[0,0\:|\: 1]\: Y_L \:j^R_k \gamma^k \\ 
&&  +i\chi_L\: m  \:S^{(1)}(x,y)\: \xi^i\: Y_L \int_x^y dz\:
[0,0\:|\: 1]\: j^R_i \\ 
&&  +\frac{i}{2}\:\chi_L\: m  \:S^{(1)}(x,y)\: \int_x^y dz\:
[0,1\:|\: 0]\: Y_L \:\gamma^j \gamma^k F^R_{jk} \\ 
&&  -\frac{i}{2}\:\chi_L\: m  \:S^{(1)}(x,y)\: Y_L \int_x^y dz\:
[0,0\:|\: 0]\: \gamma^j \gamma^k F^R_{jk} \\ 
&&  +\chi_L\: m  \:S^{(1)}(x,y)\: \xi \slsh\: \xi_i \xi^j \int_x^y dz_1\:
[0,2\:|\: 1]\: F^L_{kj} \int_{z_1}^y dz_2\:
[0,2\:|\: 0]\: F_L^{ki} \int_{z_2}^y
[0,0\:|\: 0]\: \gamma^l \:(D_l Y_L) \\
&&  +\chi_L\: m  \:S^{(1)}(x,y)\: \xi \slsh\: \xi^j \int_x^y dz_1\:
[0,1\:|\: 1]\: F^L_{mj} \int_{z_1}^y dz_2\:
[0,1\:|\: 0]\: \gamma_k F_L^{mk} \:Y_L \\ 
&&  +\chi_L\: m  \:S^{(1)}(x,y)\: \xi \slsh\: \xi_i \xi^j \int_x^y dz_1\:
[0,2\:|\: 1]\: F^L_{kj} \int_{z_1}^y dz_2\:
[0,2\:|\: 0]\: \gamma^l \:(D_l Y_L) \int_{z_2}^y
[0,1\:|\: 0]\: F_R^{ki} \\
&&  +\chi_L\: m  \:S^{(1)}(x,y)\: \xi \slsh\: \xi^i \int_x^y dz_1\:
[0,1\:|\: 1]\: F^L_{ki} \int_{z_1}^y dz_2\:
[0,1\:|\: 0]\: Y_L \:\gamma_j F_R^{kj} \\ 
&&  +\chi_L\: m  \:S^{(1)}(x,y)\: \xi \slsh\: \xi_i \int_x^y dz_1\:
[0,1\:|\: 1]\: \gamma^j F^L_{kj} \:Y_L \int_{z_1}^y dz_2\:
[0,1\:|\: 0]\: F_R^{ki} \\ 
&&  +\chi_L\: m  \:S^{(1)}(x,y)\: \xi \slsh\: \xi_i \xi^j \int_x^y dz_1\:
[0,3\:|\: 0]\: \gamma^l \:(D_l Y_L) \int_{z_1}^y dz_2\:
[0,1\:|\: 1]\: F^R_{kj} \int_{z_2}^y
[0,1\:|\: 0]\: F_R^{ki} \\ 
&&  +\chi_L\: m  \:S^{(1)}(x,y)\: \xi \slsh\: \xi_i \xi^j \int_x^y dz_1\:
[0,2\:|\: 1]\: \gamma^l \:(D_l Y_L) \int_{z_1}^y dz_2\:
[0,3\:|\: 0]\: F^R_{kj} \int_{z_2}^y
[0,1\:|\: 0]\: F_R^{ki} \\ 
&&  +\chi_L\: m  \:S^{(1)}(x,y)\: \xi \slsh\: \xi_i \int_x^y dz_1\:
[0,1\:|\: 1]\: Y_L \:\gamma^j F^R_{mj} \int_{z_1}^y dz_2\:
[0,1\:|\: 0]\: F_R^{mi} \\ 
&&  -2\chi_L\: m  \:S^{(1)}(x,y)\: \xi_i \xi^j \:Y_L \int_x^y dz_1\:
[0,1\:|\: 1]\: F^R_{kj} \int_{z_1}^y dz_2\:
[0,1\:|\: 0]\: F_R^{ki} \;\;\; .
\end{eqnarray*}

The higher orders in the mass matrices are treated similarly. To 
the order ${\cal{O}}((y-x)^2)$ on the light cone, only the terms up to 
fourth order in $m \:Y_{L\!/\!R}$ contribute (see (\ref{l:29n})).
One gets the formulas
\begin{eqnarray*}
\lefteqn{ \chi_L \:m^2\: \sum_{n_1, n_2, n_3=0}^\infty
((-s \:\Aslsh_L)^{n_1} \:s\: Y_L \:s\: (-\Aslsh_R \:s)^{n_2} \:Y_R\:s\:
(-\Aslsh_R \:s)^{n_3})(x,y) } \\
&&\!\!\!\!\!\!\!\!\!\!\!\!
\stackrel{\mbox{\footnotesize{phase-free}}}{\asymp}\;
\frac{i}{2}\: \chi_L\: m^2 \:S^{(0)}(x,y)\: \xi \slsh\: \int_x^y dz\:
[0,0\:|\: 0]\: Y_L \:Y_R \\ 
&&  -\frac{i}{2}\: \chi_L\: m^2 \:S^{(1)}(x,y)\: \xi \slsh\: \int_x^y dz_1\:
[0,1\:|\: 0]\: \gamma^j (D_j Y_L)
\int_{z_1}^y dz_2\: [0,0\:|\: 0]\: \gamma^k (D_k Y_R) \\ 
&&  +\frac{i}{2}\: \chi_L\: m^2 \:S^{(1)}(x,y)\: \xi \slsh\: \int_x^y dz\:
[0,0\:|\: 1]\: Y_L \:(\Delta Y_R) \\ 
&&  +i \chi_L\: m^2 \:S^{(1)}(x,y)\: \xi \slsh\: \int_x^y dz\:
[0,0\:|\: 1]\: (D_i  Y_L) \:(D^i  Y_R) \\ 
&&  +\frac{i}{2}\: \chi_L\: m^2 \:S^{(1)}(x,y)\: \xi \slsh\: \int_x^y dz\:
[0,0\:|\: 1]\: (\Delta Y_L)\: Y_R \\ 
&&  +i \chi_L\: m^2 \:S^{(1)}(x,y)\: \int_x^y dz\:
[0,1\:|\: 0]\: Y_L \:\gamma^j (D_j Y_R) \\ 
&&  +i \chi_L\: m^2 \:S^{(1)}(x,y)\: \int_x^y dz\:
[0,1\:|\: 0]\: \gamma^j (D_j Y_L)\: Y_R \\ 
&&  -i \chi_L\: m^2 \:S^{(1)}(x,y)\: Y_L \int_x^y dz\:
[0,0\:|\: 0]\: \gamma^j (D_j Y_R) \\ 
&&  -\frac{1}{2}\: \chi_L\: m^2 \:S^{(1)}(x,y)\: \xi \slsh\: \xi^i \int_x^y dz_1\:
[0,1\:|\: 1]\: j^L_i \int_{z_1}^y dz_2\: [0,0\:|\: 0]\: Y_L \:Y_R \\ 
&&  + \chi_L\: m^2 \:S^{(1)}(x,y)\: \xi \slsh\: \xi^i \int_x^y dz_1\:
[0,1\:|\: 1]\: F^L_{ji} 
\int_{z_1}^y dz_2\: [0,1\:|\: 0]\: Y_L \:(D^j  Y_R) \\ 
&&  + \chi_L\: m^2 \:S^{(1)}(x,y)\: \xi \slsh\: \xi^i \int_x^y dz_1\:
[0,1\:|\: 1]\: F^L_{ji} \int_{z_1}^y dz_2\: [0,1\:|\: 0]\: (D^j  
Y_L) \:Y_R \\ 
&&  +\frac{1}{4}\: \chi_L\: m^2 \:S^{(1)}(x,y)\: \xi \slsh\: \int_x^y dz_1\:
[0,1\:|\: 0]\: \gamma^j \gamma^k F^L_{jk} 
\int_{z_1}^y dz_2\: [0,0\:|\: 0]\: Y_L \:Y_R \\ 
&&  + \chi_L\: m^2 \:S^{(1)}(x,y)\: \xi^i \int_x^y dz_1\:
[0,2\:|\: 0]\: \gamma^j F^L_{ji} 
\int_{z_1}^y dz_2\: [0,0\:|\: 0]\: Y_L \:Y_R \\ 
&&  -\frac{1}{2}\: \chi_L\: m^2 \:S^{(1)}(x,y)\: \xi \slsh\: \xi^i \int_x^y dz_1\:
[0,2\:|\: 0]\: Y_L \:Y_R \int_{z_1}^y dz_2\: [0,0\:|\: 1]\: j^L_i \\ 
&&  -\frac{1}{2}\: \chi_L\: m^2 \:S^{(1)}(x,y)\: \xi \slsh\: \xi^i \int_x^y dz_1\:
[0,1\:|\: 1]\: Y_L \:Y_R \int_{z_1}^y dz_2\: [0,2\:|\: 0]\: j^L_i \\ 
&&  + \chi_L\: m^2 \:S^{(1)}(x,y)\: \xi \slsh\: \xi^i \int_x^y dz_1\:
[0,1\:|\: 1]\: Y_L \:(D^j  Y_R) \int_{z_1}^y dz_2\: [0,1\:|\: 0]\: F^L_{ji} \\ 
&&  + \chi_L\: m^2 \:S^{(1)}(x,y)\: \xi \slsh\: \xi^i \int_x^y dz_1\:
[0,1\:|\: 1]\: (D^j  Y_L)\: Y_R \int_{z_1}^y dz_2\:
[0,1\:|\: 0]\: F^L_{ji} \\ 
&&  +\frac{1}{4}\: \chi_L\: m^2 \:S^{(1)}(x,y)\: \xi \slsh\: \int_x^y dz_1\:
[0,1\:|\: 0]\: Y_L \:Y_R \int_{z_1}^y dz_2\:
[0,0\:|\: 0]\: \gamma^j \gamma^k F^L_{jk} \\ 
&&  + \chi_L\: m^2 \:S^{(1)}(x,y)\: \xi^i \int_x^y dz_1\:
[0,2\:|\: 0]\: Y_L \:Y_R \int_{z_1}^y dz_2\: [0,1\:|\: 0]\: \gamma^j 
F^L_{ji} \:+\: {\cal{O}}((y-x)^2) \\
\lefteqn{ \chi_L \:m^3\:\sum_{n_1, n_2, n_3, n_4=0}^\infty
((-s \:\Aslsh_L)^{n_1} \:s\: Y_L \:s\: (-\Aslsh_R \:s)^{n_2} \:Y_R\:s\:
(-\Aslsh_L \:s)^{n_3} \:Y_L\:s\: (-\Aslsh_R \:s)^{n_4})(x,y) } \\
&&\!\!\!\!\!\!\!\!\!\!\!\!
\stackrel{\mbox{\footnotesize{phase-free}}}{\asymp}\;
\chi_L\: m^3 \:S^{(1)}(x,y)\: Y_L \int_x^y dz\:
[0,0\:|\: 0]\: Y_R \:Y_L \\ 
&&  +\frac{1}{2}\: \chi_L\: m^3 \:S^{(1)}(x,y)\: \xi \slsh\: \int_x^y dz_1\:
[0,1\:|\: 0]\: Y_L \:Y_R \int_{z_1}^y dz_2\: [0,0\:|\: 0]\: \gamma^j 
(D_j Y_L) \\ 
&&  +\frac{1}{2}\: \chi_L\: m^3 \:S^{(1)}(x,y)\: \xi \slsh\: \int_x^y dz_1\:
[0,1\:|\: 0]\: \gamma^j (D_j Y_L) \int_{z_1}^y dz_2\: [0,0\:|\: 0]\:
Y_R \:Y_L \:+\: {\cal{O}}((y-x)^2) \\ 
\lefteqn{ \chi_L \:m^4\:\sum_{n_1, \ldots, n_5=0}^\infty
((-s \:\Aslsh_L)^{n_1} \:s\: Y_L \:s\: (-\Aslsh_R \:s)^{n_2} } \\
&& \hspace*{3cm} \times\:Y_R\:s\:(-\Aslsh_L \:s)^{n_3} \:Y_L\:s\:
(-\Aslsh_R \:s)^{n_4}\:Y_R \:s\: (-\Aslsh_L \:s)^{n_5})(x,y) \\
&&\!\!\!\!\!\!\!\!\!\!\!\!
\stackrel{\mbox{\footnotesize{phase-free}}}{\asymp}\;
\frac{i}{2}\: \chi_L\: m^4 \:S^{(1)}(x,y)\: \xi \slsh\:
\int_x^y dz_1\: [0,1\:|\: 0]\: Y_L \:Y_R 
\int_{z_1}^y dz_2\: [0,0\:|\: 0]\: Y_L \:Y_R \:+\: {\cal{O}}((y-x)^2)
\end{eqnarray*}
These classes of Feynman diagrams completely characterize the Green's 
functions to the order ${\cal{O}}((y-x)^2)$ on the light cone; notice 
that we only get a finite number of contributions.

\end{document}